\documentclass{article}

\usepackage{paper_style}
\usepackage{iftex}
\ifPDFTeX
  \usepackage[T1]{fontenc}
  \usepackage{times}
\else
  \usepackage{fontspec}
\fi
\usepackage{amsmath,amsfonts,bm}

\def\eqref#1{equation~\ref{#1}}
\def\Eqref#1{Equation~\ref{#1}}
\def\1{\bm{1}}

\DeclareMathAlphabet{\mathsfit}{\encodingdefault}{\sfdefault}{m}{sl}
\SetMathAlphabet{\mathsfit}{bold}{\encodingdefault}{\sfdefault}{bx}{n}

\usepackage{amssymb,mathtools}
\newtheorem{proposition}{Proposition}
\usepackage{booktabs,multirow}
\usepackage{algorithm}
\usepackage{algpseudocode}
\algrenewcommand{\algorithmicrequire}{\textbf{Inputs:}}
\algrenewcommand{\algorithmicensure}{\textbf{Outputs:}}
\usepackage{graphicx}
\usepackage{xcolor}
\usepackage{wrapfig}
\usepackage{placeins}
\usepackage{needspace}
\usepackage{tikz}
\usepackage{pgfplots}
\usepackage{pgfplotstable}
\usepgfplotslibrary{groupplots}
\pgfplotsset{compat=1.18}
\usepackage{xspace}
\usepackage{hyperref}
\usepackage{url}
\input{method_name.tex}
\ExplSyntaxOn
\cs_new:Npn \__paper_missing_metric_p:n #1
  {
    \str_if_eq_p:ee
      { \tl_trim_spaces_apply:nN {#1} \tl_to_str:n } { -- }
  }
\cs_new:Npn \__paper_result_row:w #1 & #2 & #3 & #4 & #5 & #6 & #7 & #8 \q_stop
  {
    \bool_lazy_all:nTF
      {
        { \__paper_missing_metric_p:n {#2} }
        { \__paper_missing_metric_p:n {#3} }
        { \__paper_missing_metric_p:n {#4} }
        { \__paper_missing_metric_p:n {#5} }
        { \__paper_missing_metric_p:n {#6} }
        { \__paper_missing_metric_p:n {#8} }
      }
      { }
      { #1 & #2 & #3 & #4 & #5 & #6 & #7 & #8 \\ }
  }
\NewExpandableDocumentCommand{\ResultRow}{m}
  { \__paper_result_row:w #1 \q_stop }
\cs_new:Npn \__paper_missing_runtime_p:n #1
  {
    \bool_lazy_or_p:nn
      { \__paper_missing_metric_p:n {#1} }
      {
        \str_if_eq_p:ee
          { \tl_trim_spaces_apply:nN {#1} \tl_to_str:n } { N/A }
      }
  }
\cs_new:Npn \__paper_runtime_row:w #1 & #2 & #3 & #4 & #5 \q_stop
  {
    \bool_lazy_all:nTF
      {
        { \__paper_missing_runtime_p:n {#2} }
        { \__paper_missing_runtime_p:n {#3} }
        { \__paper_missing_runtime_p:n {#4} }
        { \__paper_missing_runtime_p:n {#5} }
      }
      { }
      { #1 & #2 & #3 & #4 & #5 \\ }
  }
\NewExpandableDocumentCommand{\RuntimeRow}{m}
  { \__paper_runtime_row:w #1 \q_stop }
\ExplSyntaxOff


\newcommand{\papertitle}{\methodname{}: Joint Token Watermarking for Diffusion Language Models via Multi-Marginal Optimal Transport}
\title{\papertitle}
\author{%
  Omer Luxembourg\\
  \texttt{omerlux@post.bgu.ac.il}
  \And
  Dor Tsur\\
  \texttt{dortsur93@gmail.com}
  \And
  Haim H. Permuter\\
  \texttt{haimp@bgu.ac.il}
}

\begin{document}
\maketitle

\begin{abstract}

Watermarks for autoregressive language models are typically applied one token at a time. Diffusion language models (DLMs) generate text in parallel by successively updating multiple token positions. Most existing DLM watermarks focus on side-information generation, leaving the other watermarking components similar to their autoregressive counterparts. In this work, we leverage the parallel nature of DLMs to jointly watermark multiple tokens. By jointly watermarking token blocks, we improve detection without increasing the watermark's distortion.
To do that, we introduce \methodname (\textbf{T}ransport-based \textbf{I}nteraction for \textbf{D}istortion-free \textbf{E}mbedding of water\textbf{mark}s), which leverages multi-marginal optimal transport. \methodname{} is a distortion-free watermark whose score function builds on intra-block token interactions. \methodname{} offer a computationally tractable joint watermark by solving an entropic MOT over sparse computational graphs that encode token interactions in the block. Consequently, \methodname's computational complexity scales linearly with the DLM's block size. 
We demonstrate \methodname's superior performance in several generation tasks, such as summarization, question answering, and code generation. For example, on Multi-News, \methodname{} attains up to $89\%$-$99\%$ TPR@5\% while keeping ROUGE-L within $0.6$ points of the unwatermarked baseline, while on LCC it reaches up to $82\%$-$92\%$ TPR@5\% while maintaining higher code similarity. 
Code: \url{https://github.com/omerlux/TIDEmark}.
\end{abstract}

\raggedbottom
\begingroup
\setlength{\intextsep}{0pt}
\setlength{\textfloatsep}{12pt}
\section{Introduction}
\label{sec:introduction}

Text watermarking embeds a detectable signal in model-generated text
\citep{kirchenbauer2023watermark,kuditipudi2024robust}.
It has gained practical importance for large-scale language-model services, as illustrated by SynthID-Text in Google's Gemini
\citep{dathathri2024scalable} and its announced adoption for future versions
of Anthropic's Claude \citep{anthropic2026watermark}.
In this scheme, a watermark key is used to generate common randomness, which is privately shared between the embedder and detector.
Embedding a stronger watermark signal can change which tokens the model selects and therefore change the content it generates. In code generation, for example, a token change can alter what a program does even when the code remains syntactically valid. A main goal in watermark design is therefore making watermarked text easily detectable by the detector using pre-shared side information, while remaining indistinguishable from unwatermarked text to general observers of the watermarked text. Such indistinguishability often reflects task performance maintenance in various benchmarks.

Diffusion language models can update several token positions in a single
inference step, rather than generating text in a strictly autoregressive manner \citep{sahoo2024mdlm,shi2024simplified,nie2025llada,duo}.
The model predicts token distributions, and a planner selects which
token positions to commit. The committed positions are then sampled to obtain token IDs
\citep{trainplan,benhamu2025ebsampler,luxembourg2026dus}.

\begin{figure}[!t]
\centering
\includegraphics[width=\linewidth]{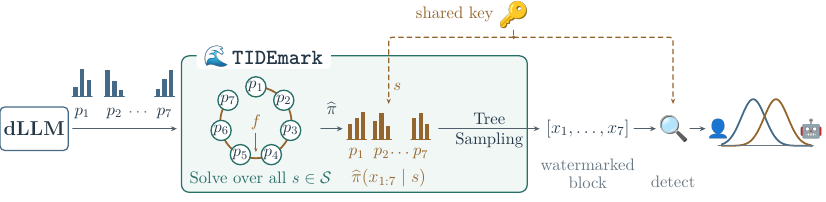}
\caption{%
\textbf{\methodname{}: joint watermarking and detection.}
The dLLM supplies recorded token distributions $p_1,\ldots,p_7$.
\methodname{} uses Sinkhorn with BP to compute a joint coupling
across all watermark states, favoring high watermark scores.
A key-derived state $s$ selects the conditional distribution
used for joint tree sampling.
The detector reuses the key and the block's graph structure.
The blue and gold curves illustrate unwatermarked and watermarked
score distributions, respectively.%
}
\label{fig:tidemark-overview}
\end{figure}

Standard approaches focus on unmasking strategies, while remasking and token-editing operations provide additional mechanisms for revising earlier token choices and improving generation quality
\citep{wang2025remasking,bie2026llada21speedingtextdiffusion,google2026diffusiongemma}.

\Needspace{3\baselineskip}
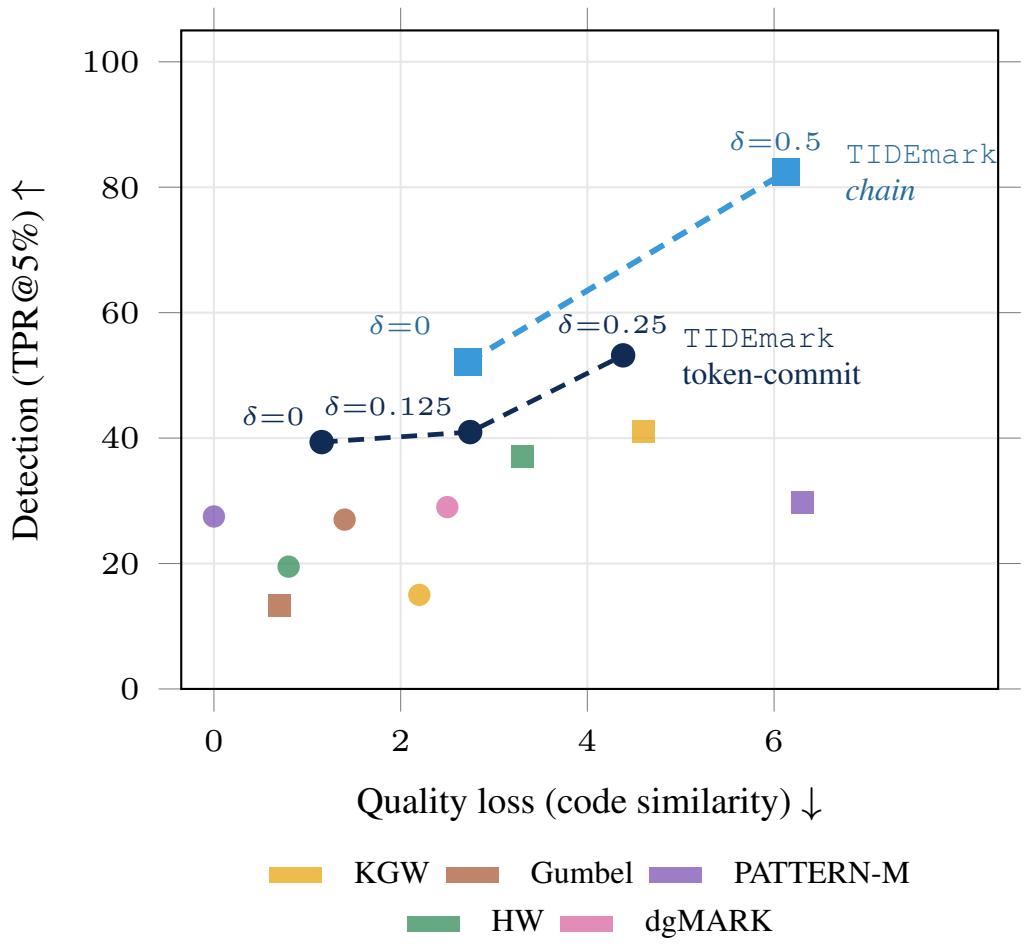
\begin{wrapfigure}{R}{0.50\textwidth}
\vspace{-0.5\baselineskip}
\centering
\begingroup
\definecolor{lccKGW}{HTML}{E69F00}
\definecolor{lccGumbel}{HTML}{A5502C}
\definecolor{lccDG}{HTML}{D65B9B}
\definecolor{lccPattern}{HTML}{7545AF}
\definecolor{lccHW}{HTML}{23834B}
\definecolor{lccChain}{HTML}{3A9AD9}
\definecolor{lccCommit}{HTML}{102C54}
\newcommand{\lccLegendItem}[2]{%
  \textcolor{#1!70!white}{\rule{.35cm}{.10cm}}\hspace{6pt}#2}
\begin{tikzpicture}
\begin{axis}[
  width=\linewidth,height=0.85\linewidth,
  font=\scriptsize,tick label style={font=\tiny},
  label style={font=\scriptsize},
  xlabel={Quality loss (code similarity) $\downarrow$},
  ylabel={Detection (TPR@5\%) $\uparrow$},
  xmin=-0.35,xmax=8.4,xtick={0,2,4,6},
  ymin=0,ymax=105,ytick={0,20,40,60,80,100},
  scaled ticks=false,grid=major,major grid style={gray!18},
  axis background/.style={fill=white},tick align=outside,
  baseline/.style={
    only marks,mark size=2.1pt,forget plot,
    mark options={fill=#1,fill opacity=0.7,draw opacity=0}
  },
  clip=false
]
\addplot[
  color=lccCommit,line width=0.9pt,densely dashed,mark=*,mark size=2.3pt,
  mark options={solid,fill=lccCommit,draw opacity=0},forget plot
] coordinates {
  (1.1542553191,39.3617021277)
  (2.7446808511,40.9574468085)
  (4.3829787234,53.1914893617)
};
\addplot[
  color=lccChain,line width=1.1pt,densely dashed,no marks,forget plot
] coordinates {
  (2.7180851064,52.1276595745)
  (6.1276595745,82.4468085106)
};
\addplot[
  only marks,color=lccChain,mark=square*,mark size=2.6pt,forget plot,
  mark options={fill=lccChain,draw opacity=0}
] coordinates {
  (2.7180851064,52.1276595745)
  (6.1276595745,82.4468085106)
};
\addplot[baseline=lccKGW,mark=*] coordinates {(2.2,15.0)};
\addplot[baseline=lccKGW,mark=square*] coordinates {(4.6,41.0)};
\addplot[baseline=lccGumbel,mark=*] coordinates {(1.4,27.0)};
\addplot[baseline=lccGumbel,mark=square*] coordinates {(0.7,13.3)};
\addplot[baseline=lccPattern,mark=*] coordinates {(0.0,27.5)};
\addplot[baseline=lccPattern,mark=square*] coordinates {(6.3,29.8)};
\addplot[baseline=lccHW,mark=square*] coordinates {(3.3,37.0)};
\addplot[baseline=lccHW,mark=*] coordinates {(0.8,19.5)};
\addplot[baseline=lccDG,mark=*] coordinates {(2.5,29.0)};

\node[font=\fontsize{6}{7}\selectfont,text=lccChain!70!black,
  anchor=west,align=left,xshift=11pt,inner sep=0pt]
  at (axis cs:6.1276595745,82.4468085106) {\methodname{}\\\textit{chain}};
\node[font=\fontsize{6}{7}\selectfont,text=lccCommit,
  anchor=west,align=left,xshift=11pt,inner sep=0pt]
  at (axis cs:4.3829787234,53.1914893617) {\methodname{}\\token-commit};
\node[font=\fontsize{5}{6}\selectfont,text=lccCommit,anchor=south east,
  xshift=-2pt,yshift=2pt,inner sep=1pt]
  at (axis cs:2.7446808511,40.9574468085) {$\delta{=}0.125$};
\node[font=\fontsize{5}{6}\selectfont,text=lccCommit,anchor=south,
  xshift=-2pt,yshift=3pt,inner sep=1pt]
  at (axis cs:4.3829787234,53.1914893617) {$\delta{=}0.25$};
\node[font=\fontsize{5}{6}\selectfont,text=lccChain!70!black,anchor=east,inner sep=1pt]
  at (axis cs:2.4,58) {$\delta{=}0$};
\node[font=\fontsize{5}{6}\selectfont,text=lccCommit,anchor=south east,
  xshift=-2pt,yshift=2pt,inner sep=1pt]
  at (axis cs:1.1542553191,39.3617021277) {$\delta{=}0$};
\node[font=\fontsize{5}{6}\selectfont,text=lccChain!70!black,anchor=south,
  xshift=-2pt,yshift=3pt,inner sep=1pt]
  at (axis cs:6.1276595745,82.4468085106) {$\delta{=}0.5$};
\node[anchor=north,inner sep=0pt,font=\tiny] at (rel axis cs:0.5,-0.26) {%
  \shortstack[c]{%
    \lccLegendItem{lccKGW}{KGW}\enspace
    \lccLegendItem{lccGumbel}{Gumbel}\enspace
    \lccLegendItem{lccPattern}{PATTERN-M}\\[2pt]
    \lccLegendItem{lccHW}{HW}\enspace
    \lccLegendItem{lccDG}{dgMARK}}};
\end{axis}
\end{tikzpicture}
\setlength{\abovecaptionskip}{3pt}
\setlength{\belowcaptionskip}{0pt}
\caption{\textbf{Detection--quality tradeoff on LCC (LLaDA-Instruct).}
Lines connect tested \methodname{} settings. Numbers denote post-MOT
tilt strength $\delta$. Markers $\bullet$ and $\blacksquare$ indicate
token-commit and block-ready insertion, respectively.
Loss is the decrease in code similarity from the clean reference of $58.0$.
Upper left is better.}
\label{fig:lcc-detection-quality}
\endgroup
\end{wrapfigure}
Different sequences of model predictions and planner decisions can
produce the same final text. Therefore, a detector cannot uniquely recover the generation trajectory from the text alone. Consequently, a watermark must be agnostic to the aforementioned intermediate generation steps and operations.

% \begingroup

Token-level optimal-transport watermarks, such as KGW, Gumbel or HeavyWater (HW), couple each token separately with watermark information
\citep{kirchenbauer2023watermark,tsur2025heavywater}.
These couplings can be constructed for each token distribution separately in the block once the block is complete.
Pairwise OT exploits the choices available at each position.
However, within a block, there's variability in each position's entropy: some positions may admit several plausible tokens, whereas others are tightly constrained.
This raises the question of whether jointly watermarking token sets can use the available flexibility across token positions more effectively,
leading to better block detection without sacrificing text quality and task performance.
% \par
% Figure~\ref{fig:joint-watermark-motivation} compares joint and separate OT on two fixed token distributions, with per-token scores derived by averaging a common joint score. In this example, joint OT achieves higher expected score gain at comparable conditional KL divergence from the original distributions
See Figure~\ref{fig:joint-watermark-motivation}, where we show that joint-block scores yield higher expected gain at comparable distributional divergence.
% \par

In this paper we propose \methodname, a watermark for diffusion language models that jointly watermarks token blocks (and sub-blocks) to improve the detectability-quality tradeoff. \methodname uses multi-marginal optimal transport (MOT) to design a watermark distribution for joint token distributions, rather than applying a separate watermark at each position. 
% In the ideal coupling, averaging
% over the watermark randomness recovers each recorded token distribution.
% A detector checks the resulting signal using the key, and final generated tokens.
To alleviate the exponential complexity of vanilla MOT, 
% A dense MOT coupling represents every combination of candidate tokens across the block, which is exponentially complex in block size.
we use tree-structured token-interaction graphs, which, due to their sparsity, result in efficient Sinkhorn iterations over simplified latent structures
\citep{haasler2021multimarginal,benamou2015iterative, tsur2025neural}.
Unlike existing diffusion watermarks that embed step-by-step as tokens are committed (\emph{token-commit}), we introduce \emph{block-ready} watermarking, which applies the watermark once upon block completion to avoid repeated overhead and prevent signal erasure during iterative token revisions. We evaluate \methodname{} under both timings and extend this block-ready scheme to existing token-level baselines.

Empirically across summarization, QA, and code generation on LLaDA-Instruct and LLaDA 2.1, \methodname{} improves the detectability--quality tradeoff, achieving up to $89.0\%$--$99.0\%$ TPR@5\% on Multi-News with minimal ROUGE-L degradation (within $0.6$ points of clean references) and up to $82.4\%$--$92.6\%$ TPR@5\% on LCC while preserving code similarity (within $6.1$ points).
Figure~\ref{fig:tidemark-overview} outlines the joint watermarking
and detection procedure of \methodname.
% \par
% \endgroup

% {
\Needspace{5\baselineskip}
\begingroup
\setlength{\parskip}{0pt}
\setlength{\topsep}{1pt}
\setlength{\partopsep}{0pt}
\setlength{\parsep}{0pt}
\setlength{\itemsep}{1pt}
The main contributions of this work are as follows:
\begin{enumerate}
    \item We propose \methodname, a new watermark for diffusion language models. \methodname extends token-level watermarks to joint watermarking of token blocks or token groups selected during denoising, including token revisions.
    \item We prove that \methodname is a distortion-free watermark when applied over generated distribution blocks.
    \item We integrate MOT over sparse graphical structures into \methodname to obtain an efficient implementation with polynomial complexity. Focusing on tree graphs, the algorithm scales linearly in block length.
\end{enumerate}

% \input{figures/joint_separate_comparison.tex}
% \noindent\begin{minipage}[t]{\linewidth}
% \begin{enumerate}
%     \setcounter{enumi}{2}
%     \item We integrate MOT over sparse graphical structures into \methodname to results with an efficient implementation with polynomial complexity. Focusing on tree graphs, the algorithm scales linearly in block length.
% \end{enumerate}
% \end{minipage}
% \par

% Section~\ref{sec:background} reviews token-level watermarking, optimal
% transport, and diffusion generation.
% Section~\ref{sec:method} introduces joint block-level watermarking and its
% ideal marginal-preservation guarantee.
% Section~\ref{sec:method-factor-graphs} develops the graph-structured
% formulation, efficient sampling, and detection procedure.
% Section~\ref{sec:results} presents the experimental setup and benchmark results.
% Section~\ref{sec:discussion} discusses the design tradeoffs and concludes
% the paper. The appendices provide related work, proofs, algorithmic details,
% reproducibility information, additional results, and limitations.
% \par
% \WFclear
% }
In the remainder of this paper, we review the relevant background (Section~\ref{sec:background}), formalize \methodname{} and its algorithm (Sections~\ref{sec:method}-\ref{sec:method-factor-graphs}), report its empirical detection--quality tradeoffs (Section~\ref{sec:results}), and conclude in Section~\ref{sec:conclusion}.
\par
\endgroup

\section{Background}
\label{sec:background}

\subsection{Token-level watermarking and optimal transport}
\label{sec:background-ar}
% {
A text watermarking scheme consists of an embedder, whose goal is to reweigh the next token distribution $p\in\Delta(\mathcal{X})$ to embed a detectable statistical signal. This signal is based on a given key, which is shared with the detector, whose goal is to detect this signal given the generated token sequence and the privately shared key \citep{kirchenbauer2023watermark,kuditipudi2024robust}.

Following the construction in \citep{tsur2025heavywater}, we divide the watermarking procedure into the following three components: 1) a score function $f$, whose purpose is to assign a score for each token-side information pair, 2) side-information $S$, which is often characterized by its distribution $P_S$ and a seed generator, and 3) a watermark transformation, denoted by the conditional distribution $P_{X|S}$. We take $P_S$ to be uniform over a predetermined finite set \(\mathcal S\). In practice, key-derived seeds
allow the generator and detector to reproduce exact samples of $P_S$.
The watermark transformation uses \(s\) to modify \(p\) before sampling
a token \(x\in\mathcal X\). The detector
averages the scores \(f(x,s)\) over the generated sequence of tokens and side information samples.

In token-level watermarks, a common proxy for output quality preservation
is \textit{distortion}, which measures the change in the token distribution
after averaging over the side information \citep{tsur2025heavywater}.
Formally, let \(X\sim P_{X\mid S}\) denote the watermarked token obtained from a next token distribution $p$ and its corresponding distribution. A watermark is \textit{distortion-free} when \(\mathbb{E}_S[P_{X\mid S}]=p\).
% }

% {
Given a score
function \(f:\mathcal X\times\mathcal{S}\to\mathbb{R}\) and a next token distribution $p$, finding a distortion-free watermark can be done by solving an optimal transport problem. We denote by \(\Pi_{X,S}\) the set of couplings of $(p,P_S)$, i.e., the set of joint
distributions \(P_{XS}\) whose marginals are \(P_X\) and \(P_S\) w.r.t $S$ and $X$, respectively. We take \(P_S\) to be uniform over \(\mathcal S\). The OT problem is given by
\begin{equation}\label{eq:ot}
P_{XS}^{\star}\in
\operatorname*{arg\,max}_{P_{XS}\in\Pi_{X,S}}
\left\{
\mathbb E_{P_{XS}}[f(X,S)]
-
\mathbb E_{P_XP_S}[f(X,S)]
\right\}.
\end{equation}
Here \(P_XP_S\) denotes the product of the two marginal distributions.
In practice, a solution to \eqref{eq:ot} is obtained through Sinkhorn's algorithm, which solves an entropy-regularized version
of this problem \citep{cuturi2013sinkhorn,peyre2019computational}.
Given side information \(s\in\mathcal{S}\), the watermarked token is sampled from
\(P_{X\mid S=s}(x)=|\mathcal S|P_{XS}^{\star}(x,s)\).
% \par
% }
In this paper, we consider the HW-based score function, which is constructed by sampling a heavy-tailed continuous distribution.\footnote{Following
\citet{tsur2025heavywater}, sample a
\(\operatorname{LogNormal}(0,1)\) distribution.}
% \par
% }

\subsection{Diffusion generation and watermarking}
\label{sec:background-dllm}

% {
Generation in diffusion language models often starts from masked positions in absorbing-state
models or uniformly random vocabulary tokens in uniform-state models
\citep{lou2024discrete,sahoo2024mdlm,duo}.
A block groups token positions processed over successive inference steps, with the denoiser predicting token values and the planner selecting
positions to update
\ifPDFTeX
  \mbox{\citep{nie2025llada,trainplan,benhamu2025ebsampler,luxembourg2026dus}}%
\else
  \citep{nie2025llada,trainplan,benhamu2025ebsampler,luxembourg2026dus}%
\fi.
Standard unmasking keeps committed tokens fixed, whereas uniform-state
sampling and remasking or token-editing procedures allow further revisions
\citep{duo,wang2025remasking,bie2026llada21speedingtextdiffusion,google2026diffusiongemma}.
% \par
% }

% {
For positionwise watermarking, the watermark transformation acts separately on each selected token distribution, even when several positions are processed in parallel. The detector computes the score \(f(x,s)\) from each final token \(x\) and its corresponding side information \(s\), reproduced through the specified seed-generation procedure.
% \par
% }
% {
Section~\ref{sec:method} extends this token-level formulation to completed
diffusion blocks, coupling their token distributions with the side
information through multi-marginal optimal transport.
% \par
% }

\section{Jointly Watermarking Diffusion Language Models}
\label{sec:method}
\label{sec:method-joint-mmot}

\begin{wrapfigure}{R}{0.40\textwidth}
\vspace{-0.5\baselineskip}
\centering
\begingroup
\definecolor{separateOT}{HTML}{41464B}
\definecolor{jointOT}{HTML}{3A9AD9}
\begin{tikzpicture}
\begin{axis}[
    width=0.96\linewidth,
    height=0.235125\textwidth,
    font=\scriptsize,
    tick label style={font=\scriptsize},
    xlabel={Conditional KL divergence},
    ylabel={Expected score gain},
    xmode=log,
    log basis x=10,
    xmin=1e-6, xmax=1e-2,
    ymin=-0.018, ymax=0.75,
    xtick={1e-6,1e-5,1e-4,1e-3,1e-2},
    xticklabels={$10^{-6}$,{},$10^{-4}$,{},$10^{-2}$},
    ytick={0,0.2,0.4,0.6},
    scaled ticks=false,
    grid=major,
    major grid style={gray!22},
    tick align=outside,
    legend columns=2,
    legend style={
        at={(0.5,1.03)}, anchor=south,
        font=\scriptsize, draw=none, fill=none,
        inner sep=0pt,
        cells={anchor=west}
    },
    clip=true,
    unbounded coords=jump
]
\addplot+[
    restrict expr to domain={\coordindex}{1:6},
    color=separateOT, line width=0.8pt,
    mark=*, mark size=1.4pt,
    mark options={solid, fill=separateOT},
    error bars/.cd,
    x dir=both, x explicit,
    y dir=both, y explicit,
    error bar style={line width=0.5pt},
    error mark=|,
    error mark options={rotate=0, mark size=1.8pt, line width=0.5pt}
] table[
    col sep=comma,
    x=separate_kl, y=separate_gain,
    x error minus=separate_xminus, x error plus=separate_xplus,
    y error minus=separate_yminus, y error plus=separate_yplus
] {data/joint_separate_comparison.csv};
\addlegendentry{Separate scores}
\addplot+[
    restrict expr to domain={\coordindex}{1:4},
    color=jointOT, line width=0.8pt,
    mark=*, mark size=1.4pt,
    mark options={solid, fill=jointOT},
    error bars/.cd,
    x dir=both, x explicit,
    y dir=both, y explicit,
    error bar style={line width=0.5pt},
    error mark=|,
    error mark options={rotate=0, mark size=1.8pt, line width=0.5pt}
] table[
    col sep=comma,
    x=joint_kl, y=joint_gain,
    x error minus=joint_xminus, x error plus=joint_xplus,
    y error minus=joint_yminus, y error plus=joint_yplus
] {data/joint_separate_comparison.csv};
\addlegendentry{Joint score}
\end{axis}
\end{tikzpicture}
\endgroup
\setlength{\abovecaptionskip}{3pt}
\setlength{\belowcaptionskip}{6pt}
\caption{\textbf{Joint versus separate scores.}
At comparable mean conditional KL divergence, joint scores achieve
substantially higher expected score gain over unchanged sampling
than their separate per-token projections.}
\label{fig:joint-watermark-motivation}
\end{wrapfigure}
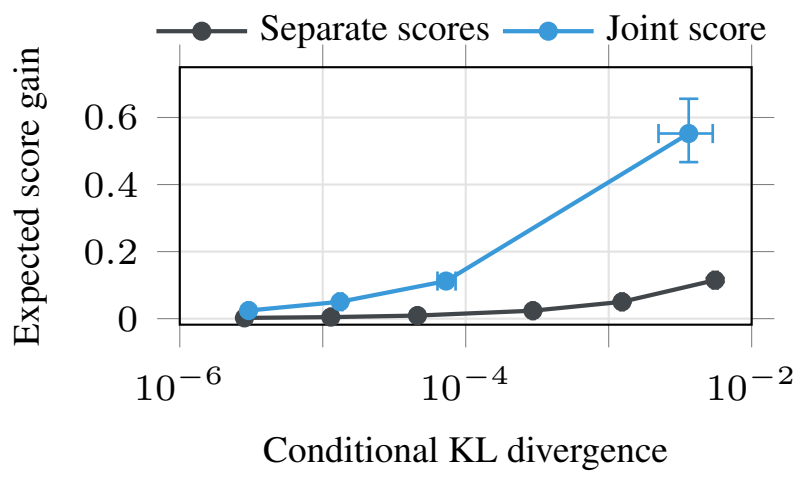

% {
With the goal of watermark diffusion-model outputs, one straightforward approach is applying a token-level watermarking scheme by adapting its seed-generation procedure
to diffusion-style generation
\citep{wu2025dmark,gloaguen2026wmdlm,raban2026lrdwm,hong2026dgmark}.
In this approach, each token position is watermarked separately.
Detection would follows the score-based approach used in autoregressive models: the detector would reproduce the side information from the shared key, from which they would calculate scores to be compared against the predetermined threshold.

In diffusion language models, a block of tokens is generated jointly.
This joint generation of token probabilities offers an opportunity to
devise a watermark mapping that is non-uniform across positions.
Specifically, we seek to maximize the expected joint watermark score
subject to the prescribed token and side-information marginals
(\Eqref{eq:method-mot-objective}).
By jointly scoring token choices within a block, we can achieve larger
expected scores at comparable mean conditional KL divergence from the
original unwatermarked token distributions. For each side-information
state, KL is computed from the corresponding watermarked token marginals
to their original distributions, and averaged over token positions and
states. Figure~\ref{fig:joint-watermark-motivation} illustrates the
resulting advantage of joint scores over separate per-token scores.
Next, we explain how to construct a joint block score and solve the
resulting optimization through MOT.

% We propose a joint block-level watermark based on an extension of OT.
% Instead of scoring each token separately, the joint formulation scores
% combinations of tokens against the side information.
% We investigate whether this flexibility strengthens detection while
% preserving each specified input token distribution after averaging over
% the side information.

\subsection{Block-Level watermark through multi-marginal optimal transport}
% {
Consider a completed token block \((t_1,\ldots,t_b)\) before watermarking, generated by the denoising and planning procedure described in Section~\ref{sec:background-dllm},
with corresponding token distributions \((p_1,\ldots,p_b)\), each over
\(\mathcal X\). Our goal is to jointly watermark this set of distributions.\footnote{%
Resampling a completed block from its recorded token distributions may
alter temporal token dependencies. Section~\ref{sec:method-commit-time}
discusses this issue and the token-commit alternative.}
% }
% {
As we look for an block-level extension of single-block OT-based watermarks, we resolve to MOT, which is the extension
of OT when more than two marginal distributions are considered.
Formally, given \(b\) marginal distributions \((p_1,\ldots,p_b)\), MOT seeks a joint distribution over \(\mathcal X^b\) that preserves these marginals while minimizing a given expected cost function.

In our setting, the MOT problem is
defined on the set \(\mathcal{X}^b\times\mathcal S\).
That is, our MOT considers \(b+1\) marginal distributions: the \(b\) token distributions and a single uniform side-information distribution \(P_S\). Each token distribution
\(p_i\) corresponds to a random variable \(X_i\), with \(X^b=(X_1,\ldots,X_b)\).
% }
% {
For a given score function
\(f:\mathcal X^b\times\mathcal S\to\mathbb R\), the MOT problem is
given by the following optimization over joint probability
distributions \(\pi\) of \((X^b,S)\):
\begin{equation}
  \begin{aligned}
    \pi^\star
      &\in\operatorname*{arg\,min}_{\pi}
        \mathbb E_{\pi}[-f(X^b,S)],\\
    \text{subject to}\quad
      &\pi_{i}=p_i\quad(i=1,\ldots,b),\qquad \pi_{b+1}=P_S,
  \end{aligned}
  \label{eq:method-mot-objective}
\end{equation}
where $\pi_{i}$ is the marginalization of the $(b+1)$-dimensional coupling $\pi$ w.r.t. all coordinates other than the $i$th.
We write $\pi_S:=\pi_{b+1}$ for the side-information marginal.
For heavy-tailed scores such as those used in
HW \citep{tsur2025heavywater}, restricting \(f\) to \([0,1]\) would require explicit normalization of the scores defining the
transport cost \(-f\).

The output of the optimization in \eqref{eq:method-mot-objective}
is a joint distribution \(\pi^\star\) over
\(\mathcal X^b\times\mathcal S\). For a given realization of the side-information \(s\in\mathcal{S}\), the watermarked block $x^b$ is jointly sampled from the conditional
distribution, given by
\begin{equation}
  P_{X^b\mid S=s}(x^b)
  =\frac{\pi^\star(x^b,s)}{P_S(s)}
  =|\mathcal S|\,\pi^\star(x^b,s).
  \label{eq:method-mot-conditional}
\end{equation}
% \par
In practice, $\pi^\star$ is a $(b+1)$-dimensional tensor. Thus, the corresponding watermarked token distributions can be obtained by marginalizing this conditional distribution over the other token
positions.
% }

% \par
% {
We refer to the resulting watermarking procedure as \textbf{\methodname{}}
(Transport-based Interaction for Distortion-free Embedding of watermarks).
To watermark a generated token block, we collect the generated token distributions in each position, which act as the set of marginal distributions. We then solve the corresponding MOT over the set of $(b+1)$ distributions.
To solve the MOT, we use Sinkhorn's algorithm, which solves the entropy-regularized version of the proposed problem.
% \par
% The secret key and side-information rule determine the state used to sample
% the block's tokens jointly.
% The same key allows the detector to reproduce this state and reconstruct the watermark scores.
Detection in \methodname remains lightweight: It aggregates the token block scores and compares it against a
threshold, without any requirement to re-solve the MOT problem.
In Section~\ref{sec:method-factor-graphs} we discuss the complexity of \methodname and propose a computationally efficient solution.
% \par
% }

% \WFclear
% {
% \textcolor{red}{Stopped here}
% For fixed recorded token distributions, 
Following the optimization problem's construction, when operating over a completed block \methodname is a \textit{distortion-free} watermark:
% 's optimal joint coupling
% preserves each input marginal after averaging over the prescribed
% side-information distribution.
% This fixed-input guarantee does not establish marginal preservation during
% adaptive token-commit generation or preservation of the full text distribution.

\begin{proposition}[\methodname{} is distortion free]
\label{prop:method-distortion-free}
Fix the recorded token distributions \(p_1,\ldots,p_b\) and a coupling
\(\pi^\star\) satisfying the token and side-information marginal
constraints in \eqref{eq:method-mot-objective}. Draw one shared state
\(S\sim P_S\), then sample \(X^b\) from the corresponding conditional
distribution in \eqref{eq:method-mot-conditional}.
Averaging the position-\(i\) marginal of the sampled block distribution
over the shared side information recovers the recorded token distribution:
\begin{equation}
  \mathbb E_{S\sim P_S}\!
      \sum_{\substack{x^b\in\mathcal X^b\\x_i=x}}
      \frac{\pi^\star(x^b,S)}{P_S(S)}
  =p_i(x),
  \qquad i=1,\ldots,b,\quad x\in\mathcal X.
  \label{eq:method-marginal-preservation}
\end{equation}
\end{proposition}
The proof of Proposition~\ref{prop:method-distortion-free} is given in Appendix~\ref{sec:appendix-marginal-preservation}. It follows by the fact that \methodname optimizes over couplings.
% \par
% }

\section{Efficient block-level watermarking}
\label{sec:method-factor-graphs}

% \subsection{The cost of block-level watermarks}
\methodname offers a strong alternative to token-level watermarks. However, it relies on MOT, whose complexity scales exponent with the block size $b$.
The reason is the the corresponding Sinkhorn iteration involves operations on a $(b+1)$-dimensional tensor representing all of the joint token-state combinations. 
While top-p or top-k sampling significantly reduces the number of candidate tokens at each position with a small effect on performance \cite{tsur2025heavywater}, the exponential dependence on $b$ limits implementation to very small block sizes.
In what follows, we propose a modification of \methodname that reduces the exponential dependence in $b$, unlocking block-level watermarks at practical block sizes.

\subsection{Complexity reduction through graph decompositions}
% {
To address the exponential dependence in \(b\), we utilize a simplification of Sinkhorn's algorithm that builds upon sparse graph structures \cite{altschuler2023polynomial}.

To that end, we decompose the score function into pairwise components. That is, we represent the score through the sum of components:
\begin{equation}
  f(X^b,S)
  =\sum_{(i,j)\in\mathcal I}f(X_i,X_j,S),
\end{equation}
\ifPDFTeX
  \par
  \Needspace{3\baselineskip}
  \begin{wrapfigure}{R}{0.32\textwidth}
\vspace{-0.5\baselineskip}
\definecolor{treeTeal}{HTML}{286F68}
\definecolor{treeTealWash}{HTML}{F0F7F5}
\centering
\resizebox{0.198\textwidth}{!}{%
\begin{tikzpicture}[
  x=1cm,y=1cm,text=black,
  token/.style={
    circle,draw=treeTeal,fill=treeTealWash,line width=.75pt,
    minimum size=5.6mm,inner sep=0pt,outer sep=1pt,
    font=\small
  },
  interaction/.style={draw=black,line width=.55pt,->,>=stealth},
  panel/.style={font=\small\bfseries}
]
\path[use as bounding box] (-1.8,.28) rectangle (1.8,10.2);

\begin{scope}[yshift=6.4cm]
  \node[token] (a1) at ( 0,    3.350) {$X_1$};
  \node[token] (a2) at ( .860, 2.936) {$X_2$};
  \node[token] (a3) at (1.072, 2.005) {$X_3$};
  \node[token] (a4) at ( .477, 1.259) {$X_4$};
  \node[token] (a5) at (-.477, 1.259) {$X_5$};
  \node[token] (a6) at (-1.072,2.005) {$X_6$};
  \node[token] (a7) at (-.860, 2.936) {$X_7$};
  \foreach \u/\v in {1/2,2/3,3/4,4/5,5/6,6/7}
    \draw[interaction] (a\u) to[bend left=12] (a\v);
  \node[panel,anchor=east,overlay] at (-1.8,2.25) {(a)};
\end{scope}

\begin{scope}[yshift=3.2cm]
  \node[token]      (b1) at (-.953,2.80) {$X_1$};
  \node[token]      (b2) at ( 0,   3.35) {$X_2$};
  \node[token]      (b3) at ( .953,2.80) {$X_3$};
  \node[token]      (b4) at ( 0,   2.25) {$X_4$};
  \node[token]      (b5) at ( .953,1.70) {$X_5$};
  \node[token]      (b6) at ( 0,   1.15) {$X_6$};
  \node[token]      (b7) at (-.953,1.70) {$X_7$};
  \foreach \v in {1,2,3,5,6,7}
    \draw[interaction] (b4) -- (b\v);
  \node[panel,anchor=east,overlay] at (-1.8,2.25) {(b)};
\end{scope}

\begin{scope}
  \node[token]      (c1) at (-1.40,1.15) {$X_1$};
  \node[token]      (c2) at (-.82, 2.25) {$X_2$};
  \node[token]      (c3) at (-.46, 1.15) {$X_3$};
  \node[token]      (c4) at ( 0,   3.35) {$X_4$};
  \node[token]      (c5) at ( .46, 1.15) {$X_5$};
  \node[token]      (c6) at ( .82, 2.25) {$X_6$};
  \node[token]      (c7) at ( 1.40,1.15) {$X_7$};
  \foreach \u/\v in {4/2,4/6,2/1,2/3,6/5,6/7}
    \draw[interaction] (c\u) -- (c\v);
  \node[panel,anchor=east,overlay] at (-1.8,2.25) {(c)};
\end{scope}

\end{tikzpicture}%
}

\caption{Token graphs for a fixed state: (a) \textit{chain},
(b) \textit{middle-star}, (c) \textit{balanced tree}.
Arrows show parent-to-child sampling.}
\label{fig:method-tree-structures}
\end{wrapfigure}
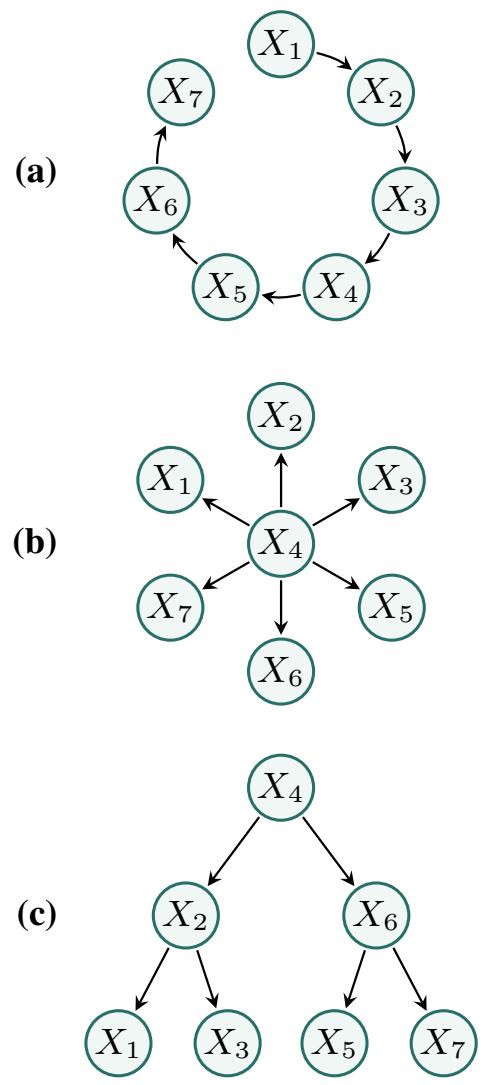
\fi
where \(\mathcal I\) is a subset of the pairs of distinct indices in
\(\{1,\ldots,b\}\), each included once.
Such a pairwise-decomposable score
can be equivalently represented with a computational graph \(\mathcal G=(\mathcal V,\mathcal E)\).\footnote{%
Our Sinkhorn implementation uses a tree-structured token interaction
graph for each fixed side-information value to compute the required
marginals efficiently \cite{haasler2021multimarginal}.
Section~\ref{sec:method-fitting} describes how we solve the MOT problem.}

\ifPDFTeX\else
  \Needspace{3\baselineskip}
  \begin{wrapfigure}{R}{0.32\textwidth}
\vspace{-0.5\baselineskip}
\definecolor{treeTeal}{HTML}{286F68}
\definecolor{treeTealWash}{HTML}{F0F7F5}
\centering
\resizebox{0.198\textwidth}{!}{%
\begin{tikzpicture}[
  x=1cm,y=1cm,text=black,
  token/.style={
    circle,draw=treeTeal,fill=treeTealWash,line width=.75pt,
    minimum size=5.6mm,inner sep=0pt,outer sep=1pt,
    font=\small
  },
  interaction/.style={draw=black,line width=.55pt,->,>=stealth},
  panel/.style={font=\small\bfseries}
]
\path[use as bounding box] (-1.8,.28) rectangle (1.8,10.2);

\begin{scope}[yshift=6.4cm]
  \node[token] (a1) at ( 0,    3.350) {$X_1$};
  \node[token] (a2) at ( .860, 2.936) {$X_2$};
  \node[token] (a3) at (1.072, 2.005) {$X_3$};
  \node[token] (a4) at ( .477, 1.259) {$X_4$};
  \node[token] (a5) at (-.477, 1.259) {$X_5$};
  \node[token] (a6) at (-1.072,2.005) {$X_6$};
  \node[token] (a7) at (-.860, 2.936) {$X_7$};
  \foreach \u/\v in {1/2,2/3,3/4,4/5,5/6,6/7}
    \draw[interaction] (a\u) to[bend left=12] (a\v);
  \node[panel,anchor=east,overlay] at (-1.8,2.25) {(a)};
\end{scope}

\begin{scope}[yshift=3.2cm]
  \node[token]      (b1) at (-.953,2.80) {$X_1$};
  \node[token]      (b2) at ( 0,   3.35) {$X_2$};
  \node[token]      (b3) at ( .953,2.80) {$X_3$};
  \node[token]      (b4) at ( 0,   2.25) {$X_4$};
  \node[token]      (b5) at ( .953,1.70) {$X_5$};
  \node[token]      (b6) at ( 0,   1.15) {$X_6$};
  \node[token]      (b7) at (-.953,1.70) {$X_7$};
  \foreach \v in {1,2,3,5,6,7}
    \draw[interaction] (b4) -- (b\v);
  \node[panel,anchor=east,overlay] at (-1.8,2.25) {(b)};
\end{scope}

\begin{scope}
  \node[token]      (c1) at (-1.40,1.15) {$X_1$};
  \node[token]      (c2) at (-.82, 2.25) {$X_2$};
  \node[token]      (c3) at (-.46, 1.15) {$X_3$};
  \node[token]      (c4) at ( 0,   3.35) {$X_4$};
  \node[token]      (c5) at ( .46, 1.15) {$X_5$};
  \node[token]      (c6) at ( .82, 2.25) {$X_6$};
  \node[token]      (c7) at ( 1.40,1.15) {$X_7$};
  \foreach \u/\v in {4/2,4/6,2/1,2/3,6/5,6/7}
    \draw[interaction] (c\u) -- (c\v);
  \node[panel,anchor=east,overlay] at (-1.8,2.25) {(c)};
\end{scope}

\end{tikzpicture}%
}

\caption{Token graphs for a fixed state: (a) \textit{chain},
(b) \textit{middle-star}, (c) \textit{balanced tree}.
Arrows show parent-to-child sampling.}
\label{fig:method-tree-structures}
\end{wrapfigure}
\fi
Each node in $\mathcal{G}$ represents a token position, and each edge corresponds to a pair
\((i,j)\in\mathcal I\). The pair ordering specifies the arguments of \(f(X_i,X_j,S)\).
% \par
% \par
To construct \(f(x_i,x_j,s)\), we follows \citep{tsur2025heavywater} and sample a heavy-tailed distribution. 
% Reproducible heavy-tailed draws assign a
% score to each token pair and side-information value.
For each token pair, scores are standardized to have zero mean and
unit variance over the side-information states.
The set of scores is generated prior to the generation of the text using the shared secret key, and is shared between the watermark embedder and detector. 
At the embedder, the standardized scores are additionally normalized
to $[0,1]$ before solving the MOT problem
(see Appendix~\ref{sec:appendix-experimental-settings} for more details).
% This score construction is shared by optimization and detection.
% Optimization additionally range-normalizes each edge's retained candidate-pair/state table, as detailed in Appendix~\ref{sec:appendix-experimental-settings}.
% \par

The complexity of the corresponding Sinkhorn iteration is directly connected to the graph's structure \citep{altschuler2023polynomial,tsur2025neural}. Specific graph structures
introduce sparsity in the cost tensor, allowing the required tensor operations to be computed without materializing the full tensor. Consequently, these tensor operations reduce to matrix-vector and matrix-matrix operations. Specifically, for tree graphical structures, the Sinkhorn iteration complexity becomes polynomial in $b$ (see Appendix~\ref{sec:appendix-graph-complexity} for an analysis).
% \par
% }
% \WFclear
% {

The tree restriction still leaves a choice of which token pairs are being scored.
We consider three graph structures: a \textit{chain}, a \textit{middle-star}, and a \textit{balanced tree} (Figure~\ref{fig:method-tree-structures}).
These graphs are defined by token positions and are not inferred from the model's original dependencies.
% \par
The \textit{chain} connects consecutive positions, with edges directed from the
first position toward the last.
The \textit{middle-star} connects the middle position to every other position.
The \textit{balanced tree} is rooted at the middle token position.
Its left and right subtrees are constructed recursively, using the
middle position of each remaining interval as the subtree root.
When there are two middle positions, the earlier one is used.
% \par

Each structure connects all \(b\) positions through \(b-1\) edges.
% Changing the graph changes the scored token pairs while keeping the token and shared-state marginal constraints fixed.
These choices let us compare local interactions in the \textit{chain},
centralized interactions in the \textit{middle-star}, and hierarchical
interactions in the \textit{balanced tree}.
They also have different depths: \(O(b)\) for the \textit{chain},
\(O(\log b)\) for the \textit{balanced tree}, and \(O(1)\) for the
\textit{middle-star}. Each sweep computes the marginals by traversing the
tree from the leaves to the root and back, so deeper trees require more
sequential rounds.
Graph structures connecting more widely separated token positions may
better preserve text coherence by placing watermark scores on pairs
with lower native mutual information \citep{luxembourg2026dus}.
% \par
% }/
% \WFclear
\subsection{Solving MOT and sampling watermarked tokens}
\label{sec:method-fitting}
To obtain a joint distribution for watermarking the block, the solver applies
Sinkhorn iterations to the entropy-regularized MOT problem on the graph from
Section~4.1. The target marginals are the recorded token distributions
\(p_1,\ldots,p_b\) and the side-information distribution \(P_S\).

\noindent\textbf{Belief Propagation and Sinkhorn.}
Sparse Sinkhorn uses belief propagation (BP) to compute the marginals needed
for each update.
For each state \(s\in\mathcal S\), BP computes a weight
for each candidate token by summing over the other token positions one
branch at a time
\citep{yedidia2003understanding}.
The solver scales the weights for each state by its current scaling factor.
The token marginals come from summing the scaled token weights over states
and normalizing. The side-information marginal comes from normalizing the
scaled total weights of the states.
Using these marginals, Sinkhorn updates the token scaling factors
\(u_i(x_i)\) and the side-information scaling factors \(v(s)\)
\citep{haasler2021multimarginal,benamou2015iterative}.
After each update, BP recomputes the marginals using the updated scaling factors.
The graph and pair scores remain fixed, and the solver stops when the marginal error reaches the specified tolerance
or the iteration budget is exhausted.

The full objective is given in Appendix~\ref{sec:appendix-entropic-mot}.
For regularization strength \(\varepsilon>0\), the optimal coupling on
the prescribed supports admits the factorized form
\citep{haasler2021multimarginal}
\begin{equation}
  \pi_\varepsilon^\star(x^b,s)
  =v(s)\prod_{i=1}^{b}u_i(x_i)
  \prod_{(i,j)\in\mathcal I}
  \exp\!\left(\frac{f(x_i,x_j,s)}{\varepsilon}\right),
  \label{eq:method-entropic-factorization}
\end{equation}
where \(u_i(x_i)\) scales candidate token \(x_i\) at position \(i\), \(v(s)\)
scales side-information state \(s\), and \(f(x_i,x_j,s)\) is the pair score for
that edge and state. The factors returned by Sinkhorn define the computed
coupling \(\widehat\pi\) used for sampling.

\begingroup
\setlength{\columnsep}{0.03\textwidth}
\setlength{\intextsep}{0pt}
\setlength{\parskip}{1pt}
\begin{wrapfigure}[25]{r}{0.42\textwidth}
\vspace*{-3pt}
\begin{minipage}{\linewidth}
\color{black}
\setlength{\intextsep}{0pt}
\begin{algorithm}[H]
\caption{Generating a watermarked block with \methodname{}}
\label{alg:method-joint-generation}
\footnotesize
\begin{algorithmic}[1]
  \Require Distributions \(p_1,\ldots,p_b\), secret key \(\kappa\)
  \Ensure Watermarked block \(x_1,\ldots,x_b\)
  \State \(\widehat\pi \gets \Call{Sinkhorn-BP}{p_1,\ldots,p_b}\)
  \State \(s \gets \Call{SideInformation}{\kappa}\)
  \State \((x_1,\ldots,x_b) \gets \Call{TreeSample}{\widehat\pi,s}\)
  \State \Return \(x_1,\ldots,x_b\)
\end{algorithmic}
\end{algorithm}
\par\vspace{3pt}
\begin{algorithm}[H]
\caption{Detecting block \methodname{}}
\label{alg:method-detection}
\footnotesize
\begin{algorithmic}[1]
  \Require Observed generated tokens \(x\), secret key \(\kappa\)
  \Ensure Score \(z_{\mathrm{eq}}\) and detection decision
  \State \(G\gets |x|,\quad L\gets\lceil G/b\rceil\)
  \State \(\{x^{(k)}\}_{k=1}^{L}\gets\Call{Partition}{x,b}\)
  \For{\(k=1,\ldots,L\)}
    \State \(s_k\gets\Call{SideInformation}{\kappa}\)
    \State \(r_k\gets\displaystyle\sum_{(c,p)\in\mathcal E_k}
      f(x_{k,c},x_{k,p},s_k)\)
  \EndFor
  \State \(T_{\mathrm{obs}}\gets G^{-1}\sum_{k=1}^{L}r_k\)
  \State \(p\gets p_0(T_{\mathrm{obs}};x)\)\(\,^{\dagger}\)
  \State \(z_{\mathrm{eq}}\gets\Phi^{-1}(1-p)\)
  \State \Return \(z_{\mathrm{eq}},\ (p<\alpha)\)
\end{algorithmic}
\par
\noindent\(\dagger\) \(p_0(t;x)\) is the analytical upper-tail probability
under the declared null hypothesis.
Details in Appendix~\ref{sec:appendix-analytical-detection}.
\end{algorithm}
\end{minipage}
\end{wrapfigure}
\noindent\textbf{Tree sampling.}
The tree structure allows joint sampling without constructing the full
table of token combinations \citep{haasler2021multimarginal}.
In the selected base configurations, input temperature and candidate
restriction define the target token distributions before solving MOT.
After solving MOT, the key and side-information rule determine the block's
state \(s\). To sample from \(\widehat\pi(X^b\mid S=s)\), the sampler uses
the tree factors returned by the solver, first drawing the root token from
its marginal given \(s\). It then proceeds from root to leaves, drawing each
child token conditional on its parent's sampled token and \(s\).
Appendix~\ref{sec:appendix-tree-algorithms} illustrates sampling and
detection on a four-token \textit{chain}.

In Algorithm~\ref{alg:method-joint-generation}, \(p_i\) is the recorded
distribution at token position \(i=1,\ldots,b\).
In Algorithms~\ref{alg:method-joint-generation} and~\ref{alg:method-detection},
\(\kappa\) denotes the secret key. The side-information generator uses this
key with the text context specified by the seeding procedure.
Appendix~\ref{sec:appendix-tree-algorithms} details Sinkhorn-BP,
BP-Marginals, and tree sampling in
Algorithms~\ref{alg:appendix-sinkhorn-bp},
\ref{alg:appendix-bp-marginals}, and~\ref{alg:appendix-tree-sampling},
respectively.

\subsection{Detecting the watermark}
\label{sec:method-sampling-detection}
{
\nopagebreak[4]
For block-ready detection, the detector splits the observed
generated tokens into blocks using the known completion starting position
and block length \(b\).
\par

The graph structure is a public detector setting.
For block \(k\), the detector uses the same positional rule as generation
to construct the edge set \(\mathcal E_k\).
The watermark key and the specified side-information rule reproduce the
state for each block.
\par

Following HW's score construction
\citep{tsur2025heavywater}, optimization and detection use the same underlying
pair scores, standardized across states.
Optimization additionally range-normalizes each candidate-pair score table
for the transport optimization.
Detection sums the standardized scores at the reproduced block states
and divides by the number of generated tokens.
Appendix~\ref{sec:appendix-experimental-settings} explains these two
normalization stages.
\par

Let \(G>0\) be the number of observed generated tokens, \(b>1\) the block length,
and \(L=\lceil G/b\rceil\) the number of blocks.
Write \(x_{k,i}\) for token \(i\) in block \(k\) and \(s_k\) for the reproduced
watermark state of that block.
In each edge \((c,p)\in\mathcal E_k\), \(c\) is the child position and \(p\)
is its parent.
The score is
\begingroup
\setlength{\abovedisplayskip}{4pt}
\setlength{\belowdisplayskip}{4pt}
\setlength{\abovedisplayshortskip}{\abovedisplayskip}
\setlength{\belowdisplayshortskip}{\belowdisplayskip}
\begin{equation}
  T_{\mathrm{obs}}
  =
  \frac{1}{G}
  \sum_{k=1}^{L}
  \underbrace{
    \sum_{(c,p)\in\mathcal E_k}
    f(x_{k,c},x_{k,p},s_k)
  }_{\text{sum of pair scores in block }k}.
  \label{eq:method-detection-score}
\end{equation}
\endgroup
\par
\WFclear
\ifPDFTeX
  \Needspace{4\baselineskip}
\else
  \Needspace{5\baselineskip}
\fi

The analytical null model assumes that block states are uniform, mutually
independent, and independent of the observed tokens, with the score construction
held fixed. The detector rejects the null when
\(p_0(T_{\mathrm{obs}};x)<\alpha\), where \(p_0\) is the upper-tail probability
under this null model and \(\alpha\) is the chosen significance level.
\par
{
Appendix~\ref{sec:appendix-analytical-detection} and
Algorithm~\ref{alg:method-detection} describe this block-ready analytical
test under the stated null assumptions.
For the benchmark comparisons, detection thresholds are calibrated using
scores from unwatermarked texts.
Appendix~\ref{sec:appendix-experimental-settings} describes the calibration
procedure.
}

\par
At each token-commit update, the watermark scores pairs of adjacent
positions. The detector uses the same adjacency rule to build its scoring
\textit{chain} directly from the observed token sequence, including eligible
predecessor pairs at prompt and block boundaries. It scores those pairs using
the current block's watermark state.
Appendix~\ref{sec:appendix-commit-time} provides further details.
\par
}

\endgroup
\par

\subsection{Beyond block-ready watermarking}
\label{sec:method-commit-time}
\WFclear

In addition to choosing whether to watermark tokens jointly or
separately, we can choose when to apply the watermark.
We consider two insertion strategies:
% \begin{enumerate}
    % \item

    \textbf{Block-ready.} During denoising, the probability
    distribution used when each token is last written or revised is
    saved. Once the block is complete, the watermark is applied to
    these saved distributions, and the tokens are resampled before
    generation proceeds to the next block. Resampling requires no
    additional DLM call. Tokens from the current block
    are released only after watermarking and resampling are complete.

    % \item
    \textbf{Token-commit.} At each denoising step, watermarking
    is applied after the planner selects the positions to update and
    before the tokens at those positions are sampled and written.
    Subsequent denoising steps can use the watermarked tokens as context.
% \end{enumerate}

Block-ready watermarking leaves no further within-block updates to
restore coherence after watermarking. We therefore also apply
\methodname{} using token-commit insertion. The fixed-input
distortion-free guarantee does not cover this adaptive token-commit
setting.
At each step, we solve MOT on a contiguous \textit{chain} spanning
the selected positions and their immediate neighboring context.
We condition on unselected written tokens, marginalize unselected
masked positions, and write back only the selected positions,
including newly unmasked tokens and revisions.

\noindent\textbf{Post-MOT tilting.}
For either insertion timing, tilting favors higher-scoring samples by
reweighting the sampling distribution after solving MOT.
The marginal-preservation guarantee does not extend to this reweighting.
Appendix~\ref{sec:appendix-post-fit-tilting} gives the reweighting rule.
\endgroup

\setcounter{topnumber}{1}
\begingroup
\setlength{\textfloatsep}{5pt plus 2pt minus 2pt}
\newcommand{\bestvalue}[1]{{\bfseries\boldmath #1}}
\newcommand{\secondvalue}[1]{\underline{#1}}
\newcommand{\newvalue}[1]{#1}
\newcommand{\ppltoce}[1]{%
  \pgfmathparse{ln(#1)}%
  \pgfmathprintnumber[fixed,precision=3,fixed zerofill]{\pgfmathresult}%
}
\section{Results}
\label{sec:results}
This section examines the detection--quality tradeoff across models, tasks,
and insertion timings. Appendix~\ref{sec:appendix-experiments} presents
additional HumanEval and Finance QA results, examines how input temperature
and post-MOT tilting affect the detection--quality tradeoff, and reports
robustness to editing attacks and generation runtime.

{
\noindent\textbf{Experimental setup.}
The experiments use LLaDA 1.0-Instruct \citep{nie2025llada} and LLaDA 2.1-mini
\citep{bie2026llada21speedingtextdiffusion} on four benchmarks:
LCC \citep{guo2023longcoder} and HumanEval \citep{chen2021codex} for code
generation, Finance QA for question answering, and Multi-News
\citep{fabbri2019multinews} for summarization (the last two from WaterBench,
\citealp{tu2023waterbench}).
Denoiser and planner parameters are aligned, subject to method-specific
differences.

\noindent\textbf{Baselines.} The baselines include position-seeded KGW and Gumbel
\citep{kirchenbauer2023watermark,gloaguen2026wmdlm,wu2025dmark,bagchi2025position}.
Additional baselines are dgMARK \citep{hong2026dgmark},
PATTERN-MARK \citep{chen2025patternmark}, and HW
\citep{tsur2025heavywater}.
The tables report token-commit and block-ready results for KGW, Gumbel,
PATTERN-MARK, and HW, with coverage varying by model and task.
dgMARK is evaluated in token-commit mode.
These four baselines watermark each token position separately under both
insertion timings. In block-ready mode, they resample each position from
its recorded distribution after watermarking it, once the block is complete.
In token-commit mode, they watermark the positions selected for the current
update just before sampling.
}

{
The evaluation covers token-commit \methodname{} (Section~\ref{sec:method-commit-time})
and three block-ready configurations: \textit{chain}, \textit{middle-star},
and \textit{balanced tree}.
Unwatermarked generations provide a reference for generation quality.
Tables~\ref{tab:experiments-performance-compact} and
\ref{tab:experiments-multinews-compact} present the LCC and Multi-News results.
\par
}

{
\noindent\textbf{Evaluation:}
Following the setup in Appendix~\ref{sec:appendix-experimental-settings},
we report detection using AUROC and true-positive rates at empirical thresholds
targeting 1\%, 5\%, and 10\% false-positive rates (TPR@1, TPR@5, and TPR@10).
The same set of unwatermarked generations is used to select these thresholds
and measure their false-positive rates. Separately, each text's empirical $p$-value is based
on how many reference clean scores equal or exceed its detector score.
The mean $-\log_{10}(p)$ is computed over successfully scored watermarked
texts, with higher values indicating stronger evidence of watermarking.

Generation quality is measured by completion-only cross-entropy (CE),
scored by a quantized LLaMA-3.3-70B-Instruct model and averaged in nats per
token. Lower CE indicates better generation quality. Downstream task
performance is evaluated using pass@1 on HumanEval, code similarity on LCC,
and ROUGE-L F1 on Finance QA and Multi-News.
\par
}

\begin{table}[t]
\centering
\begingroup
\setlength{\tabcolsep}{2pt}
\renewcommand{\arraystretch}{1.10}
\caption{\textbf{LCC detection and code quality.}
\methodname{} variants are (Co) commit, (Ch) \textit{chain}, (S) \textit{middle-star}, and
(B) \textit{balanced tree}. Co+tilt and Ch+tilt use post-MOT strengths
$\delta=0.25$ and $\delta=0.5$, respectively, on both models.
Best and second-best watermarked values are bold and
underlined.}
\label{tab:experiments-performance-compact}
\begin{minipage}[t]{0.49\linewidth}
\centering
\vspace{0pt}
\textbf{(a) LLaDA-Instruct}\par\vspace{2pt}
\resizebox{\linewidth}{!}{%
\begin{tabular}{@{}lccccc@{\hspace{3pt}}cc@{}}
\toprule
Method & \shortstack{TPR@1\\(\%)} & \shortstack{TPR@5\\(\%)} &
\shortstack{TPR@10\\(\%)} & AUC & \shortstack{Mean\\$\text{-}\log_{10}p$} & CE & \shortstack{Code\\sim.} \\
\midrule
\ResultRow{Clean reference & -- & -- & -- & -- & -- & \ppltoce{2.026} & 58.0}
\midrule
\multicolumn{8}{@{\hspace{3pt}}l}{\emph{Token-commit}} \\
\ResultRow{KGW & 5.0 & 15.0 & 28.5 & 0.652 & 0.814 & \ppltoce{2.653} & 55.8}
\ResultRow{PATTERN-M & 16.0 & 27.5 & 38.0 & 0.693 & 0.853 &
\ppltoce{2.345} & \bestvalue{58.5}}
\ResultRow{dgMARK & 13.5 & 29.0 & 40.5 & 0.770 & 0.975 &
\bestvalue{\ppltoce{2.132}} & 55.5}
\ResultRow{Gumbel & 12.0 & 27.0 & 39.0 & 0.761 & 1.260 &
\secondvalue{\ppltoce{2.320}} & 56.6}
\ResultRow{HW & 13.0 & 19.5 & 29.0 & 0.705 & 1.014 &
\ppltoce{2.445} & 57.2}
\ResultRow{\methodname{} (Co) & 4.8 & 39.4 & 54.3 &
0.837 & 1.130 & 0.948 & 56.8}
\ResultRow{\methodname{} (Co+tilt) & 18.1 & 53.2 &
\secondvalue{67.0} & \secondvalue{0.893} & \secondvalue{1.408} &
1.117 & 53.6}
\midrule
\multicolumn{8}{@{\hspace{3pt}}l}{\emph{Block-ready}} \\
\ResultRow{KGW & \secondvalue{30.3} & 41.0 &
56.9 & 0.858 & 1.262 &
\ppltoce{4.615} & 53.4}
\ResultRow{Gumbel & 6.4 & 13.3 & 30.9 &
0.641 & 0.702 & \ppltoce{2.498} &
\secondvalue{57.3}}
\ResultRow{PATTERN-M & 11.2 & 29.8 & 40.4 & 0.741 & 0.875 &
\ppltoce{9.405} & 51.7}
\ResultRow{HW & 17.0 & 37.0 & 50.5 &
0.766 & 1.319 & -- & 54.7}
\ResultRow{\methodname{} (Ch) & 17.0 &
52.1 & 64.4 &
0.846 & 1.311 & 1.280 & 55.3}
\ResultRow{\methodname{} (Ch+tilt) & \bestvalue{52.1} &
\bestvalue{82.4} & \bestvalue{89.4} & \bestvalue{0.959} &
\bestvalue{1.819} & 1.672 & 51.9}
\ResultRow{\methodname{} (S) & 16.0 & 41.0 & 54.8 &
0.805 & 1.148 & 1.271 & 54.9}
\ResultRow{\methodname{} (B) & 26.6 &
\secondvalue{56.9} & 63.3 &
0.856 & 1.387 & 1.321 & 54.5}
\bottomrule
\end{tabular}%
}
\end{minipage}%
\hfill
\begin{minipage}[t]{0.49\linewidth}
\centering
\vspace{0pt}
\textbf{(b) LLaDA 2.1}\par\vspace{2pt}
\resizebox{\linewidth}{!}{%
\begin{tabular}{@{}lccccc@{\hspace{3pt}}cc@{}}
\toprule
Method & \shortstack{TPR@1\\(\%)} & \shortstack{TPR@5\\(\%)} &
\shortstack{TPR@10\\(\%)} & AUC & \shortstack{Mean\\$\text{-}\log_{10}p$} & CE & \shortstack{Code\\sim.} \\
\midrule
\ResultRow{Clean reference & -- & -- & -- & -- & -- & \ppltoce{1.872} & 40.5}
\midrule
\multicolumn{8}{@{\hspace{3pt}}l}{\emph{Token-commit}} \\
\ResultRow{KGW & 51.1 & 85.6 & 88.8 & 0.960 & 1.849 &
\ppltoce{3.075} & \secondvalue{40.9}}
\ResultRow{Gumbel & 2.1 & 21.3 & 30.9 & 0.676 & 0.762 &
\bestvalue{\ppltoce{1.911}} & \bestvalue{42.3}}
\ResultRow{dgMARK & -- & -- & -- & -- & -- & -- & --}
\ResultRow{PATTERN-M & 45.2 & 64.4 & 75.0 &
0.900 & 1.475 & 1.791 & 35.5}
\ResultRow{HW & \bestvalue{98.4} & \bestvalue{98.4} & \bestvalue{98.4} &
\bestvalue{0.993} & \bestvalue{2.155} & \ppltoce{130.916} & 28.1}
\ResultRow{\methodname{} (Co) & 8.5 & 36.7 & 51.6 & 0.788 &
1.112 & \secondvalue{0.677} & 40.6}
\ResultRow{\methodname{} (Co+tilt) & 24.5 & 50.5 & 68.6 & 0.856 &
1.379 & 0.823 & 40.1}
\midrule
\multicolumn{8}{@{\hspace{3pt}}l}{\emph{Block-ready}} \\
\ResultRow{KGW & 82.4 & \secondvalue{93.6} & \secondvalue{94.7} &
\secondvalue{0.980} & 2.039 &
\ppltoce{11.934} & 31.3}
\ResultRow{Gumbel & 12.8 & 36.2 & 47.3 & 0.766 & 1.022 & \ppltoce{2.703} & 37.1}
\ResultRow{PATTERN-M & 84.0 & 87.8 & 89.4 &
0.963 & 1.913 & 3.074 & 25.1}
\ResultRow{HW & \secondvalue{96.3} & \bestvalue{98.4} & \bestvalue{98.4} &
\bestvalue{0.993} & \secondvalue{2.146} & \ppltoce{161.694} & 26.5}
\ResultRow{\methodname{} (Ch) & 13.8 & 29.8 & 43.1 & 0.757 &
1.005 & 0.705 & 39.7}
\ResultRow{\methodname{} (Ch+tilt) & 86.2 & 92.6 & 93.1 & 0.971 &
2.038 & 1.357 & 37.3}
\ResultRow{\methodname{} (S) & 38.3 & 50.5 & 56.9 & 0.848 &
1.262 & \ppltoce{2.389} & 38.9}
\ResultRow{\methodname{} (B) & 21.8 & 40.4 & 53.7 & 0.804 & 1.136 &
\ppltoce{2.500} & 39.5}
\bottomrule
\end{tabular}%
}
\end{minipage}
\endgroup
\end{table}

\begin{table}[t]
\centering
\begingroup
\setlength{\tabcolsep}{2pt}
\renewcommand{\arraystretch}{1.08}
\caption{\textbf{Multi-News detection and text quality.}
\methodname{} variants are (Co) commit, (Ch) \textit{chain}, (S) \textit{middle-star}, and
(B) \textit{balanced tree}. Best and second-best watermarked values are bold and
underlined. Co+tilt and Ch+tilt use post-MOT strengths $\delta=0.25$ and
$\delta=0.5$, respectively, on both models.}
\label{tab:experiments-multinews-compact}
\begin{minipage}[t]{0.49\linewidth}
\centering
\vspace{0pt}
\textbf{(a) LLaDA-Instruct}\par\vspace{2pt}
\resizebox{\linewidth}{!}{%
\begin{tabular}{@{}lccccc@{\hspace{3pt}}cc@{}}
\toprule
Method & \shortstack{TPR@1\\(\%)} & \shortstack{TPR@5\\(\%)} &
\shortstack{TPR@10\\(\%)} & AUC & \shortstack{Mean\\$\text{-}\log_{10}p$} & CE & \shortstack{ROUGE\\L} \\
\midrule
\ResultRow{Clean reference & -- & -- & -- & -- & -- & \ppltoce{1.815} & 26.9}
\midrule
\multicolumn{8}{@{\hspace{3pt}}l}{\emph{Token-commit}} \\
\ResultRow{KGW & 39.5 & 59.5 & 65.5 & 0.864 & 1.387 & \ppltoce{2.581} & 26.1}
\ResultRow{Gumbel & 63.5 & 79.5 & 83.0 & 0.931 & 1.770 & \ppltoce{2.358} & 26.2}
\ResultRow{dgMARK & \secondvalue{83.0} & \secondvalue{91.0} &
\secondvalue{95.0} & 0.974 & \secondvalue{1.998} &
\ppltoce{2.687} & 26.3}
\ResultRow{PATTERN-M & 14.5 & 56.5 & 60.5 & 0.834 & 1.235 &
\ppltoce{2.415} & 26.5}
\ResultRow{HW & -- & -- & -- & -- & -- & -- & --}
\ResultRow{\methodname{} (Co) & 72.0 & 78.0 & 86.0 &
0.941 & 1.834 & 1.040 & 26.0}
\ResultRow{\methodname{} (Co+tilt) & 80.5 & 84.5 & 88.5 &
0.953 & 1.942 & 1.129 & 24.6}
\midrule
\multicolumn{8}{@{\hspace{3pt}}l}{\emph{Block-ready}} \\
\ResultRow{KGW & 47.0 & 65.0 & 77.5 & 0.905 & 1.531 & \ppltoce{2.727} & 25.6}
\ResultRow{Gumbel & 20.5 & 31.0 & 40.5 & 0.740 & 0.948 &
\bestvalue{\ppltoce{1.970}} & \bestvalue{26.8}}
\ResultRow{PATTERN-M & 20.5 & 43.0 & 57.0 & 0.820 & 1.165 &
\ppltoce{5.568} & 23.5}
\ResultRow{HW & \bestvalue{97.0} & \bestvalue{98.0} & \bestvalue{98.0} & \bestvalue{0.994} &
\bestvalue{2.162} & \ppltoce{5.915} & 23.8}
\ResultRow{\methodname{} (Ch) & 51.5 & 63.5 & 69.5 &
0.885 & 1.533 & 0.772 & 26.4}
\ResultRow{\methodname{} (Ch+tilt) & \secondvalue{83.0} & 89.0 & 93.0 &
\secondvalue{0.975} & 1.997 & 0.930 & 26.3}
\ResultRow{\methodname{} (S) & 38.0 & 47.0 & 57.5 &
0.812 & 1.295 & \secondvalue{0.759} & 26.2}
\ResultRow{\methodname{} (B) & 43.5 & 64.5 & 72.5 &
0.895 & 1.549 & 0.815 & \secondvalue{26.6}}
\bottomrule
\end{tabular}%
}
\end{minipage}%
\hfill
\begin{minipage}[t]{0.49\linewidth}
\centering
\vspace{0pt}
\textbf{(b) LLaDA 2.1}\par\vspace{2pt}
\resizebox{\linewidth}{!}{%
\begin{tabular}{@{}lccccc@{\hspace{3pt}}cc@{}}
\toprule
Method & \shortstack{TPR@1\\(\%)} & \shortstack{TPR@5\\(\%)} &
\shortstack{TPR@10\\(\%)} & AUC & \shortstack{Mean\\$\text{-}\log_{10}p$} & CE & \shortstack{ROUGE\\L} \\
\midrule
\ResultRow{Clean reference & -- & -- & -- & -- & -- & \ppltoce{4.128} & 24.7}
\midrule
\multicolumn{8}{@{\hspace{3pt}}l}{\emph{Token-commit}} \\
\ResultRow{KGW & 96.0 & \bestvalue{99.5} & \bestvalue{100.0} &
\secondvalue{0.999} & 2.181 & \ppltoce{5.581} & 24.4}
\ResultRow{Gumbel & 13.5 & 50.5 & 60.5 & 0.839 & 1.184 &
\bestvalue{\ppltoce{4.205}} & \bestvalue{24.8}}
\ResultRow{dgMARK & -- & -- & -- & -- & -- & -- & --}
\ResultRow{PATTERN-M & -- & -- & -- & -- & -- & -- & --}
\ResultRow{HW & -- & -- & -- & -- & -- & -- & --}
\ResultRow{\methodname{} (Co) & 61.0 & 78.0 & 89.5 & 0.952 & 1.796 &
1.561 & \secondvalue{24.7}}
\ResultRow{\methodname{} (Co+tilt) & 84.5 & 94.5 & 97.5 & 0.991 &
2.067 & 1.655 & 24.5}
\midrule
\multicolumn{8}{@{\hspace{3pt}}l}{\emph{Block-ready}} \\
\ResultRow{KGW & \bestvalue{99.5} & \bestvalue{99.5} & \bestvalue{100.0} &
\bestvalue{1.000} & \bestvalue{2.200} & \ppltoce{24.521} & 23.7}
\ResultRow{Gumbel & 26.5 & 56.0 & 67.0 & 0.885 & 1.359 & \ppltoce{5.765} &
24.6}
\ResultRow{PATTERN-M & -- & -- & -- & -- & -- & -- & --}
\ResultRow{HW & -- & -- & -- & -- & -- & -- & --}
\ResultRow{\methodname{} (Ch) & 19.5 & 50.5 & 64.0 &
0.856 & 1.249 & \secondvalue{1.489} & 24.6}
\ResultRow{\methodname{} (Ch+tilt) & 97.5 & \secondvalue{99.0} &
\secondvalue{99.5} & 0.997 & 2.180 & 1.819 & 24.4}
\ResultRow{\methodname{} (S) & \secondvalue{99.0} & \bestvalue{99.5} &
\secondvalue{99.5} & \secondvalue{0.999} & \secondvalue{2.191} &
\ppltoce{16.530} & 23.7}
\ResultRow{\methodname{} (B) & 62.0 & 71.0 & 82.5 & 0.935 & 1.652 &
\ppltoce{7.095} & 24.6}
\bottomrule
\end{tabular}%
}
\end{minipage}
\endgroup
\end{table}

\subsection{Detection--Quality Tradeoff}
\label{sec:results-frontier}

{
\noindent\textbf{Code Generation Results:}
Code generation is a challenging domain for highly detectable watermarks
because token distributions can have low entropy \citep{tsur2025heavywater}.
We focus on LCC, where task performance is
measured by code similarity.
Appendix~\ref{sec:appendix-full-results} reports HumanEval results using
pass@1.
}

{
As seen in Table~\ref{tab:experiments-performance-compact}, \methodname{}
offers a favorable trade-off between watermark detectability and code
quality on LCC.
With LLaDA-Instruct, \textit{chain} achieves higher TPR@5\%
(52.1\% vs.\ 41.0\%), higher code similarity (55.3 vs.\ 53.4), and lower
CE than block-ready KGW.
Some baselines detect more strongly but incur larger quality losses.
The comparison between \methodname{} insertion timings depends on the
model: on LLaDA-Instruct, \textit{chain} improves TPR@5\% over token-commit
at the cost of code similarity, while on LLaDA 2.1, token-commit improves
both TPR@5\% (36.7\% vs.\ 29.8\%) and code similarity (40.6 vs.\ 39.7).
\par

On LCC with LLaDA-Instruct, tilting \textit{chain} raises TPR@5\% from
52.1\% to 82.4\% while reducing code similarity from 55.3 to 51.9.
With LLaDA 2.1, tilted \textit{chain} nearly matches block-ready KGW's
TPR@5\% (92.6\% vs.\ 93.6\%) while retaining higher code similarity
(37.3 vs.\ 31.3).
Appendix~\ref{sec:appendix-post-fit-tilting} compares detection and
generation quality across tilt strengths.
\par
}

{
\noindent\textbf{Summarization.}
On Multi-News with LLaDA-Instruct (Table~\ref{tab:experiments-multinews-compact}),
\textit{chain} nearly matches block-ready KGW in TPR@5\% (63.5\% vs.\ 65.0\%),
with higher ROUGE-L and lower CE.
Among the token-commit methods, dgMARK outperforms both \methodname{}
variants on TPR@5\%, ROUGE-L, and CE.
\ifPDFTeX\else\par\fi
}
\setlength{\intextsep}{3pt}
\setlength{\parskip}{3pt}
Tilting \textit{chain} raises TPR@5\% to 89.0\% on LLaDA-Instruct and 99.0\%
on LLaDA 2.1, with small decreases in ROUGE-L. On LLaDA 2.1, tilting also
raises token-commit TPR@5\% from 78.0\% to 94.5\%, but increases CE for both
insertion timings.

Token-commit KGW reaches 99.5\% TPR@5\% at the same
reported ROUGE-L as tilted \textit{chain} and with lower CE.
Appendix~\ref{sec:appendix-full-results} reports the Finance QA results.
\par

\subsection{Robustness to edits}
\label{sec:results-robustness}

\begin{wraptable}[\ifPDFTeX26\else25\fi]{R}{0.42\textwidth}
% Align the caption with the subsection heading above the wrap anchor.
\vspace*{-21pt}
\begingroup
\setlength{\tabcolsep}{1.5pt}
\renewcommand{\arraystretch}{1.10}
\setlength{\abovecaptionskip}{0pt}
\setlength{\belowcaptionskip}{4pt}
\caption{\textbf{Robustness to edits on LLaDA-Instruct.}
TPR (\%) is reported at fixed cutoffs targeting 5\% FPR on unedited
matched-clean text. FPR may change after editing. Random deletion,
insertion, and substitution, as well as BERT word substitution, use a
20\% edit rate. Back-trans. denotes English-German-English back-translation,
and Rewrite uses Qwen2.5-1.5B-Instruct. \methodname{}$^\dagger$ uses
block-index-seeded \textit{chain} with post-MOT tilt $\delta=1$.}
\label{tab:attack-main-summary}

\begin{minipage}[t]{\linewidth}
{\fontsize{6.6}{8}\selectfont
\begin{tabular*}{\linewidth}{@{\extracolsep{\fill}}lrrrrrr@{}}
\toprule
Method & Del. & Ins. & Subst. & \shortstack{BERT\\sub.}
& \shortstack{Back-\\trans.} & Rewrite \\
\midrule
\multicolumn{7}{@{}c@{}}{\textbf{(a) Finance QA}} \\
\multicolumn{7}{@{\hspace{2pt}}l}{\emph{Token-commit}} \\
LR-DWM & 21.0 & 31.0 & 25.0 & 26.5 & 23.5 & 2.5 \\
DMark & \textbf{91.5} & \textbf{90.5} & 82.5 & 86.0 & \textbf{71.0} & 15.5 \\
dgMARK & 21.0 & 30.5 & 33.0 & 32.5 & 30.0 & 10.5 \\
\midrule
\multicolumn{7}{@{\hspace{2pt}}l}{\emph{Block-ready}} \\
\methodname{}$^\dagger$ & 74.5 & 80.5 & \textbf{95.5} & \textbf{94.0} &
62.0 & \textbf{20.5} \\
\midrule
\multicolumn{7}{@{}c@{}}{\textbf{(b) Multi-News}} \\
\multicolumn{7}{@{\hspace{2pt}}l}{\emph{Token-commit}} \\
LR-DWM & 20.5 & 29.5 & 22.0 & 28.5 & 20.5 & 1.0 \\
DMark & \textbf{82.5} & \textbf{83.0} & 79.5 & 81.0 & 70.0 & 11.0 \\
dgMARK & 30.0 & 35.5 & 46.5 & 48.0 & 26.0 & 14.5 \\
\midrule
\multicolumn{7}{@{\hspace{2pt}}l}{\emph{Block-ready}} \\
\methodname{}$^\dagger$ & 70.0 & 79.5 & \textbf{97.5} & \textbf{98.0} &
\textbf{72.0} & \textbf{18.0} \\
\bottomrule
\end{tabular*}}
\end{minipage}
\par\vspace{4pt}
\endgroup
\end{wraptable}

One way to improve robustness to edits is to design shared randomness that
the detector can recover from edited text \citep{kirchenbauer2024reliability}.
\methodname{}'s watermark design is agnostic to the choice of
shared-randomness distribution and hashing scheme.
Here, we derive each block's shared state from its index and the watermark key.
To assess robustness to text edits, we compare \methodname{} with three
baselines under random edits, BERT word substitution, back-translation,
and rewriting. Table~\ref{tab:attack-main-summary} shows that block-ready
\methodname{} achieves the highest TPR under both substitution attacks,
retaining 95.5\% on Finance QA and 97.5\% on Multi-News after random
substitution.
DMark, however, leads under deletion and insertion. Rewriting reduces TPR
to at most 20.5\% across the compared configurations.
In Appendix~\ref{sec:appendix-attack-robustness}, we provide additional
implementation details and robustness results.
\par

\makeatletter
\@startsection{section}{1}{\z@}{-0.8ex}{0.4ex}{\large\sc\raggedright}{Conclusions}
\makeatother
\label{sec:discussion}
\label{sec:conclusion}

In this paper, we addressed the problem of watermarking DLMs by exploiting
their parallel generation mechanism. We introduced \methodname{}, a novel
watermarking scheme, leveraging MOT over sparse graphical structures to
jointly and efficiently watermark multiple tokens at once rather than
watermarking each token independently. We proved that the proposed
block-level watermark is distortion-free. Our empirical results demonstrate
that \methodname{} can offer a favorable detectability-quality trade-off
compared to existing watermarking methods. Furthermore, our investigation
of token-commit and block-ready watermarking highlights insertion timing
as an additional design dimension for DLMs, affecting both watermark
performance and computational overhead. Overall, our findings highlight
the potential of joint token watermarking as an alternative to conventional
token-level watermarking schemes.

\WFclear
\endgroup

\setcounter{topnumber}{2}

\clearpage
\flushbottom
\bibliographystyle{paper_references}
\begingroup
\setlength{\linewidth}{\textwidth}
\bibliography{main}
\endgroup

\appendix
\clearpage
\pdfbookmark[0]{Appendix contents}{appendix-contents}
\section*{Appendix contents}
\begingroup

\makeatletter
\setcounter{tocdepth}{3}
\newcommand{\AppendixContentsSection}[2]{%
  \par\addvspace{0.8em}\noindent
  \hyperref[#1]{\textbf{\makebox[1.5em][l]{\ref*{#1}}#2}}%
  \hfill\hyperref[#1]{\textbf{\pageref*{#1}}}\par
}
\newcommand{\AppendixContentsEntry}[2]{%
  \@dottedtocline{2}{1.5em}{2.8em}%
    {\hyperref[#1]{\numberline{\ref*{#1}}#2}}%
    {\hyperref[#1]{\pageref*{#1}}}%
}
\newcommand{\AppendixContentsSubentry}[2]{%
  \@dottedtocline{3}{3em}{3.8em}%
    {\hyperref[#1]{\numberline{\ref*{#1}}#2}}%
    {\hyperref[#1]{\pageref*{#1}}}%
}
\makeatother
\AppendixContentsSection{sec:related-work}{Related Work}
\AppendixContentsSection{sec:appendix-theory}{Theoretical Details}
\AppendixContentsEntry{sec:appendix-notation}{Notation}
\AppendixContentsEntry{sec:appendix-marginal-preservation}{Proof of marginal preservation}
\AppendixContentsEntry{sec:appendix-entropic-mot}{Entropy-regularized multi-marginal optimal transport}
\AppendixContentsEntry{sec:appendix-tree-algorithms}{Sinkhorn-BP and conditional tree sampling}
\AppendixContentsEntry{sec:appendix-analytical-detection}{Analytical detection and numerical evaluation}
\AppendixContentsEntry{sec:appendix-commit-time}{Token-commit generation and detection}
\AppendixContentsEntry{sec:appendix-graph-complexity}{Operation counts for tree-structured MOT}
\AppendixContentsSection{sec:appendix-experiments}{Experimental Setup, Ablations, and Limitations}
\AppendixContentsEntry{sec:appendix-experimental-settings}{Experimental configurations}
\AppendixContentsEntry{sec:appendix-full-results}{Additional benchmark results}
\AppendixContentsEntry{sec:appendix-deferred-resampling}{Deferred-resampling control}
\AppendixContentsEntry{sec:appendix-threshold-transfer}{Threshold transfer}
\AppendixContentsEntry{sec:appendix-state-size}{State-size sensitivity}
\AppendixContentsEntry{sec:appendix-temperature-sensitivity}{Temperature sensitivity}
\AppendixContentsEntry{sec:appendix-post-fit-tilting}{Post-MOT tilting}
\AppendixContentsEntry{sec:appendix-eb-sampler}{EB-sampler}
\AppendixContentsEntry{sec:appendix-commit-graph-scope}{Token-commit graph scope}
\AppendixContentsEntry{sec:appendix-attack-robustness}{Robustness to attacks}
\AppendixContentsSubentry{sec:attack-protocol}{Attack protocol and state construction}
\AppendixContentsSubentry{sec:attack-detection-quality}{Detection and quality under attacks}
\AppendixContentsSubentry{sec:attack-state-edge-recovery}{State and edge recovery under edits}
\AppendixContentsSubentry{sec:attack-topology-sensitivity}{Topology sensitivity}
\AppendixContentsSubentry{sec:attack-llada21-transfer}{Transfer to LLaDA2.1}
\AppendixContentsSubentry{sec:attack-seeding-comparison}{Seeding comparison}
\AppendixContentsEntry{sec:appendix-regularization-tradeoff}{Regularization tradeoff}
\AppendixContentsEntry{sec:results-cost}{Runtime cost}
\AppendixContentsEntry{sec:limitations}{Limitations and Scope}
\endgroup
\clearpage
\section{Related Work}
\label{sec:related-work}

{
\noindent\textbf{Coupling-based watermarking.}
Watermarking can be formulated as a coupling between a generated token
and secret side information, with prescribed marginals controlling
the token distribution \citep{tsur2025optimized}.
HeavyWater and SimplexWater develop this approach for low-entropy
next-token distributions through score-based optimal transport
\citep{tsur2025heavywater}.
They use different watermark score constructions to compute
token-side-information couplings and detect the resulting dependence
between the generated text and the secret.
Both methods apply these couplings token by token during
autoregressive generation.
\par
}

{
\noindent\textbf{Context-aware watermarking.}
Several diffusion watermarks adapt KGW-style green-list biasing to
partially revealed contexts \citep{kirchenbauer2023watermark}.
DMark constructs keyed token-pair preferences from revealed neighbors,
with variants that also use predictions for masked neighbors
\citep{wu2025dmark}.
LR-DWM derives separate green-list biases from revealed left and right
neighbors, using distinct keys for the two directions
\citep{raban2026lrdwm}.
WM-DLM accounts for unresolved context by averaging context-dependent
watermark biases over context uncertainty \citep{gloaguen2026wmdlm}.
\par
}

{
\noindent\textbf{Watermark signal construction.}
Watermark signals can take the form of vocabulary-color patterns or
pseudorandom sampling perturbations.
PATTERN-MARK uses a key-generated sequence of vocabulary colors to
assign watermark preferences to output positions, allowing tokens to
be generated in arbitrary order \citep{chen2025patternmark}.
Pseudorandom Gumbel perturbations indexed by token position guide token
sampling during diffusion \citep{bagchi2025position}.
SAC-Copula introduces dependence between these perturbations through
a Gaussian copula, producing a spatially correlated Gumbel field
rather than independent noise across positions \citep{li2026saccopula}.
Rather than modifying token logits, dgMARK embeds the signal through
the token-commit schedule, selecting positions using parity matching
between token identities and positions \citep{hong2026dgmark}.
\par
}

{
\noindent\textbf{Candidate-selection watermarking.}
Watermarking can also operate through selection among generated chunks.
WaterMax retains candidates with strong keyed watermark scores
\citep{giboulot2024watermax}, while WaterSearch generates continuations
under different watermark seeds and selects among them using both
fidelity and watermark strength \citep{lin2025watersearch}.
\par
}

{
\noindent\textbf{Graph-based Optimal Transport.}
Graph-structured costs enable multimarginal optimal transport algorithms
that combine marginal scaling with BP, including tree-based methods
and junction-tree extensions \citep{haasler2021multimarginal,fan2022complexity}.
Polynomial-time algorithms have also been established for structured
classes of multimarginal transport problems
\citep{altschuler2023polynomial}.
A complementary approach, Neural Entropic MOT (NEMOT), uses neural
networks trained on mini-batches to estimate entropic multimarginal
transport, with non-asymptotic error bounds \citep{tsur2025neural}.
\par
}

\section{Theoretical Details}
\label{sec:appendix-theory}

\subsection{Notation}
\label{sec:appendix-notation}
We use uppercase letters for random variables and lowercase letters for
their realizations. The following table summarizes the notation used in
the formulation, solver, and detector.
\suppressfloats[t]
\begin{table}[p]
\centering
\caption{\textbf{Notation.} Symbols used in the formulation, solver, and detector.}
\label{tab:notation}
\begingroup
\small
\setlength{\tabcolsep}{5pt}
\renewcommand{\arraystretch}{1.08}
\begin{tabular}{@{}p{0.23\linewidth}p{0.73\linewidth}@{}}
\toprule
Symbol & Meaning \\
\midrule
\multicolumn{2}{@{}l}{\textbf{Tokens and watermark coupling}} \\
\midrule
$\mathcal X$ & Token vocabulary. \\
$b$ & Number of token positions in the block formulation; nominal block length when partitioning a sequence. \\
$t_i$ & Token at position $i$ before watermarking. \\
$p_i$ & Recorded token distribution at position $i$. \\
$X_i,\ x_i$ & Watermarked token random variable and its realization. \\
$X^b,\ x^b$ & Tuple of $b$ watermarked token variables and its realization. \\
$A_i=\operatorname{supp}(p_i)$ & Candidate support at position $i$. \\
$\mathcal S,\ S,\ s$ & Finite side-information set, state random variable, and realized state. \\
$P_S$ & Prescribed side-information distribution, uniform over $\mathcal S$. \\
$\kappa$ & Secret watermark key. \\
$\pi,\ \pi^\star$ & Joint token-state coupling and an optimizer of the unregularized MOT objective. \\
$\pi_\varepsilon^\star$ & Optimal entropy-regularized coupling. \\
$\widehat\pi$ & Numerical coupling computed by the solver. \\
$\varepsilon$ & Entropy-regularization strength. \\
$H(\pi)$ & Entropy of the joint coupling. \\
\midrule
\multicolumn{2}{@{}l}{\textbf{Graph and solver}} \\
\midrule
$\mathcal G=(\mathcal V,\mathcal E)$ & Token interaction graph, its nodes, and its edges. \\
$\mathcal I$ & Index pairs included in the score decomposition. \\
$f(x^b,s)$ & Block watermark score. \\
$f(x_i,x_j,s)$ & Token-pair score contributing to the block score. \\
$\psi_{ij}(x_i,x_j,s)$ & Pair kernel $\exp(f(x_i,x_j,s)/\varepsilon)$. \\
$u_i,\ v$ & Token and side-information scaling factors. \\
$q_i,\ q_S$ & Token and side-information marginals of the current coupling. \\
$\operatorname{ne}(i)$ & Neighbors of node $i$. \\
$m_{i\to j}(x_j,s)$ & BP message from node $i$ to node $j$. \\
$\delta,\ \tau$ & Current maximum marginal error and stopping tolerance. \\
\midrule
\multicolumn{2}{@{}l}{\textbf{Detection}} \\
\midrule
$x,\ G=|x|$ & Observed generated sequence and its token count. \\
$L=\lceil G/b\rceil$ & Number of observed blocks. \\
$b_k$ & Observed token count in block $k$. \\
$\mathcal E_k$ & Scored edges in block $k$. \\
$s_k$ & Reproduced watermark state for block $k$. \\
$h_k(s;x)$ & Block-$k$ pair-score sum evaluated at state $s$. \\
$r_k$ & Block-$k$ pair-score sum evaluated at the reproduced state. \\
$T_{\mathrm{obs}}$ & Observed sequence statistic, $G^{-1}\sum_k r_k$. \\
$T_0$ & Corresponding random statistic under the analytical null. \\
$p_0(t;x)$ & Inclusive analytical upper-tail probability at threshold $t$, conditional on $x$. \\
$\alpha$ & Detection significance threshold. \\
$\Phi,\ z_{\mathrm{eq}}$ & Standard normal CDF and Gaussian-equivalent score $z_{\mathrm{eq}}=\Phi^{-1}(1-p_0)$. \\
\bottomrule
\end{tabular}
\endgroup
\end{table}

\begingroup

\subsection{Proof of marginal preservation}
\label{sec:appendix-marginal-preservation}

{
This section restates the marginal-preservation guarantee in
Equation~\ref{eq:method-marginal-preservation} and proves it for the
conditional sampling procedure of \methodname.

\noindent\textbf{Proposition~\ref{prop:method-distortion-free} (restated).}
Fix the recorded token distributions \(p_1,\ldots,p_b\) and a coupling
\(\pi^\star\) satisfying the token and side-information marginal
constraints in \eqref{eq:method-mot-objective}. Draw one shared state
\(S\sim P_S\), then sample \(X^b\) from the corresponding conditional
distribution in \eqref{eq:method-mot-conditional}.
Averaging the position-\(i\) marginal of the sampled block distribution
over the shared side information recovers the recorded token distribution:
\[
  \mathbb E_{S\sim P_S}\!
    \left[
      \sum_{\substack{x^b\in\mathcal X^b\\x_i=x}}
      \frac{\pi^\star(x^b,S)}{P_S(S)}
    \right]
  =p_i(x),
  \qquad i=1,\ldots,b,\quad x\in\mathcal X.
\]

\noindent\textit{Proof.}
Hold the recorded distributions \(p_1,\ldots,p_b\) and the feasible coupling
\(\pi^\star\) fixed before drawing the shared state \(S\sim P_S\).
Since \(P_S\) is uniform on finite nonempty \(\mathcal S\),
\(P_S(s)=1/|\mathcal S|>0\).

The conditional in Equation~\ref{eq:method-mot-conditional} is
\[
  P_{X^b\mid S=s}(x^b)
  =
  \frac{\pi^\star(x^b,s)}{P_S(s)}.
\]
The state-marginal constraint ensures normalization:
\[
  \sum_{x^b\in\mathcal X^b}
  \frac{\pi^\star(x^b,s)}{P_S(s)}
  =
  \frac{P_S(s)}{P_S(s)}
  =
  1.
\]

For any position \(i\) and token \(x\):
\[
\begin{aligned}
  \Pr(X_i=x)
  &=
  \mathbb E_{S\sim P_S}
  \left[
    \sum_{\substack{x^b\in\mathcal X^b\\x_i=x}}
    \frac{\pi^\star(x^b,S)}{P_S(S)}
  \right]
  \\[6pt]
  &=
  \sum_{s\in\mathcal S} P_S(s)
  \sum_{\substack{x^b\in\mathcal X^b\\x_i=x}}
  \frac{\pi^\star(x^b,s)}{P_S(s)}
  \\[6pt]
  &=
  \sum_{s\in\mathcal S}
  \sum_{\substack{x^b\in\mathcal X^b\\x_i=x}}
  \pi^\star(x^b,s)
  \\[6pt]
  &=p_i(x).
\end{aligned}
\]
The factors \(P_S(s)\) cancel, and the final equality follows from
the token-marginal constraint in Equation~\ref{eq:method-mot-objective}.
Thus, averaging over the shared side information recovers each
recorded token distribution:
\[
  \Pr(X_i=x)=p_i(x),
  \qquad i=1,\ldots,b,\quad x\in\mathcal X.
\]
}
\endgroup

\begingroup

\subsection{Entropy-regularized multi-marginal optimal transport}
\label{sec:appendix-entropic-mot}

For a general score \(f:\mathcal X^b\times\mathcal S\to\mathbb R\) and
\(\varepsilon>0\), the entropy-regularized MOT problem is
\citep{benamou2015iterative,tsur2025heavywater}
\begin{equation}
  \begin{aligned}
    \pi_\varepsilon^\star
      &=\operatorname*{arg\,min}_{\pi}
        \left\{
          \mathbb E_{\pi}[-f(X^b,S)]-\varepsilon H(\pi)
        \right\},\\
    \text{subject to}\quad
      &\pi_i=p_i\quad(i=1,\ldots,b),\qquad \pi_S=P_S,
  \end{aligned}
  \label{eq:appendix-entropic-mot}
\end{equation}
where the optimization is over joint probability distributions and
\begin{equation}
  H(\pi)
  =
  -\sum_{x^b\in\mathcal X^b}\sum_{s\in\mathcal S}
  \pi(x^b,s)\log\pi(x^b,s).
  \label{eq:appendix-joint-entropy}
\end{equation}
Regularization changes the objective, not the marginal constraints.
\par
\endgroup

\subsection{Sinkhorn-BP and conditional tree sampling}
\label{sec:appendix-tree-algorithms}

{
Algorithm~\ref{alg:appendix-sinkhorn-bp} solves the graph-based OT problem:
BP computes the current token and side-information marginals, and Sinkhorn
updates the scaling factors to bring them toward their targets,
\(p_1,\ldots,p_b\) and \(P_S\)
\citep{haasler2021multimarginal,benamou2015iterative}.
Algorithm~\ref{alg:appendix-bp-marginals} implements the BP computation:
for the current scaling factors, it passes messages along the tree for
each side-information value and combines the results to obtain those
marginals \citep{yedidia2003understanding,haasler2021multimarginal}.
\par
}

{
\noindent\textbf{Factorized coupling.}
Work on the finite supports \(A_i=\operatorname{supp}(p_i)\), with
\(p_i(x_i)>0\) for \(x_i\in A_i\), and the uniform state distribution \(P_S\).
Let \(\varepsilon>0\), assume finite pair scores, and define positive,
fixed pair kernels
\begin{equation}
  \psi_{ij}(x_i,x_j,s)
  =
  \exp\!\left(\frac{f(x_i,x_j,s)}{\varepsilon}\right),
  \qquad \{i,j\}\in\mathcal E.
  \label{eq:appendix-pair-kernel}
\end{equation}
The token graph is a connected tree for each fixed state.
The lookup \(\psi_{ji}(x_j,x_i,s)=\psi_{ij}(x_i,x_j,s)\) reads the same stored
table with its axes exchanged.

For current positive scalings \(u_i\) and \(v\), write
\begin{equation}
  W_{u,v}(x^b,s)
  =
  v(s)\prod_{i=1}^{b}u_i(x_i)
  \prod_{\{i,j\}\in\mathcal E}\psi_{ij}(x_i,x_j,s),
  \qquad
  \widehat\pi_{u,v}=
  \frac{W_{u,v}}{\sum_{x^b,s}W_{u,v}(x^b,s)},
  \label{eq:appendix-current-tree-coupling}
\end{equation}
Normalization can equivalently be absorbed into \(v\).
The scaling vectors enforce marginal constraints, they are not
additional watermark scores.
\par
}

{
\noindent\textbf{Sinkhorn updates.}
For the current normalized coupling \(\widehat\pi=\widehat\pi_{u,v}\),
let \(q_i=\widehat\pi_i\) and \(q_S=\widehat\pi_S\) denote its token
and side-information marginals. The update
\(u_i(a)\leftarrow u_i(a)p_i(a)/q_i(a)\) gives
\begin{equation}
  \widehat\pi'(x^b,s)
  =
  \widehat\pi(x^b,s)\frac{p_i(x_i)}{q_i(x_i)}.
  \label{eq:appendix-coordinate-scaling}
\end{equation}
The updated distribution is normalized and its position-\(i\)
marginal is \(p_i\). The state update is analogous.
BP recomputes the marginals after each coordinate update.
Full BP calls make this ordering explicit; unchanged
messages can be reused without substituting stale marginals.
A maximum iteration count limits solver latency, even if the
prescribed tolerance has not been reached.
The stopping test uses the current marginal error
\[
  \delta
  =
  \max\left\{\max_i\|q_i-p_i\|_\infty,\,
                 \|q_S-P_S\|_\infty\right\}.
\]
Algorithm~\ref{alg:appendix-sinkhorn-bp} adapts tree-structured Sinkhorn
scaling to the prescribed token and side-information marginals
\citep{haasler2021multimarginal,benamou2015iterative}. In the algorithm,
\(q_{1:b}=(q_1,\ldots,q_b)\) denotes the current token marginals.
\par
}

\begin{algorithm}[H]
\caption{Sinkhorn-BP solver}
\label{alg:appendix-sinkhorn-bp}
\small
\begin{algorithmic}[1]
  \Require Token marginals \(p_1,\ldots,p_b\), state marginal \(P_S\)
  \Ensure Computed coupling \(\widehat\pi\)
  \State Initialize \(u_i\gets1\) for each \(i\), and \(v\gets1\)
  \Repeat
    \For{\(i=1,\ldots,b\)}
      \State \((q_{1:b},q_S)\gets\Call{BP-Marginals}{u,v}\)
      \State \(u_i\gets u_i\odot(p_i/q_i)\)
    \EndFor
    \State \((q_{1:b},q_S)\gets\Call{BP-Marginals}{u,v}\)
    \State \(v\gets v\odot(P_S/q_S)\)
  \Until{current marginal error \(\leq\tau\) or the iteration budget is reached}
  \State \Return normalized coupling \(\widehat\pi\) defined by the computed factors
\end{algorithmic}
\end{algorithm}

{
\noindent\textbf{BP marginal computation.}
Let \(\operatorname{ne}(i)\) be the neighbors of token node \(i\).
The index \(\ell\in\operatorname{ne}(i)\) runs over these neighboring nodes.
In the message equation, \(x_i\in A_i\) and \(x_j\in A_j\) are candidate
token values at the sending and receiving nodes, respectively.
For each fixed state \(s\), the message from \(i\) to \(j\) is
\begin{equation}
  m_{i\to j}(x_j,s)
  =
  \sum_{x_i\in A_i}
  u_i(x_i)\psi_{ij}(x_i,x_j,s)
  \prod_{\ell\in\operatorname{ne}(i)\setminus\{j\}}
  m_{\ell\to i}(x_i,s).
  \label{eq:appendix-bp-message}
\end{equation}
An empty product equals one. Computing messages toward a chosen root
and then away from it yields all directed messages
\citep{haasler2021multimarginal}.
The resulting node weights are
\(\beta_i(x_i,s)=u_i(x_i)\prod_{\ell\in\operatorname{ne}(i)}m_{\ell\to i}(x_i,s)\).
They are unnormalized weights, not the final token marginals.
BP computes the token weights separately for each side-information value.
Algorithm~\ref{alg:appendix-bp-marginals} applies standard sum-product
belief propagation to the state-conditioned trees
\citep{yedidia2003understanding,haasler2021multimarginal}.
It combines these results using each state's current weight to obtain
the token and side-information marginals
(the target side-information distribution \(P_S\) is uniform).
\par
}

\begin{algorithm}[H]
\caption{BP-Marginals}
\label{alg:appendix-bp-marginals}
\small
\begin{algorithmic}[1]
  \Statex \textbf{Fixed:} Tree \(\mathcal G=(\mathcal V,\mathcal E)\), supports \(A_1,\ldots,A_b\), kernels \(\psi\), root \(r\)
  \Require Current scaling factors \(u,v\)
  \Ensure Current token marginals \(q_{1:b}\) and side-information marginal \(q_S\)
  \State Root the tree at \(r\); define \(\operatorname{parent}(i)\) and \(\operatorname{child}(i)\)
  \For{\(s\in\mathcal S\)}
    \For{each child node \(i\) in leaves-to-root order}
      \State Compute \(m_{i\to\operatorname{parent}(i)}(\cdot,s)\) by Equation~\ref{eq:appendix-bp-message}
        \Comment{Subtree to parent}
    \EndFor
    \For{each parent node \(i\) in root-to-leaves order}
      \State Compute \(m_{i\to j}(\cdot,s)\), \(j\in\operatorname{child}(i)\), by Equation~\ref{eq:appendix-bp-message}
        \Comment{Parent to children}
    \EndFor
    \State For every \(i\in\mathcal V\) and \(x_i\in A_i\), set
    \Statex \hspace{\algorithmicindent}\(\beta_i(x_i,s)\gets u_i(x_i)\prod_{\ell\in\operatorname{ne}(i)}m_{\ell\to i}(x_i,s)\)
      \Comment{Unnormalized token weights}
    \State \(Z_s\gets\sum_{x_r\in A_r}\beta_r(x_r,s)\)
      \Comment{Total weight at fixed \(s\)}
  \EndFor
  \State \(Z\gets\sum_{s\in\mathcal S}v(s)Z_s\)
  \State For every \(i\in\mathcal V\) and \(x_i\in A_i\), set
  \Statex \hspace{\algorithmicindent}\(q_i(x_i)\gets Z^{-1}\sum_{s\in\mathcal S}v(s)\beta_i(x_i,s)\)
  \State \(q_S(s)\gets v(s)Z_s/Z\), for every \(s\)
  \State \Return \(q_{1:b},q_S\)
\end{algorithmic}
\end{algorithm}

{
\noindent\textbf{Sampling from the computed conditional distribution.}
Given a side-information value \(s\), the final tree factors define the
conditional distribution \(\widehat\pi(X^b\mid S=s)\).
The factor \(v(s)\) cancels when conditioning on this state.
Scores are attached to child-to-parent edges, so the root's weights
combine its token scaling with the incoming BP messages without an
additional root-only score. Each child is then sampled conditional on
its parent's token, with weights combining its token scaling, the
connecting pair factor and messages from its own children - normalizing
these weights gives the categorical probabilities used to sample the
output tokens. Algorithm~\ref{alg:appendix-tree-sampling} summarizes this
conditional tree-sampling procedure.
\par
}

\begin{algorithm}[H]
\caption{Conditional tree sampling}
\label{alg:appendix-tree-sampling}
\small
\begin{algorithmic}[1]
  \Statex \textbf{Fixed:} Traversal root \(r\)
  \Require Computed factors of \(\widehat\pi\), side-information value \(s\)
  \Ensure A sample \(x_1,\ldots,x_b\) from \(\widehat\pi(X^b\mid S=s)\)
  \State Root the tree at \(r\); define \(\operatorname{parent}(i)\) and \(\operatorname{child}(i)\)
  \State Compute leaves-to-root messages for state \(s\) from the computed factors using Equation~\ref{eq:appendix-bp-message}
  \State \(w_r(x_r)\gets u_r(x_r)\prod_{\ell\in\operatorname{child}(r)}m_{\ell\to r}(x_r,s)\), for every \(x_r\in A_r\)
  \State Draw \(x_r\sim w_r/\sum_{x_r\in A_r}w_r(x_r)\)
  \For{each child node \(j\) in root-to-leaves order, with \(i=\operatorname{parent}(j)\)}
    \State \(w_j(x_j)\gets u_j(x_j)\psi_{ij}(x_i,x_j,s)\prod_{\ell\in\operatorname{child}(j)}m_{\ell\to j}(x_j,s)\), for every \(x_j\in A_j\)
    \State Draw \(x_j\sim w_j/\sum_{x_j\in A_j}w_j(x_j)\)
  \EndFor
  \State \Return \(x_1,\ldots,x_b\)
\end{algorithmic}
\end{algorithm}

{
\noindent\textbf{Example: sampling and detecting a four-token block.}
Consider a block-ready watermark with four positions connected as a chain:
\[
X_1\rightarrow X_2\rightarrow X_3\rightarrow X_4.
\]
Four positions are used only for illustration.
After solving MOT across all states, the key and side-information rule
determine the block's state \(s_k\).
\par
\noindent\textbf{Sampling.}
For the selected state \(s_k\), the computed \textit{chain} distribution factorizes as
\[
\widehat\pi(x_1,x_2,x_3,x_4\mid s_k)
=q_1(x_1\mid s_k)\prod_{i=2}^{4}q_i(x_i\mid x_{i-1},s_k).
\]
Here, \(q_1\) is the computed root marginal and \(q_i\) is the computed
conditional distribution of token \(i\) given its parent.
Tokens are drawn in order:
\[
\begin{aligned}
x_1&\sim q_1(\cdot\mid s_k),&
x_2&\sim q_2(\cdot\mid x_1,s_k),\\
x_3&\sim q_3(\cdot\mid x_2,s_k),&
x_4&\sim q_4(\cdot\mid x_3,s_k).
\end{aligned}
\]
BP computes these probabilities from the computed factors, including
messages from positions not yet sampled.
The pair scores \(f\) influence these probabilities, but tokens are
sampled, not chosen by maximizing \(f\).
\par
\noindent\textbf{Detection.}
Given the final tokens, the key and the public block settings, the detector
reconstructs \(s_k\) and the three pairs connecting positions
\((1,2)\), \((2,3)\) and \((3,4)\).
It adds their scores using the declared argument order.
It does not need the saved model predictions, computed factors or
denoising history.
\par
}

\subsection{Analytical detection and numerical evaluation}
\label{sec:appendix-analytical-detection}

{
This section describes scoring and analytical calibration
without clean-reference texts.
\par
\noindent\textbf{The finite-state null.}
Fix the observed generated tokens \(x\), block graphs, and pair-score
construction. Let \(G=|x|>0\), \(b>1\) be the block length, and
\(L=\lceil G/b\rceil\) the number of blocks. Define the pair-score sum
for block \(k\) as
\begin{equation}
  h_k(s;x)=
  \sum_{(c,p)\in\mathcal E_k}
  f(x_{k,c},x_{k,p},s).
  \label{eq:appendix-block-pair-score}
\end{equation}
Under the analytical null, the block states are conditionally independent
and uniform over \(\mathcal S\). For the \(L\) observed blocks,
\begin{equation}
  T_0=\frac{1}{G}\sum_{k=1}^{L}h_k(S_k;x),
  \qquad
  S_k\overset{\mathrm{iid}}{\sim}\operatorname{Unif}(\mathcal S).
  \label{eq:appendix-null-statistic}
\end{equation}
The shared state retains dependence among edge scores within each block;
independence is assumed only across block states.
\par

{
\noindent\textbf{Sequence score.}
Let \(b_k\) be the number of observed tokens in block \(k\), with
\(\sum_k b_k=G\).
The sequence score weights each block's per-token score by its token count:
\begin{equation}
  T_{\mathrm{obs}}
  =
  \sum_{k=1}^{L}\frac{b_k}{G}\frac{h_k(s_k;x)}{b_k}
  =
  \frac{1}{G}\sum_{k=1}^{L}h_k(s_k;x).
  \label{eq:appendix-token-normalized-score}
\end{equation}
Thus, every scored pair contributes to the total, which is divided by the
generated-token count \(G\).
\par
}

{
\noindent\textbf{Moment-standardized score.}
For each block, the mean and variance of its score are calculated across
all states:
\begin{equation}
  \mu_k=\frac{1}{|\mathcal S|}\sum_s h_k(s;x),
  \qquad
  v_k=\frac{1}{|\mathcal S|}\sum_s\bigl(h_k(s;x)-\mu_k\bigr)^2.
  \label{eq:appendix-null-score-moments}
\end{equation}
Under the independent-block-state null, for \(\sum_k v_k>0\), the sequence
score is standardized as
\begin{equation}
  Z=\frac{\sum_k h_k(s_k;x)-\sum_k\mu_k}
  {\sqrt{\sum_k v_k}}.
  \label{eq:appendix-moment-standardized-score}
\end{equation}
The block variances retain dependence between edges sharing a state.
\par
}

\noindent\textbf{Analytical p-value.}
The observed statistic \(T_{\mathrm{obs}}\) is defined in
Section~\ref{sec:method-sampling-detection}.
Its analytical p-value is the probability that the null score is at least
as large as the observed score:
\begin{equation}
  p_0(T_{\mathrm{obs}};x)
  =
  \frac{1}{|\mathcal S|^L}
  \sum_{s_1,\ldots,s_L\in\mathcal S}
  \mathbf 1\!\left\{
    \frac{1}{G}\sum_{k=1}^{L}h_k(s_k;x)
    \geq T_{\mathrm{obs}}
  \right\}.
  \label{eq:appendix-finite-state-tail}
\end{equation}
For a fixed text with positive null variance, evaluating
this tail using \(Z\) gives the same p-value as using the original score.
\par

\noindent\textbf{Numerical evaluation.}
For each block, evaluating \(h_k(s;x)\) over all states gives its finite
score distribution under the uniform-state null.
Convolving the block distributions and dividing the sum by \(G\) gives
the distribution of \(T_0\).

\noindent\textbf{Gaussian-equivalent significance.}
For \(0<p_0<1\), the equivalent-normal score is
\begin{equation}
  z_{\mathrm{eq}}=\Phi^{-1}(1-p_0),
  \label{eq:appendix-equivalent-z}
\end{equation}
Here, \(\Phi\) is the standard-normal cumulative distribution
function.
This transformation reports the calculated tail probability on a normal
scale.
\par

}

\subsection{Token-commit generation and detection}
\label{sec:appendix-commit-time}

{
Token-commit \methodname applies the coupling during denoising rather than
after block completion. In a parallel unmasking update, newly revealed
tokens receive predictions from the same pre-update context
\citep{sahoo2024mdlm}. No newly sampled token in that update has yet
served as context for another selected token's prediction. Replacing
these proposals therefore does not invalidate within-update conditioning
on their newly sampled values. Adaptive schedules, including
entropy-budget unmasking, vary which positions are updated together.
Subsequent model updates can condition on the watermarked tokens,
without implying independence in the target text distribution or
guaranteeing unchanged generation quality \citep{benhamu2025ebsampler,luxembourg2026dus,wu2025fastdllm}.

Earlier watermarked tokens can carry information about the reused state.
Given later predictions, some state values may therefore be more likely
than others.
Consequently, enforcing the marginal constraints in each new coupling does
not establish preservation of the actual sampled token marginals during
adaptive token-commit generation.

{\noindent\textbf{Token-commit span.}
For a nonempty set \(C\) selected by the native sampler, let
\(b_{\mathrm{end}}\) be the last position in the current live block. The
token-commit span configuration solves MOT on a contiguous rightward \textit{chain} over
\[
V_{\mathrm{span}} =
\{\min(C)-1,\ldots,\min(\max(C)+1,b_{\mathrm{end}})\}.
\]
The predecessor may lie in the prompt or preceding block. The graph includes
the selected positions and any written or masked context in the span. We solve
MOT using current-forward token distributions and the current
block's shared state. During conditional sampling, unselected written
positions are clamped to their actual tokens, unselected masked positions
are marginalized, and only the selected positions are written back.} MOT optimization
and conditional sampling follow Appendix~\ref{sec:appendix-tree-algorithms};
rightward \textit{chain} orientation does not change the bidirectional BP
computation. LLaDA 1.0 updates previously masked positions, whereas
LLaDA 2.1 also permits revisions of already visible tokens. For a revision,
the model may have used the token's old value when predicting its replacement.

\noindent\textbf{Token-commit block.}
For the token-commit block comparator, each update solves MOT on a \textit{chain} over the
entire current live block and its predecessor. As in token-commit span,
previously written positions that are not selected for revision stay at
their observed tokens and are not resampled. Unselected masked positions
are marginalized, and only positions selected by the native sampler are
written back. The two configurations differ in the scope of the graph
used for MOT. Both apply the coupling during denoising.

\noindent\textbf{Singleton updates.}
For a singleton update, the same span includes the selected position's
predecessor and, when available within the live block, its successor.
The conditioning and marginalization rules above apply without a
preview-and-discard step.

\noindent\textbf{Public detection.}
The final text does not reveal the commit set or the graphs used for individual updates.
Using the key, the public detector scores adjacent token pairs, including
predecessor edges at prompt and block boundaries, with the current block's
watermark state. It requires neither the commit set nor the denoising
history. The detector excludes the first terminal token and all subsequent
positions.

\noindent\textbf{Cost.}
The per-sweep bound in Appendix~\ref{sec:appendix-graph-complexity} applies
to each update's span \textit{chain}. Total optimization work sums these costs over
updates and solver sweeps. The public predecessor-chain detector scores a
linear number of adjacent edges in the generated length.
\par
}
\begingroup
\definecolor{commitTeal}{HTML}{286F68}
\definecolor{commitTealWash}{HTML}{F0F7F5}
\definecolor{commitGray}{HTML}{A3A7AA}
\definecolor{commitGrayWash}{HTML}{ECEEEF}
\definecolor{commitGold}{HTML}{B8860B}
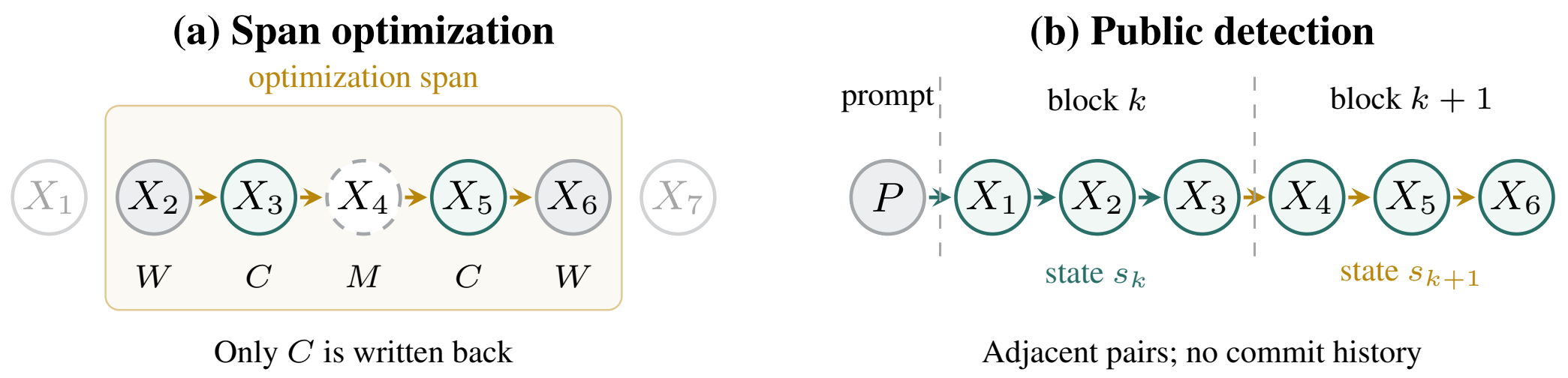
\begin{figure}[t]

\centering
\resizebox{\linewidth}{!}{%
\begin{tikzpicture}[
  x=1cm,y=1cm,text=black,
  token/.style={
    circle,draw=commitTeal,fill=commitTealWash,line width=.75pt,
    minimum size=5.6mm,inner sep=0pt,outer sep=1pt,font=\small
  },
  written/.style={
    token,draw=commitGray,fill=commitGrayWash
  },
  masked/.style={
    token,draw=commitGray,dashed,fill=white
  },
  outside/.style={
    token,draw=commitGray!50,fill=white,text=black!35
  },
  fitting/.style={draw=commitGold,line width=.8pt,->,>=stealth},
  scoring/.style={draw=commitTeal,line width=.8pt,->,>=stealth},
  boundary/.style={draw=commitGray,dashed,line width=.5pt},
  panel/.style={font=\small\bfseries},
  role/.style={font=\scriptsize}
]
\path[use as bounding box] (-.45,-1.5) rectangle (11.65,1.55);

\node[panel] at (2.4,1.25) {(a) Span optimization};
\filldraw[fill=commitGold!5,draw=commitGold!45,rounded corners=3pt]
  (.43,-.86) rectangle (4.37,.7);
\node[role,text=commitGold] at (2.4,.9) {optimization span};
\foreach \i in {1,7}
  \node[outside] (a\i) at ({.8*(\i-1)},0) {$X_{\i}$};
\foreach \i in {2,6}
  \node[written] (a\i) at ({.8*(\i-1)},0) {$X_{\i}$};
\foreach \i in {3,5}
  \node[token] (a\i) at ({.8*(\i-1)},0) {$X_{\i}$};
\node[masked] (a4) at (2.4,0) {$X_4$};
\foreach \i/\j in {2/3,3/4,4/5,5/6}
  \draw[fitting] (a\i.east) -- (a\j.west);
\foreach \i/\r in {2/W,3/C,4/M,5/C,6/W}
  \node[role] at ({.8*(\i-1)},-.6) {$\r$};
\node[role] at (2.4,-1.2) {Only $C$ is written back};

\begin{scope}[xshift=6.4cm]
  \node[panel] at (2.4,1.25) {(b) Public detection};
  \node[written] (b0) at (0,0) {$P$};
  \foreach \i in {1,...,6}
    \node[token] (b\i) at ({.8*\i},0) {$X_{\i}$};
  \foreach \i/\j in {0/1,1/2,2/3}
    \draw[scoring] (b\i.east) -- (b\j.west);
  \foreach \i/\j in {3/4,4/5,5/6}
    \draw[scoring,draw=commitGold] (b\i.east) -- (b\j.west);
  \draw[boundary] (.4,-.45) -- (.4,.95);
  \draw[boundary] (2.8,-.45) -- (2.8,.95);
  \node[role] at (0,.75) {prompt};
  \node[role] at (1.6,.75) {block $k$};
  \node[role] at (4,.75) {block $k+1$};
  \node[role,text=commitTeal] at (1.6,-.6) {state $s_k$};
  \node[role,text=commitGold] at (4,-.6) {state $s_{k+1}$};
  \node[role] at (2.4,-1.2) {Adjacent pairs; no commit history};
\end{scope}
\end{tikzpicture}%
}
\caption{\textbf{Token-commit span coupling and public detection.}
(a) For selected set \(C=\{3,5\}\), the optimization span includes positions
\(2\) through \(6\). Unselected written context (\(W\)) is clamped,
selected positions (\(C\)) are sampled and written back, and unselected
masked context (\(M\)) is marginalized. (b) The public detector scores
adjacent eligible pairs, including prompt and block-boundary predecessor
edges, using the current block's watermark state.}
\label{fig:commit-time-graphs}
\end{figure}
\endgroup

\begingroup

\subsection{Operation counts for tree-structured MOT}
\label{sec:appendix-graph-complexity}

{
Consider the coupling on fixed, nonempty candidate supports
\(A_i\). Assume that pair factors are already available and that BP
reuses intermediate products of incoming messages. Count scalar
arithmetic and table accesses at unit cost.

\noindent\textbf{Proposition.}
One complete bidirectional BP sweep over all side-information states
computes the token and state marginals in
\begin{equation}
  O\!\left(
    |\mathcal S|
    \left[
      \sum_{i=1}^{b}|A_i|
      +
      \sum_{(i,j)\in\mathcal E}|A_i||A_j|
    \right]
  \right)
  \label{eq:appendix-tree-sweep}
\end{equation}
operations \citep{haasler2021multimarginal}. Explicit storage of pair
factors, scaling vectors, messages, and beliefs has the same asymptotic
bound.

\noindent\textit{Proof.}
For a fixed state, a message from position \(i\) to position \(j\)
sums over \(|A_i|\) candidates for each of its \(|A_j|\) outputs.
Processing both directions of every edge over all states therefore
gives the pair-product term. Reusing incoming-message products keeps
message preparation and storage within the same asymptotic bound as
the pairwise contractions.
Pair-factor storage contributes the pair-product term, while beliefs
and scaling vectors are covered by the node term.

A tree has \(b-1\) edges, giving the uniform upper bound
\begin{equation}
  O\!\left(
    |\mathcal S|\,b
    \left(\max_{1\leq i\leq b}|A_i|\right)^2
  \right).
  \label{eq:appendix-tree-uniform}
\end{equation}
Thus, for bounded candidate and state counts, a sweep is linear in
block size. \textit{chains}, \textit{middle-stars}, and \textit{balanced trees} share this bound,
although unequal support sizes can produce different edge costs.
In contrast, explicitly storing the dense coupling requires
\(|\mathcal S|\prod_i|A_i|\) entries.

Total optimization work depends on the number of BP sweeps required by the update schedule
and monitoring procedure. Factor construction, model inference,
token sampling, and detection are extra boilerplate operations.
\par
}

\endgroup

\section{Experimental Setup, Ablations, and Limitations}
\label{sec:appendix-experiments}
\setcounter{topnumber}{3}
\setcounter{totalnumber}{3}

\subsection{Experimental configurations}
\label{sec:appendix-experimental-settings}

{
\noindent\textbf{Models and generation settings.}
The experiments use LLaDA 1.0-Instruct and LLaDA 2.1-mini.
Current block-ready \methodname{} configurations process text in blocks of 32 tokens.
For code generation, the maximum output length is 256 tokens for LLaDA 1.0
and 4,096 tokens for LLaDA 2.1; question answering and summarization use a
512-token limit for both models.
All LLaDA 1.0 \methodname{} benchmarks use low-confidence unmasking with at
most two positions per update, for both token-commit and block-ready generation.
Entropy-budget sampling is evaluated in a dedicated ablation.
For LLaDA 1.0, the requested denoising budgets are 128 steps for code
generation and 256 steps for question answering and summarization.
LLaDA 2.1 uses its native adaptive sampler, which permits token revisions.
The selected Commit and \textit{chain} configurations use unmasking and editing
confidence thresholds of $0.7$ and $0.5$, respectively, with at most
16 post-refinement steps.
Model-specific prompt templates and sampling settings are retained, with
watermark parameters and clean controls specified below.
\par
}

{
\noindent\textbf{Watermark configurations.}
All MOT solver runs use at most 300 iterations with early stopping.
For LLaDA 1.0, \methodname{} constructs token marginals using the input
temperatures specified for each dataset and watermarking mode, and samples
watermarked tokens from the computed coupling with output temperature fixed at 1.
Coupling uses 256 shared states, support restricted 20 tokens per
position, score normalization, and entropy regularization $\epsilon=0.03$.
The stopping tolerance is $10^{-5}$.
The selected LLaDA 2.1 Commit and \textit{chain} configurations also use 256 shared
states, support restricted 20 tokens per position, score normalization,
entropy regularization $\epsilon=0.03$, and output temperature 1.
No post-MOT tilt is applied to the base benchmark configurations.
For both models, tilted Token-commit and Block-ready \textit{chain} variants use original-score post-MOT
strengths $\delta=0.25$ and $\delta=0.5$, respectively, shared across datasets.
Each variant retains its corresponding base configuration's remaining settings.
The robustness comparison also applies original-score post-MOT tilt
$\delta=0.25$ to MinHash token-commit and $\delta=0.5$ to MinHash block-ready.
The stronger MinHash variants in Tables~\ref{tab:attack-20-summary}
and~\ref{tab:attack-quality} and Figure~\ref{fig:attack-edit-rates}
use $\delta=0.5$ for token-commit and $\delta=1$ for block-ready \textit{chain},
retaining their corresponding base settings, matched-clean references and
frozen unedited-clean cutoffs.
This post-MOT modification is distinct from the base coupling to which the
marginal-preservation proposition applies.
\par
}

{
\noindent\textbf{Score normalization.}
For each fixed token pair, reproducible scores are generated across watermark
states.
The mean and standard deviation are computed over those states, separately
for each pair.
Subtracting this mean and dividing by this standard deviation gives the
standardized scores used by both optimization and detection.
\par
Optimization additionally range-normalizes each retained candidate-pair/state
table:
\[
r_e=\frac{f_e-f_e^{\max}}{f_e^{\max}-f_e^{\min}}\in[-1,0].
\]
The extrema are taken over the retained candidate pairs and all states
for that edge.
A constant table contributes zero score.
This bounds the scores used in optimization and controls their scale relative to
entropy regularization.
Rescaling scores while keeping regularization fixed can change the
computed coupling.
\par
Detection uses the standardized pair scores directly, without this extra
optimization-specific scaling.
Clean reference texts are scored in the same way when setting detection
thresholds.
\par
}

{
\noindent\textbf{\methodname{} solver input temperatures.}
\begin{center}
{\fontsize{6.6}{8}\selectfont
\setlength{\tabcolsep}{5pt}
\renewcommand{\arraystretch}{1.10}
\begin{tabular}{@{}lcccc@{}}
\toprule
& \multicolumn{2}{c}{LLaDA-Instruct} & \multicolumn{2}{c}{LLaDA2.1} \\
\cmidrule(lr){2-3}\cmidrule(lr){4-5}
Dataset & Token-commit & \textit{chain} & Token-commit & \textit{chain} \\
\midrule
HumanEval & 0.4 & 0.3 & 0.4 & 0.7 \\
LCC & 0.6 & 0.6 & 0.4 & 0.7 \\
Finance QA & 0.6 & 0.3 & 0.5 & 0.7 \\
Multi-News & 0.6 & 0.5 & 0.5 & 0.7 \\
\bottomrule
\end{tabular}
}
\end{center}
\textit{chain} denotes block-ready insertion.
\textit{balanced tree} and \textit{middle-star} share \textit{chain}'s input temperature within each dataset.
Watermarked sampling temperature is fixed at 1 for the configurations listed above.
\par
}

{
\noindent\textbf{Baseline settings.}
For LLaDA 1.0 code generation, KGW, Gumbel, and token-level HW use an input
temperature of $0.5$.
PATTERN-MARK and dgMARK use greedy selection on LCC. On HumanEval, both use
the low-confidence planner with two tokens per commit and top-$k=3$ token
sampling.
LLaDA 2.1 retains its native adaptive planner and separate sampling and
revision temperatures.
All HW configurations use $\epsilon=0.05$ and per-token range normalization,
applied separately to each token-by-state score table.
\par
}

{
\noindent\textbf{Clean controls.}
Each configuration uses its designated unwatermarked reference generated
with the corresponding scheduler.
HumanEval Commit and Commit+tilt detection use the designated archived clean
reference.
Empirical detection thresholds are selected and evaluated on the same clean
cohort, without held-out FPR guarantees.
\par
}

{
\noindent\textbf{Empirical p-values and calibration.}
For the selected base \methodname{} benchmarks, each text's detector score
is compared with scores from clean texts generated for other prompts in
the same benchmark:
\[
p=\frac{1+\#\{\text{reference clean scores}\geq\text{text score}\}}
{1+\#\{\text{reference clean scores}\}}.
\]
The reported mean averages $-\log_{10}(p)$ over successfully scored
watermarked texts.
TPR thresholds are selected and evaluated on the same full clean set.
These results do not independently validate false-positive rates on new text.
\par
}

{
\noindent\textbf{Numerical accuracy.}
Numerical accuracy is evaluated for the four selected untilted
LLaDA-Instruct block-ready \textit{chain} configurations.
The marginal-constraint residual measures the largest absolute deviation of a computed
token or state marginal from its target:
\[
e_{\mathrm{fit}}=\max\left\{
\max_{i,a}\left|\widehat\pi_i(a)-p_i(a)\right|,
\max_{s\in\mathcal S}\left|\widehat\pi_S(s)-\frac{1}{|\mathcal S|}\right|
\right\}.
\]
The final-sampler error measures the largest deviation from the target
token marginals after averaging the sampler's probabilities over states:
\[
e_{\mathrm{sample}}=\max_{i,a}\left|
\frac{1}{|\mathcal S|}\sum_{s\in\mathcal S}q_i(a\mid s)-p_i(a)
\right|.
\]
Here, $q_i(a\mid s)$ is the final sampler's token probability for state $s$.
These probabilities are reconstructed from the computed distribution,
rather than estimated from sampled-token frequencies.
Table~\ref{tab:numerical-accuracy} summarizes the measured errors.
Means and maxima summarize per-run errors, including solver runs that did not
converge within the 300-iteration budget.
Corresponding marginal-constraint residual means and maxima agree at the displayed
precision.
The requested stopping tolerance is $10^{-5}$, not an achieved bound for
every solver run.
\par
}
\begin{table}[ht]
\centering
\begingroup

\setlength{\tabcolsep}{5pt}
\renewcommand{\arraystretch}{1.10}
\caption{\textbf{Numerical accuracy of selected LLaDA-Instruct \textit{chain} configurations.}
Mean iterations and error summaries include all MOT solver runs, including
nonconverged runs. The maximum budget is 300 iterations per run.
Converged runs satisfy the requested stopping tolerance of $10^{-5}$.}
\label{tab:numerical-accuracy}
{\fontsize{6.6}{8}\selectfont
\begin{tabular}{@{}lrrrr@{}}
\toprule
Dataset & \shortstack{Mean iterations\\ / 300} & Mean $e_{\mathrm{sample}}$
& Max $e_{\mathrm{sample}}$ & Converged (\%) \\
\midrule
HumanEval & 54.62 & $8.69\times10^{-6}$ & $6.76\times10^{-5}$ & 99.55 \\
LCC & 82.94 & $9.97\times10^{-6}$ & $2.87\times10^{-4}$ & 96.35 \\
Finance QA & 64.14 & $8.41\times10^{-6}$ & $1.66\times10^{-4}$ & 98.96 \\
Multi-News & 62.63 & $8.37\times10^{-6}$ & $1.88\times10^{-4}$ & 98.54 \\
\bottomrule
\end{tabular}
}
\endgroup
\end{table}

\subsection{Additional benchmark results}
\label{sec:appendix-full-results}

Tables~\ref{tab:appendix-preview-humaneval} and
\ref{tab:appendix-preview-finance-qa} report HumanEval and Finance QA results
for both evaluated models, complementing the LCC and Multi-News results from
Section~\ref{sec:results}. Each table presents detection performance,
CE, and task performance, measured by pass@1 for
HumanEval and ROUGE-L for Finance QA. Results are divided into token-commit
and block-ready watermarking.

\begingroup
\color{black}
\newcommand{\bestvalue}[1]{{\bfseries\boldmath #1}}
\newcommand{\secondvalue}[1]{\underline{#1}}
\newcommand{\recommendedvalue}[1]{\textcolor{green!50!black}{#1}}
\newcommand{\newvalue}[1]{#1}
\newcommand{\unverifiedlogp}[1]{--}
\newcommand{\ppltoce}[1]{%
  \pgfmathparse{ln(#1)}%
  \pgfmathprintnumber[fixed,precision=3,fixed zerofill]{\pgfmathresult}%
}
{
HumanEval exposes the functional-correctness cost accompanying stronger
detection (Table~\ref{tab:appendix-preview-humaneval}).
With LLaDA-Instruct, untilted \textit{balanced tree} retains 30.5\% pass@1 at
23.8\% TPR@5\%.
Token-commit dgMARK offers much stronger detection, at 81.1\% TPR,
but lower pass@1, at 22.0\%.
The choice between these operating points therefore depends on the balance
between detection and successful code generation.
\par

On LLaDA 2.1, untilted token-commit \methodname{} and \textit{chain} retain pass@1
of 77.4\% and 79.3\%, respectively, but both reach only 14.6\% TPR@5\%.
Tilting changes this balance differently across timings.
Token-commit TPR increases to 19.5\% while pass@1 decreases to 75.0\%.
\textit{chain} instead reaches 80.5\% TPR, accompanied by a substantial pass@1
decline from 79.3\% to 34.8\%.
\par
}
\begin{table}[ht]
\centering
\begingroup
\setlength{\tabcolsep}{2pt}
\renewcommand{\arraystretch}{1.10}
\caption{\textbf{HumanEval detection and code quality.}
\methodname{} variants are (Co) commit, (Ch) \textit{chain}, (S) \textit{middle-star}, and
(B) \textit{balanced tree}. Co+tilt and Ch+tilt use post-MOT strengths
$\delta=0.25$ and $\delta=0.5$, respectively, on both models.
Best and second-best watermarked values are bold and
underlined.}
\label{tab:appendix-preview-humaneval}
\begin{minipage}[t]{0.49\linewidth}
\centering
\vspace{0pt}
\textbf{(a) LLaDA-Instruct}\par\vspace{2pt}
\resizebox{\linewidth}{!}{%
\begin{tabular}{@{}lccccc@{\hspace{3pt}}cc@{}}
\toprule
Method & \shortstack{TPR@1\\(\%)} & \shortstack{TPR@5\\(\%)} &
\shortstack{TPR@10\\(\%)} & AUC & \shortstack{Mean\\$\text{-}\log_{10}p$} & CE & \shortstack{pass@1\\(\%)} \\
\midrule
\ResultRow{Clean reference & -- & -- & -- & -- & -- & \ppltoce{2.970} & 35.4}
\midrule
\multicolumn{8}{@{\hspace{3pt}}l}{\emph{Token-commit}} \\
\ResultRow{KGW & 12.8 & 43.3 & 51.8 & 0.803 & 1.024 &
1.450 & 19.5}
\ResultRow{PATTERN-M & 7.3 & 27.4 & 46.3 & 0.758 & 0.758 &
1.304 & 23.8}
\ResultRow{dgMARK & \secondvalue{65.2} & \secondvalue{81.1} &
\secondvalue{88.4} & \secondvalue{0.962} & \secondvalue{1.729} &
1.438 & 22.0}
\ResultRow{Gumbel & 37.2 & 55.5 & 62.8 & 0.869 & 1.327 &
1.348 & 24.4}
\ResultRow{HW & 2.4 & 7.3 & 9.1 & 0.505 & 0.445 &
2.967 & \bestvalue{34.8}}
\ResultRow{\methodname{} (Co) & 6.1 & 25.6 & 33.5 & 0.701 & 0.814 &
\bestvalue{1.037} & \secondvalue{30.5}}
\ResultRow{\methodname{} (Co+tilt) & 4.9 & 30.5 & 49.4 & 0.777 & 0.999 &
\secondvalue{1.091} & 26.8}
\midrule
\multicolumn{8}{@{\hspace{3pt}}l}{\emph{Block-ready}} \\
\ResultRow{KGW & \bestvalue{71.3} & \bestvalue{89.6} & \bestvalue{95.1} &
\bestvalue{0.979} & \bestvalue{1.899} & \ppltoce{9.375} & 4.9}
\ResultRow{Gumbel & 9.8 & 15.2 & 23.8 & 0.725 & 0.797 &
\ppltoce{3.648} & 20.1}
\ResultRow{PATTERN-M & 22.0 & 39.0 & 62.2 & 0.851 & 1.156 &
\ppltoce{15.512} & 1.2}
\ResultRow{HW & 0.6 & 12.8 & 22.0 & 0.689 & 0.720 &
\ppltoce{4.503} & 24.4}
\ResultRow{\methodname{} (Ch) & 2.4 & 26.2 & 36.0 &
0.730 & 0.847 & 1.277 & 26.8}
\ResultRow{\methodname{} (Ch+tilt) & 14.0 & 54.3 & 61.0 & 0.843 & 1.253 &
1.419 & 22.0}
\ResultRow{\methodname{} (S) & 5.5 & 12.8 & 29.3 &
0.699 & 0.756 & 1.300 & 28.0}
\ResultRow{\methodname{} (B) & 10.4 & 23.8 & 37.2 &
0.745 & 0.892 & 1.310 & \secondvalue{30.5}}
\bottomrule
\end{tabular}%
}
\end{minipage}%
\hfill
\begin{minipage}[t]{0.49\linewidth}
\centering
\vspace{0pt}
\textbf{(b) LLaDA 2.1}\par\vspace{2pt}
\resizebox{\linewidth}{!}{%
\begin{tabular}{@{}lccccc@{\hspace{3pt}}cc@{}}
\toprule
Method & \shortstack{TPR@1\\(\%)} & \shortstack{TPR@5\\(\%)} &
\shortstack{TPR@10\\(\%)} & AUC & \shortstack{Mean\\$\text{-}\log_{10}p$} & CE & \shortstack{pass@1\\(\%)} \\
\midrule
\ResultRow{Clean reference & -- & -- & -- & -- & -- & \ppltoce{1.888} & 84.2}
\midrule
\multicolumn{8}{@{\hspace{3pt}}l}{\emph{Token-commit}} \\
\ResultRow{KGW & 24.4 & 47.0 & 64.0 & 0.852 & 1.274 & \ppltoce{2.843} & 65.9}
\ResultRow{Gumbel & 5.5 & 12.8 & 19.5 & 0.573 & 0.590 &
\bestvalue{\ppltoce{1.919}} & \bestvalue{79.9}}
\ResultRow{dgMARK & -- & -- & -- & -- & -- & -- & --}
\ResultRow{PATTERN-M & -- & -- & -- & -- & -- & -- & --}
\ResultRow{HW & \bestvalue{94.5} & \bestvalue{95.7} & \bestvalue{97.0} &
\bestvalue{0.988} & \bestvalue{2.056} & \ppltoce{70.079} & 21.3}
\ResultRow{\methodname{} (Co) & 5.5 & 14.6 & 23.2 & 0.627 & 0.644 &
\secondvalue{0.666} & 77.4}
\ResultRow{\methodname{} (Co+tilt) & 7.9 & 19.5 & 28.0 & 0.645 &
0.730 & 0.711 & 75.0}
\midrule
\multicolumn{8}{@{\hspace{3pt}}l}{\emph{Block-ready}} \\
\ResultRow{KGW & \secondvalue{84.2} & \secondvalue{90.9} & \secondvalue{92.7} &
\secondvalue{0.971} & \secondvalue{1.948} & \ppltoce{14.334} & 12.8}
\ResultRow{Gumbel & 9.2 & 15.9 & 21.3 & 0.631 & 0.706 & \ppltoce{2.411} & 67.1}
\ResultRow{PATTERN-M & -- & -- & -- & -- & -- & -- & --}
\ResultRow{HW (input 0.7) & 6.1 & 12.2 & 24.4 & 0.617 & 0.641 &
\secondvalue{\ppltoce{1.947}} & 77.4}
\ResultRow{\methodname{} (Ch) & 4.3 & 14.6 & 25.0 & 0.627 & 0.651 &
0.713 & \secondvalue{79.3}}
\ResultRow{\methodname{} (Ch+tilt) & 67.7 & 80.5 & 86.0 & 0.936 &
1.773 & 1.217 & 34.8}
\ResultRow{\methodname{} (S) & 30.5 & 36.6 & 43.3 & 0.742 & 0.962 &
\ppltoce{2.287} & 62.8}
\ResultRow{\methodname{} (B) & 7.9 & 16.5 & 25.0 & 0.676 & 0.769 &
\ppltoce{2.318} & 66.5}
\bottomrule
\end{tabular}%
}
\end{minipage}
\endgroup
\end{table}

{
On Finance QA, untilted token-commit \methodname{} achieves stronger detection
than \textit{chain} on both models, with comparatively small differences in ROUGE-L
(Table~\ref{tab:appendix-preview-finance-qa}).
With LLaDA-Instruct, token-commit TPR@5\% reaches 88.5\% versus 75.0\%
for \textit{chain}, while their ROUGE-L scores are 17.0 and 16.9.
\textit{chain} has lower CE.
Block-ready KGW offers stronger detection than \textit{chain}, but at lower ROUGE-L
and higher CE.
\par

On LLaDA 2.1, tilting substantially narrows the detection gap to the
strongest baselines without the large task-score decline observed for
HumanEval \textit{chain}.
Tilted \textit{chain} reaches 99.0\% TPR with ROUGE-L of 20.1, compared with
100.0\% TPR and ROUGE-L of 19.3 for block-ready KGW, and also has lower CE.
This stronger detection still carries a CE cost relative to untilted \textit{chain},
increasing from 1.716 to 1.977.
Tilted token-commit \methodname{} offers a lower-CE operating point,
reaching 92.0\% TPR with CE of 1.813 and ROUGE-L of 20.0.
\par
}
\begin{table}[ht]
\centering
\begingroup
\setlength{\tabcolsep}{2pt}
\renewcommand{\arraystretch}{1.10}
\caption{\textbf{Finance QA detection and text quality.}
\methodname{} variants are (Co) commit, (Ch) \textit{chain}, (S) \textit{middle-star}, and
(B) \textit{balanced tree}. Co+tilt and Ch+tilt use post-MOT strengths
$\delta=0.25$ and $\delta=0.5$, respectively, on both models.
Best and second-best watermarked values are bold and
underlined.}
\label{tab:appendix-preview-finance-qa}
\begin{minipage}[t]{0.49\linewidth}
\centering
\vspace{0pt}
\textbf{(a) LLaDA-Instruct}\par\vspace{2pt}
\resizebox{\linewidth}{!}{%
\begin{tabular}{@{}lccccc@{\hspace{3pt}}cc@{}}
\toprule
Method & \shortstack{TPR@1\\(\%)} & \shortstack{TPR@5\\(\%)} &
\shortstack{TPR@10\\(\%)} & AUC & \shortstack{Mean\\$\text{-}\log_{10}p$} & CE & \shortstack{ROUGE\\L} \\
\midrule
\ResultRow{Clean reference & -- & -- & -- & -- & -- & \ppltoce{4.308} & 17.4}
\midrule
\multicolumn{8}{@{\hspace{3pt}}l}{\emph{Token-commit}} \\
\ResultRow{KGW & 70.0 & 85.0 & 91.0 & 0.967 & 1.906 & \ppltoce{7.294} &
\bestvalue{17.8}}
\ResultRow{Gumbel & 97.0 & 97.5 & 97.5 & 0.995 & 2.167 & \ppltoce{5.778} & 15.0}
\ResultRow{dgMARK & \secondvalue{98.0} & \secondvalue{98.5} &
\secondvalue{99.5} & \secondvalue{0.998} &
\secondvalue{2.182} & \secondvalue{\ppltoce{5.284}} & 16.6}
\ResultRow{PATTERN-M & 70.0 & 82.5 & 86.0 & 0.949 & 1.812 &
\bestvalue{\ppltoce{4.901}} & 16.6}
\ResultRow{HW & -- & -- & -- & -- & -- & -- & --}
\ResultRow{\methodname{} (Co) & 78.5 & 88.5 & 93.5 &
0.980 & 1.999 & 1.819 & 17.0}
\ResultRow{\methodname{} (Co+tilt) & 92.5 & 95.5 & 95.5 &
0.985 & 2.112 & 1.971 & \secondvalue{17.2}}
\midrule
\multicolumn{8}{@{\hspace{3pt}}l}{\emph{Block-ready}} \\
\ResultRow{KGW & 95.0 & 97.5 & 98.0 &
0.994 & 2.155 & \ppltoce{16.093} & 15.6}
\ResultRow{Gumbel & 43.5 & 64.0 & 75.0 &
0.917 & 1.519 & \ppltoce{5.756} & 17.1}
\ResultRow{PATTERN-M & 55.0 & 76.0 & 89.0 &
0.950 & 1.635 & \ppltoce{33.657} & 14.9}
\ResultRow{HW & \bestvalue{99.0} & \bestvalue{99.5} & \bestvalue{100.0} &
\bestvalue{1.000} & \bestvalue{2.200} & \ppltoce{78.942} & 14.2}
\ResultRow{\methodname{} (Ch) & 34.0 & 75.0 & 78.5 &
0.907 & 1.576 & 1.684 & 16.9}
\ResultRow{\methodname{} (Ch+tilt) & 76.5 & 91.0 & 91.5 &
0.964 & 1.964 & 1.788 & 17.0}
\ResultRow{\methodname{} (S) & 23.5 & 45.0 & 60.0 &
0.830 & 1.207 & 1.803 & 17.1}
\ResultRow{\methodname{} (B) & 49.5 & 67.0 & 79.0 &
0.921 & 1.624 & 1.742 & 16.9}
\bottomrule
\end{tabular}%
}
\end{minipage}%
\hfill
\begin{minipage}[t]{0.49\linewidth}
\centering
\vspace{0pt}
\textbf{(b) LLaDA 2.1}\par\vspace{2pt}
\resizebox{\linewidth}{!}{%
\begin{tabular}{@{}lccccc@{\hspace{3pt}}cc@{}}
\toprule
Method & \shortstack{TPR@1\\(\%)} & \shortstack{TPR@5\\(\%)} &
\shortstack{TPR@10\\(\%)} & AUC & \shortstack{Mean\\$\text{-}\log_{10}p$} & CE & \shortstack{ROUGE\\L} \\
\midrule
\ResultRow{Clean reference & -- & -- & -- & -- & -- & \ppltoce{5.035} & 20.4}
\midrule
\multicolumn{8}{@{\hspace{3pt}}l}{\emph{Token-commit}} \\
\ResultRow{KGW & 98.0 & \bestvalue{100.0} & \bestvalue{100.0} &
\bestvalue{1.000} & 2.196 & \ppltoce{6.562} & \secondvalue{20.3}}
\ResultRow{Gumbel & 15.5 & 30.5 & 39.5 & 0.802 & 1.030 &
\bestvalue{\ppltoce{5.088}} & \bestvalue{20.5}}
\ResultRow{dgMARK & -- & -- & -- & -- & -- & -- & --}
\ResultRow{PATTERN-M & -- & -- & -- & -- & -- & -- & --}
\ResultRow{HW & -- & -- & -- & -- & -- & -- & --}
\ResultRow{\methodname{} (Co) & 66.0 & 79.0 & 83.0 & 0.945 &
1.737 & \secondvalue{1.714} & 20.1}
\ResultRow{\methodname{} (Co+tilt) & 82.5 & 92.0 & 95.0 & 0.979 &
1.989 & 1.813 & 20.0}
\midrule
\multicolumn{8}{@{\hspace{3pt}}l}{\emph{Block-ready}} \\
\ResultRow{KGW & \bestvalue{100.0} & \bestvalue{100.0} & \bestvalue{100.0} &
\bestvalue{1.000} & \bestvalue{2.207} & \ppltoce{29.489} & 19.3}
\ResultRow{Gumbel & 27.5 & 41.0 & 49.0 &
0.835 & 1.203 & \ppltoce{7.412} & 20.1}
\ResultRow{PATTERN-M & -- & -- & -- & -- & -- & -- & --}
\ResultRow{HW & -- & -- & -- & -- & -- & -- & --}
\ResultRow{\methodname{} (Ch) & 30.0 & 50.5 &
55.5 & 0.844 & 1.253 & 1.716 & 20.2}
\ResultRow{\methodname{} (Ch+tilt) & 97.5 & 99.0 & \secondvalue{99.0} &
\secondvalue{0.998} & 2.172 & 1.977 & 20.1}
\ResultRow{\methodname{} (S) & \secondvalue{99.5} & \secondvalue{99.5} &
\bestvalue{100.0} & \bestvalue{1.000} &
\secondvalue{2.201} & \ppltoce{20.384} & 19.4}
\ResultRow{\methodname{} (B) & \bestvalue{100.0} & \bestvalue{100.0} &
\bestvalue{100.0} & \bestvalue{1.000} &
\bestvalue{2.207} & \ppltoce{34.386} & 19.3}
\bottomrule
\end{tabular}%
}
\end{minipage}
\endgroup
\end{table}

\endgroup

\begin{figure}[!b]

\centering
\includegraphics[width=\linewidth]{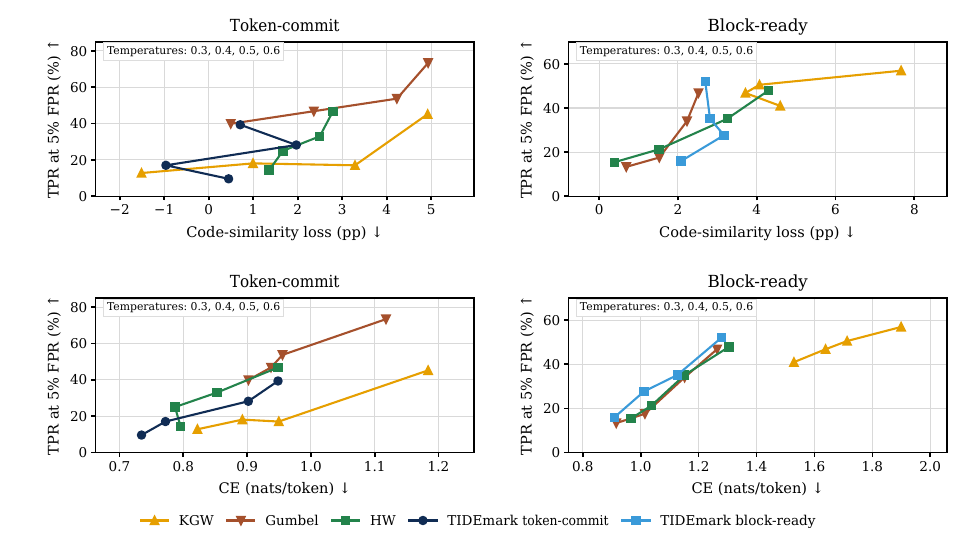}
\caption{\textbf{Temperature sensitivity on LCC.}
TPR at a target 5\% FPR versus code-similarity loss (top) and CE (bottom),
for token-commit (left) and block-ready (right) watermarking.
TIDEmark uses dark-gold circles for token-commit and light-gold squares for
block-ready insertion. Boxes list the shared temperature grid.}
\label{fig:temperature-sensitivity}
\end{figure}

\begin{figure}[!b]

\centering
\includegraphics[width=\linewidth]{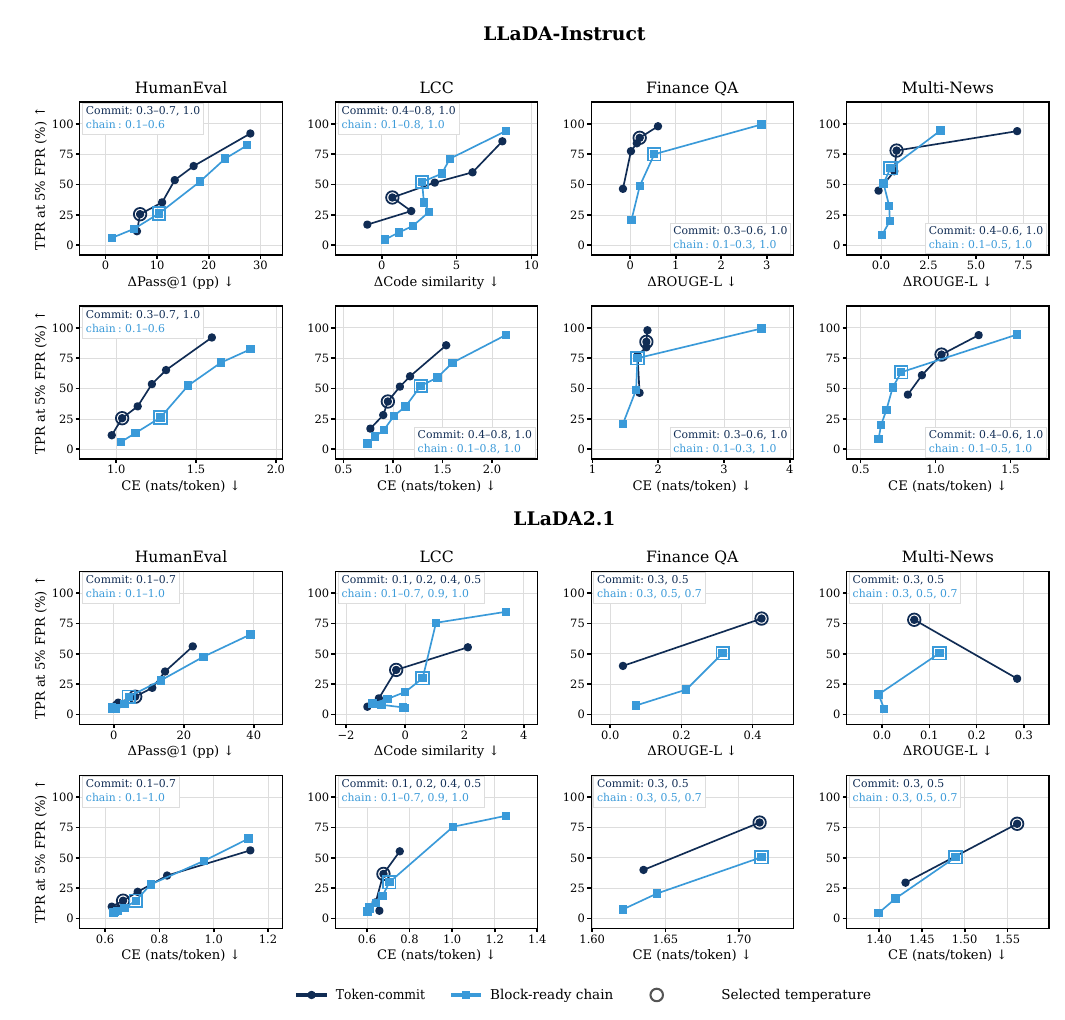}
\caption{\textbf{Temperature sensitivity of TIDEmark across models and datasets.}
The upper two rows show LLaDA-Instruct and the lower two show LLaDA2.1.
Each model reports native-clean-minus-result task-quality differences and
absolute CE against TPR at a target 5\% FPR.
White boxes list solver input temperatures, with ranges in steps of $0.1$.
Larger outlines mark selected settings.}
\label{fig:temperature-sweeps-models}
\label{fig:llada21-temperature-sensitivity}
\end{figure}

\subsection{Deferred-resampling control}
\label{sec:appendix-deferred-resampling}

{
Table~\ref{tab:deferred-resampling} separates deferred resampling from the
full watermarking procedure on LCC using LLaDA-Instruct.
The comparison includes native generation, independent deferred resampling
and block-ready \methodname{} with \textit{chain}.
The control redraws each position independently at block completion from
the recorded predictions, using input temperature $0.6$ and renormalized
top-20 support, without watermark scores or MOT optimization.
Later blocks condition on the resampled text.
Native CE excludes an empty completion.
\par
}
\begingroup\begin{table}[ht]
\centering
\begingroup

\setlength{\tabcolsep}{5pt}
\renewcommand{\arraystretch}{1.10}
\caption{\textbf{Deferred-resampling control on LCC.}}
\label{tab:deferred-resampling}
{\fontsize{6.6}{8}\selectfont
\begin{tabular}{@{}lrr@{}}
\toprule
Configuration & CE $\downarrow$ & Code similarity (\%) $\uparrow$ \\
\midrule
Native generation & 0.704 & 58.0 \\
Independent deferred resampling & 1.275 & 55.0 \\
Block-ready \methodname{} with \textit{chain} & 1.280 & 55.3 \\
\bottomrule
\end{tabular}
}
\endgroup
\end{table}
\endgroup

{
Code similarity decreases by 2.94 points under independent resampling and
by 2.69 points with block-ready \methodname{}, relative to native generation.
CE increases from 0.704 to 1.275 and 1.280, respectively.
The paired code-similarity difference between the two resampled arms is
$+0.24$ points in favor of \methodname{}.
Quality loss is therefore already present without watermark scores or MOT optimization.
\par
}

\subsection{Threshold transfer}
\label{sec:appendix-threshold-transfer}
{
This experiment tests whether a detection cutoff calibrated on one dataset
retains its target false-positive rate on another, using block-ready
\methodname{} with \textit{chain} on LLaDA-Instruct.
Each cutoff is calibrated on one dataset's clean text at a target 5\% FPR,
then applied unchanged to clean and watermarked text from the other datasets.
An additional cutoff is calibrated at the same target using pooled clean
text from all four datasets with equal per-text weighting.
The ablation uses the dataset-specific block-ready \textit{chain} generation
configurations specified in Appendix~\ref{sec:appendix-experimental-settings}.
\par
}
\begingroup\begin{table}[ht]
\centering
\begingroup

\setlength{\tabcolsep}{5pt}
\renewcommand{\arraystretch}{1.10}
\caption{\textbf{Detection-threshold transfer.}
Entries are FPR / TPR (\%) at a source-calibrated
5\% FPR target. Bold marks the lowest FPR and highest TPR separately in
each evaluation column, including ties.}
\label{tab:threshold-transfer}
{\fontsize{6.6}{8}\selectfont
\begin{tabular}{@{}lcccc@{}}
\toprule
& \multicolumn{4}{c}{Evaluated on} \\
\cmidrule(lr){2-5}
\shortstack[l]{Cutoff calibrated on\\clean text from}
& HumanEval & LCC & Finance QA & Multi-News \\
\midrule
HumanEval & 4.9 / 26.2 & 9.0 / 62.8 & 2.5 / 72.5 & 4.0 / 61.0 \\
LCC & \textbf{2.4} / 15.9 & \textbf{4.8} / 52.1
& \textbf{1.5} / 62.5 & \textbf{1.5} / 52.5 \\
Finance QA & 5.5 / \textbf{28.0} & 10.1 / \textbf{64.9}
& 5.0 / \textbf{75.0} & 5.0 / \textbf{63.5} \\
Multi-News & 5.5 / \textbf{28.0} & 9.6 / \textbf{64.9}
& 4.0 / \textbf{75.0} & 5.0 / \textbf{63.5} \\
All four datasets pooled & 4.3 / 26.2 & 9.0 / 62.8 & 2.5 / 72.5 & 4.0 / 61.0 \\
\bottomrule
\end{tabular}
}
\endgroup
\end{table}
\endgroup

{
Table~\ref{tab:threshold-transfer} shows that threshold transfer largely
retains low false-positive rates, with the largest increases occurring on LCC.
Rows identify the clean calibration source and columns identify the
evaluated dataset, with entries reporting FPR and TPR.
Diagonal entries and the pooled row evaluate clean text included in
calibration, rather than independent false-positive validation.
FPR ranges from 1.5\% to 10.1\% when calibration and evaluation datasets differ.
LCC reaches 9.0--10.1\% FPR under cutoffs calibrated on the other datasets,
compared with 4.8\% under its own cutoff.
The pooled cutoff attains 4.9\% FPR overall but 9.0\% on LCC, showing that
pooled calibration does not ensure the target rate within each dataset.
On LCC, the Finance QA cutoff raises TPR from 52.1\% to 64.9\% while
increasing FPR from 4.8\% to 10.1\%.
\par
}

\subsection{State-size sensitivity}
\label{sec:appendix-state-size}

{
State and key reuse are general considerations in keyed watermarking,
rather than concerns specific to \methodname{}.
Under low state availability, reusing a fixed watermark key across repeated
generations introduces distributional deviations, despite preserving
single-generation unbiasedness \citep{wu2026analyzing}.
The number of watermark states is distinct from the secret-key size and
the generation random seed.
Increasing this number can reduce accidental collisions between different
seeding inputs, but does not prevent state reuse for the same input.
This experiment examines how state-space size affects detection and generation
quality on Finance QA using LLaDA-Instruct.
The number of states varies from 128 to 2048 for token-commit \methodname{}
and block-ready \methodname{} with \textit{chain}.
Both use block-index-derived states, with solver input temperatures of
$0.6$ and $0.3$, respectively.
Generation settings remain fixed within each timing configuration.
\par
}
\begingroup\begin{table}[ht]
\centering
\begingroup

\setlength{\tabcolsep}{5pt}
\renewcommand{\arraystretch}{1.10}
\caption{\textbf{State-size sensitivity on Finance QA.}
Detection and generation quality across watermark state-space sizes for
token-commit and block-ready \textit{chain} configurations on LLaDA-Instruct.}
\label{tab:state-size-financeqa}
{\fontsize{6.6}{8}\selectfont
\begin{tabular}{@{}rrrrr@{}}
\toprule
States & TPR@5\% $\uparrow$ & AUC $\uparrow$ & CE $\downarrow$ & ROUGE-L $\uparrow$ \\
\midrule
\multicolumn{5}{@{}l}{\emph{Token-commit}} \\
128 & 91.0 & 0.970 & 1.760 & 17.5 \\
256 & 90.5 & 0.977 & 1.837 & 17.2 \\
512 & 88.0 & 0.971 & 1.785 & 17.5 \\
1024 & 94.0 & 0.984 & 1.850 & 17.3 \\
2048 & 89.0 & 0.970 & 1.739 & 17.2 \\
\midrule
\multicolumn{5}{@{}l}{\emph{Block-ready \textit{chain}}} \\
128 & 64.0 & 0.908 & 1.629 & 17.0 \\
256 & 75.0 & 0.907 & 1.684 & 16.9 \\
512 & 69.0 & 0.910 & 1.746 & 17.1 \\
1024 & 67.0 & 0.880 & 1.694 & 16.9 \\
2048 & 58.0 & 0.881 & 1.742 & 17.3 \\
\bottomrule
\end{tabular}
}
\endgroup
\end{table}
\endgroup

{
Table~\ref{tab:state-size-financeqa} shows no monotonic detection improvement
from increasing the state space.
Token-commit TPR ranges from 88.0\% to 94.0\%, with its highest value at 1024 states.
Block-ready \textit{chain} reaches its highest TPR of 75.0\% at 256 states,
declining to 58.0\% at 2048.
ROUGE-L varies over a narrower range, from 17.2 to 17.5 for token-commit
and from 16.9 to 17.3 for block-ready \textit{chain}.
Larger state spaces therefore do not consistently improve detection or
generation quality in this comparison.
\par
}

\subsection{Temperature sensitivity}
\label{sec:appendix-temperature-sensitivity}

{
Temperature controls different stages of watermark generation across methods.
To assess the resulting detection-quality tradeoffs, KGW, Gumbel, HW and
\methodname{} are compared on LCC with LLaDA-Instruct under both token-commit
and block-ready insertion.
Each method is evaluated at temperatures of $0.3$, $0.4$, $0.5$ and $0.6$,
holding its remaining settings fixed.
For HW and \methodname{}, the sweep changes solver input temperature.
For KGW and Gumbel, it changes sampling temperature.
The main benchmark results use the baseline settings specified in
Appendix~\ref{sec:appendix-experimental-settings}.
Code-similarity loss is measured as the matched-clean score minus the
watermarked score.
Boxes list the shared temperature grid.
\par
}
{
Figure~\ref{fig:temperature-sensitivity} shows stronger detection at higher
temperatures, accompanied by increased CE across all four methods.
From temperature $0.3$ to $0.6$, token-commit \methodname{} TPR rises from
9.6\% to 39.4\%, while code similarity changes from 57.1 to 56.8.
Gumbel reaches a higher TPR of 73.4\%, with code similarity declining
from 57.5 to 53.0.
In the block-ready setting, \methodname{} reaches 52.1\% TPR at temperature
$0.6$, compared with 47.9\% for HW, with lower CE (1.280 versus 1.304)
and higher code similarity (55.3 versus 53.3).
Temperature selection therefore changes both detection strength and the
accompanying quality cost.
\par
}

{
The model comparison varies solver input temperature for token-commit
\methodname{} and block-ready \methodname{} with \textit{chain} across HumanEval,
LCC, Finance QA and Multi-News on LLaDA-Instruct and LLaDA2.1.
Each timing configuration retains its own native-clean reference.
Task-quality differences are native clean minus watermarked output,
with lower differences and lower absolute CE indicating better quality.
\par
}
{
For the selected LLaDA 2.1 base configurations, token-commit \methodname{}
achieves higher TPR@5\% than block-ready \textit{chain} on Finance QA and
Multi-News. On HumanEval, however, the two configurations tie.
\par
}
{
Figure~\ref{fig:temperature-sweeps-models} shows that, on LLaDA2.1, increasing solver
input temperature from $0.3$ to $0.5$ raises token-commit TPR from
40.0\% to 79.0\% on Finance QA and from 29.5\% to 78.0\% on Multi-News.
For block-ready \textit{chain}, increasing temperature from $0.3$ to $0.7$ raises
TPR from 7.5\% to 50.5\% and from 4.5\% to 50.5\%, respectively.
On LLaDA-Instruct, increasing token-commit input temperature from $0.4$ to
$0.6$ raises TPR from 17.0\% to 39.4\% on LCC and from 45.0\% to 78.0\%
on Multi-News.
\par
}

\subsection{Post-MOT tilting}
\label{sec:appendix-post-fit-tilting}

\begin{figure}[ht]

\centering
\begingroup
\definecolor{tiltCommit}{HTML}{102C54}
\definecolor{tiltChain}{HTML}{3A9AD9}
\pgfplotstableread{
quality ce tpr delta dxQ dyQ dxC dyC
30.487804878 0 25.6097560976 0 0 -12 0 -13
29.2682926829 0.0324086653208 29.8780487805 0.125 4 30 30 -2
26.8292682927 0.054219214806 30.487804878 0.25 -15 -7 31 -14
23.7804878049 0.141108207422 49.3902439024 0.5 -12 -7 30 4
15.8536585366 0.301869607219 79.8780487805 1 0 -7 0 -7
}\tiltHECommit
\pgfplotstableread{
quality ce tpr delta dxQ dyQ dxC dyC
26.8292682927 0 26.2195121951 0 5 7 -5 8
21.9512195122 0.141313041041 54.2682926829 0.5 6 9 -5 10
15.8536585366 0.432269666643 85.9756097561 1 0 6 0 6
4.26829268293 0.988755447904 95.7317073171 2 0 6 0 6
0 1.87520206579 100 3 0 6 0 6
}\tiltHEChain
\pgfplotstableread{
quality ce tpr delta dxQ dyQ dxC dyC
56.8457446809 0 39.3617021277 0 0 -11 0 -13
55.2553191489 0.0739567088809 40.9574468085 0.125 -13 -4 23 -4
53.6170212766 0.168226627805 53.1914893617 0.25 -14 -5 23 6
51.7659574468 0.389195604376 76.5957446809 0.5 0 -8 0 -8
47.3670212766 0.867859930274 96.2765957447 1 0 -8 0 -8
}\tiltLCCCommit
\pgfplotstableread{
quality ce tpr delta dxQ dyQ dxC dyC
55.2819148936 0 52.1276595745 0 2 9 -4 9
51.8723404255 0.3922946353 82.4468085106 0.5 5 9 -5 9
50.1808510638 1.20904605433 98.4042553191 1 0 7 0 7
45.1861702128 2.77011266872 100 2 0 6 0 6
37.2074468085 3.42068491443 100 3 0 6 0 6
}\tiltLCCChain
\pgfplotstableread{
quality ce tpr delta dxQ dyQ dxC dyC
16.960541643 0 88.5 0 4 -10 5 -40
16.8040897197 0.139301409066 96 0.125 -2 12 35 -8
17.2213470104 0.152462080829 95.5 0.25 1 7 36 -20
16.4175965039 0.0104557957842 95 0.5 -9 -10 20 -30
16.0039590283 -0.0176424949898 99 1 0 12 0 12
}\tiltFQCommit
\pgfplotstableread{
quality ce tpr delta dxQ dyQ dxC dyC
16.9197420359 0 75 0 0 -10 0 -10
16.956353825 0.103937997693 91 0.5 10 1 -9 0
16.5796470905 0.467872273888 96 1 -12 -1 6 4
15.4089476225 1.20329987651 99.5 2 0 5 0 5
13.9751935555 1.92461433097 99.5 3 0 6 0 6
}\tiltFQChain
\pgfplotstableread{
quality ce tpr delta dxQ dyQ dxC dyC
26.0293902101 0 78 0 4 -7 20 -25
25.5338848869 0.0659871431894 84 0.125 -18 -12 25 -15
24.5662771698 0.0891207725581 84.5 0.25 -22 -23 35 -5
22.709392062 0.258359411855 94 0.5 0 -9 30 -1
18.1309533452 0.264305670308 97 1 0 -8 4 10
}\tiltMNCommit
\pgfplotstableread{
quality ce tpr delta dxQ dyQ dxC dyC
26.409165946 0 63.5 0 0 -11 0 -11
26.2528143008 0.158449857466 89 0.5 1 10 -7 12
24.5503911726 0.673178641569 99.5 1 0 7 0 7
19.4840150998 1.82633238614 99.5 2 0 6 0 6
16.7556095865 1.84883824414 100 3 0 6 0 -10
}\tiltMNChain
\pgfplotsset{
  tilt axis/.style={
    scale only axis,width=0.197\linewidth,height=0.166667\linewidth,
    font=\fontsize{6.6}{8}\selectfont,
    tick label style={font=\fontsize{5.5}{6.5}\selectfont},
    label style={font=\fontsize{6}{7}\selectfont},
    title style={font=\fontsize{7}{8}\selectfont},
    ymin=0,ymax=115,ytick={0,25,50,75,100},
    scaled ticks=false,tick align=outside,
    grid=major,major grid style={gray!18},
    axis background/.style={fill=white},clip=false,
    enlarge x limits=0.12,
    point meta=explicit symbolic,
    nodes near coords={\pgfplotspointmeta},
    every node near coord/.append style={
      font=\fontsize{5.5}{6}\selectfont,
      xshift=\tiltLabelX pt,yshift=\tiltLabelY pt,
      anchor=center,inner sep=0.6pt,fill=white,
      append after command={
        (\tikzlastnode)
        edge[line width=0.25pt,draw=\tiltLeaderColor!75,shorten >=2pt]
        ([xshift=-\tiltLabelX pt,yshift=-\tiltLabelY pt]\tikzlastnode.center)
      }
    }
  },
  tilt quality/.style={
    visualization depends on={\thisrow{dxQ} \as \tiltLabelX},
    visualization depends on={\thisrow{dyQ} \as \tiltLabelY}
  },
  tilt ce/.style={
    visualization depends on={\thisrow{dxC} \as \tiltLabelX},
    visualization depends on={\thisrow{dyC} \as \tiltLabelY}
  },
  tilt commit/.style={
    color=tiltCommit,mark=*,mark size=1.4pt,line width=0.7pt,
    /utils/exec={\def\tiltLeaderColor{tiltCommit}}
  },
  tilt chain/.style={
    color=tiltChain,mark=square*,mark size=1.5pt,line width=0.7pt,
    /utils/exec={\def\tiltLeaderColor{tiltChain}}
  }
}
\begin{tikzpicture}
\begin{groupplot}[
  tilt axis,
  group style={
    group size=4 by 2,horizontal sep=0.035\linewidth,
    vertical sep=0.08\linewidth
  }
]
\nextgroupplot[
  title={HumanEval},xlabel={$\Delta$Pass@1 (pp) $\downarrow$},
  ylabel={TPR@5\% $\uparrow$},
  legend to name=tiltSweepLegend,legend columns=2,
  legend style={draw=none,font=\scriptsize,cells={anchor=west}}
]
\addplot[tilt commit,tilt quality] table[x expr=37.19512195121951-\thisrow{quality},y=tpr,meta=delta] {\tiltHECommit};
\addlegendentry{Token-commit}
\addplot[tilt chain,tilt quality] table[x expr=37.19512195121951-\thisrow{quality},y=tpr,meta=delta] {\tiltHEChain};
\addlegendentry{Block-ready \textit{chain}}
\nextgroupplot[title={LCC},xlabel={$\Delta$Code similarity $\downarrow$},yticklabels={}]
\addplot[tilt commit,tilt quality] table[x expr=57.5531914893617-\thisrow{quality},y=tpr,meta=delta] {\tiltLCCCommit};
\addplot[tilt chain,tilt quality] table[x expr=57.973404255319146-\thisrow{quality},y=tpr,meta=delta] {\tiltLCCChain};
\nextgroupplot[title={Finance QA},xlabel={$\Delta$ROUGE-L $\downarrow$},yticklabels={}]
\addplot[tilt commit,tilt quality] table[x expr=17.168751850398085-\thisrow{quality},y=tpr,meta=delta] {\tiltFQCommit};
\addplot[tilt chain,tilt quality] table[x expr=17.448858241870703-\thisrow{quality},y=tpr,meta=delta] {\tiltFQChain};
\nextgroupplot[title={Multi-News},xlabel={$\Delta$ROUGE-L $\downarrow$},yticklabels={}]
\addplot[tilt commit,tilt quality] table[x expr=26.845659808243308-\thisrow{quality},y=tpr,meta=delta] {\tiltMNCommit};
\addplot[tilt chain,tilt quality] table[x expr=26.906863190977198-\thisrow{quality},y=tpr,meta=delta] {\tiltMNChain};
\nextgroupplot[xlabel={CE $\downarrow$},ylabel={TPR@5\% $\uparrow$}]
\addplot[tilt commit,tilt ce] table[x expr=\thisrow{ce}+1.0369393048086855,y=tpr,meta=delta] {\tiltHECommit};
\addplot[tilt chain,tilt ce] table[x expr=\thisrow{ce}+1.277376605317954,y=tpr,meta=delta] {\tiltHEChain};
\nextgroupplot[xlabel={CE $\downarrow$},yticklabels={}]
\addplot[tilt commit,tilt ce] table[x expr=\thisrow{ce}+0.9484200332161964,y=tpr,meta=delta] {\tiltLCCCommit};
\addplot[tilt chain,tilt ce] table[x expr=\thisrow{ce}+1.2796046935992564,y=tpr,meta=delta] {\tiltLCCChain};
\nextgroupplot[xlabel={CE $\downarrow$},yticklabels={}]
\addplot[tilt commit,tilt ce] table[x expr=\thisrow{ce}+1.8190309839476122,y=tpr,meta=delta] {\tiltFQCommit};
\addplot[tilt chain,tilt ce] table[x expr=\thisrow{ce}+1.6836454239493177,y=tpr,meta=delta] {\tiltFQChain};
\nextgroupplot[xlabel={CE $\downarrow$},yticklabels={}]
\addplot[tilt commit,tilt ce] table[x expr=\thisrow{ce}+1.0399393807566057,y=tpr,meta=delta] {\tiltMNCommit};
\addplot[tilt chain,tilt ce] table[x expr=\thisrow{ce}+0.771672401002015,y=tpr,meta=delta] {\tiltMNChain};
\end{groupplot}
\end{tikzpicture}
\par\vspace{2pt}
\ref*{tiltSweepLegend}
\caption{\textbf{Post-MOT tilting across datasets.}
TPR at the unedited-clean 5\% FPR target is plotted against changes in task
performance relative to native clean (top) and absolute completion CE (bottom).
Task-score differences are the corresponding native-clean value minus
the watermarked value.
Task-score changes are percentage points for HumanEval pass@1 and score
points for LCC code similarity and ROUGE-L, where lower is better.
Lower CE and higher TPR are better.
Labels identify tilt strength. Curves show all tested strengths for
Token-commit and Block-ready \textit{chain} on LLaDA-Instruct.}
\label{fig:tilt-sweeps}
\endgroup
\end{figure}
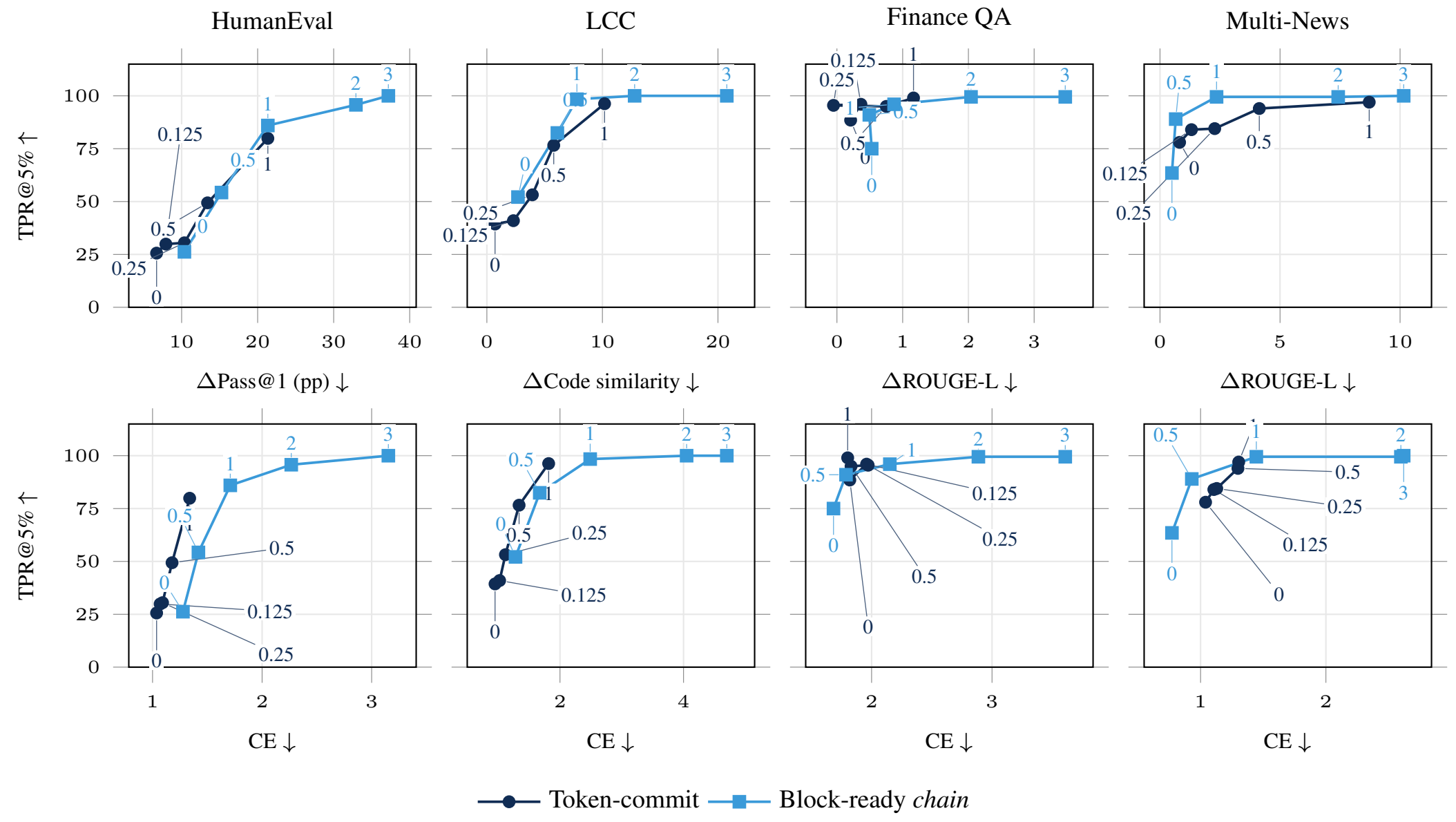

{
Post-MOT tilting reweights the conditional distribution after solving the MOT problem.
Following the approach of HeavyWater
\citep{tsur2025heavywater}, we evaluate tilting as a control over watermark
strength and generation quality for LLaDA-Instruct.
Starting from the computed conditional distribution $q_0(x^b\mid s)$,
we sample from
\begin{equation}
q_\delta(x^b\mid s)\propto
q_0(x^b\mid s)\exp\!\left(\delta f(x^b,s)\right),
\label{eq:post-fit-tilting}
\end{equation}
where $f$ is the block score defined above, using the original standardized
pair scores before optimization-table range normalization.
The computed scaling factors are not reoptimized.
Setting $\delta=0$ recovers the base conditional distribution.
Nonzero tilting changes the sampling distribution and does not inherit
the base coupling's marginal-preservation guarantee.
\par

We evaluate Token-commit and Block-ready \textit{chain} on HumanEval, LCC, Finance QA
and Multi-News. Within each dataset and timing, the input temperature and
remaining generation settings are fixed.
Token-commit uses $\delta\in\{0,0.125,0.25,0.5,1\}$, while \textit{chain} uses
$\delta\in\{0,0.5,1,2,3\}$.
\par

Figure~\ref{fig:tilt-sweeps} shows detection against task-score differences
and absolute CE, with lower values better on both horizontal axes.
Task-score differences are measured as native clean minus watermarked output.
\par
}

\subsection{EB-sampler}
\label{sec:appendix-eb-sampler}

{
EB-sampler \citep{benhamu2025ebsampler} provides an alternative to
low-confidence token selection.
The comparison evaluates KGW, Gumbel, HW, PATTERN-MARK and \methodname{}
on HumanEval using LLaDA-Instruct with entropy budget $\gamma=1$.
\methodname{} uses solver input temperatures of $0.5$ and $0.3$ for
token-commit and block-ready insertion, respectively.
The block-ready \methodname{} configuration uses \textit{chain}.
Original-score post-MOT tilting is varied within each timing configuration.
$\Delta$Pass@1 is the matched-clean score minus the watermarked score.
\par
}
\begingroup\begin{table}[ht]
\centering
\begingroup

\setlength{\tabcolsep}{3pt}
\renewcommand{\arraystretch}{1.10}
\caption{\textbf{Watermarking under EB-sampler on HumanEval.}
Detection, completion CE and functional correctness for both insertion timings.
TPR is reported at a target 5\% FPR.
Clean rows show the contextual EB reference.}
\label{tab:eb-sampler}
{\fontsize{6.6}{8}\selectfont
\begin{tabular}{@{}lrrrr@{\hspace{16pt}}lrrrr@{}}
\toprule
\multicolumn{5}{c}{Token-commit} & \multicolumn{5}{c}{Block-ready} \\
\cmidrule(lr){1-5}\cmidrule(lr){6-10}
Method & $\delta$ & TPR (\%) & CE & Pass@1 (\%)
& Method & $\delta$ & TPR (\%) & CE & Pass@1 (\%) \\
\midrule
Clean reference & -- & -- & 0.711 & 42.1
& Clean reference & -- & -- & 0.711 & 42.1 \\
\midrule
KGW & -- & 32.9 & 1.080 & 29.3 & KGW & -- & 81.1 & 1.900 & 1.8 \\
Gumbel & -- & 54.9 & 0.937 & 35.4
& Gumbel & -- & 23.8 & 1.231 & 18.9 \\
PATTERN-M & -- & 40.9 & 1.621 & 9.8
& PATTERN-M & -- & 34.1 & 2.578 & 0.0 \\
HW & -- & 1.8 & 0.996 & 33.5 & HW & -- & 20.7 & 1.551 & 18.3 \\
\midrule
\methodname{} & 0 & 29.3 & 0.799 & 34.8
& \methodname{} & 0 & 23.2 & 1.206 & 15.2 \\
\methodname{} & 0.125 & 29.3 & 0.812 & 34.1
& \methodname{} & 0.5 & 45.7 & 1.297 & 12.8 \\
\methodname{} & 0.25 & 38.4 & 0.858 & 31.1
& \methodname{} & 1 & 76.2 & 1.445 & 10.4 \\
\methodname{} & 0.5 & 59.8 & 0.976 & 28.0
& \methodname{} & 2 & 98.8 & 2.107 & 1.8 \\
\methodname{} & 1 & 97.6 & 1.390 & 12.8
& & & & & \\
\bottomrule
\end{tabular}
}
\endgroup
\end{table}
\endgroup

{
Table~\ref{tab:eb-sampler} shows different detection-quality tradeoffs
across insertion timings.
Token-commit \methodname{} achieves 34.8\% pass@1 with CE of 0.799,
while Gumbel reaches 35.4\% pass@1 and higher TPR, at 54.9\% versus 29.3\%.
Block-ready \methodname{} reaches 23.2\% TPR, close to Gumbel's 23.8\%,
with lower CE but lower pass@1.
KGW provides the strongest block-ready detection among the baselines at
81.1\% TPR, accompanied by 1.8\% pass@1.
\par
}
\begin{figure}[ht]

\centering
\includegraphics[width=\linewidth]{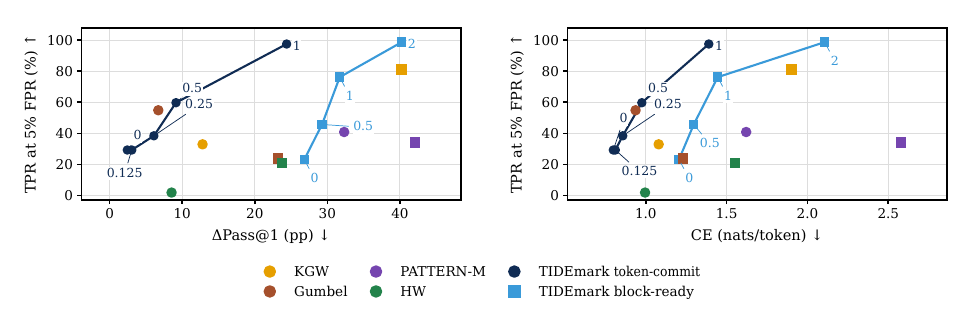}
\caption{\textbf{Detection-quality tradeoffs under EB-sampler.}
TPR at a target 5\% FPR versus $\Delta$Pass@1 (left) and CE (right).
Circles and squares denote token-commit and block-ready insertion.
TIDEmark point labels indicate post-MOT tilt strength $\delta$.}
\label{fig:eb-tradeoffs}
\end{figure}

{
Figure~\ref{fig:eb-tradeoffs} shows that post-MOT tilting strengthens
detection under EB-sampler at the cost of functional correctness.
For token-commit \methodname{}, $\delta=0.5$ raises TPR from 29.3\% to
59.8\%, while pass@1 decreases from 34.8\% to 28.0\%.
For block-ready \textit{chain}, $\delta=1$ raises TPR from 23.2\% to 76.2\%,
while pass@1 decreases from 15.2\% to 10.4\%.
At the strongest completed settings, TPR reaches 97.6\% for token-commit
and 98.8\% for block-ready, with pass@1 of 12.8\% and 1.8\%, respectively.
These results reveal a clear empirical Pareto frontier,
with token-commit \methodname{} contributing several operating points
across both quality measures.
Post-MOT tilting extends this frontier toward stronger detection,
providing a controllable tradeoff between watermark strength and
generation quality.
\par
}

\subsection{Token-commit graph scope}
\label{sec:appendix-commit-graph-scope}

{
Token-commit graph scope determines how much context participates in each
MOT optimization.
Let $C$ denote the positions selected by EB-sampler and let
$[b_{\mathrm{start}},b_{\mathrm{end}}]$ denote the current live block.
The selected-interval configuration solves MOT on a \textit{chain} over
\begin{equation}
V_{\mathrm{interval}} =
\{\min(C)-1,\ldots,\min(\max(C)+1,b_{\mathrm{end}})\},
\label{eq:graph-scope-interval}
\end{equation}
whereas the whole-block configuration solves MOT on a \textit{chain} over
\begin{equation}
V_{\mathrm{block}} =
\{b_{\mathrm{start}}-1,\ldots,b_{\mathrm{end}}\}.
\label{eq:graph-scope-block}
\end{equation}
Both include predecessor context.
The selected interval additionally includes the successor of its last
selected position when available within the block.
Unselected written tokens remain fixed, unselected masked positions are
marginalized, and only positions in $C$ are committed.
The comparison tests whether retaining the entire live block improves
detection or generation quality under the same EB-sampler.
Solver input temperature varies from $0.1$ to $1.0$.
As in the EB comparison, $\Delta$Pass@1 is the matched-clean score minus
the watermarked score.
\par
}
\begin{figure}[ht]

\centering
\includegraphics[width=\linewidth]{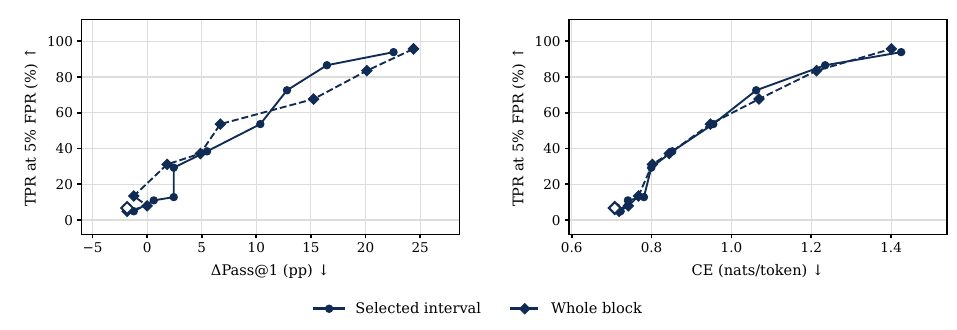}
\caption{\textbf{Token-commit graph-scope comparison on HumanEval.}
Detection-quality tradeoffs for selected-interval and whole-block MOT optimization
under EB-sampler. Successive points follow solver input temperatures
$\{0.1, 0.2, 0.3, 0.4, 0.5, 0.6, 0.7, 0.8, 0.9, 1.0\}$
along each curve. Hollow markers identify $0.1$.}
\label{fig:commit-graph-scope}
\end{figure}

{
Figure~\ref{fig:commit-graph-scope} shows that retaining the entire live
block does not consistently improve the detection-quality tradeoff.
At input temperature $0.5$, whole-block MOT optimization yields 31.1\% TPR and
35.4\% pass@1, compared with 29.3\% and 34.8\% for the selected interval.
At $0.8$, the selected interval instead achieves both higher TPR
(72.6\% versus 67.7\%) and higher pass@1 (24.4\% versus 22.0\%).
The selected interval therefore provides competitive operating points
without including the remaining live-block context in every optimization,
reducing the graph-dependent computational work per BP sweep.
\par
}

\setlength{\floatsep}{6pt plus 1pt minus 1pt}
\setlength{\textfloatsep}{8pt plus 2pt minus 2pt}
\setlength{\intextsep}{5pt plus 1pt minus 1pt}
\setlength{\abovecaptionskip}{4pt}
\setcounter{topnumber}{3}
\setcounter{totalnumber}{3}
\subsection{Robustness to attacks}
\label{sec:appendix-attack-robustness}

{
We evaluate robustness through comparisons across watermarking methods,
attack severities, graph topologies and models.
We additionally measure state agreement and edge recall to examine how
edits affect the structures used by detection.
\par
}

\subsubsection{Attack protocol and state construction}
\label{sec:attack-protocol}

{
The attack comparison uses random deletion, insertion, and substitution
at 20\%, BERT word substitution at 20\%, English-German-English
back-translation, and rewriting with Qwen2.5-1.5B-Instruct.
The severity sweeps vary random edits over 10\%, 20\% and 30\%, and
BERT word substitution over 10\%, 20\%, 30\% and 50\%.
Back-translation uses the OPUS-MT models
\texttt{Helsinki-NLP/opus-mt-en-de} and \texttt{Helsinki-NLP/opus-mt-de-en}
for English-to-German and German-to-English, respectively.
\par

For previous-block MinHash seeding, side information is constructed from
the preceding block.
This choice aims to limit seed disruption under local edits
\citep{kirchenbauer2024reliability}.
Specifically, for the previous block $B_{k-1}$,
\begin{equation*}
x^\star_{k-1}=\operatorname*{arg\,min}_{x\in B_{k-1}}h(x),
\qquad
s_k=g(x^\star_{k-1}),
\end{equation*}
where $h$ assigns hash values to token identities and $g$ maps the selected
token to the current block's watermark state.
Both functions are determined by the watermark key.
Different predecessor blocks can select the same token and therefore produce
the same state.
The independent-uniform block-state assumptions of
Equation~\ref{eq:appendix-finite-state-tail} are not established for this
MinHash construction.
The topology comparison instead derives the watermark state from the
block index and key, sharing that state across tokens within the block.
These are separate state constructions.
\par

Detection uses only the edited text, without access to the original block
alignment or a search over possible boundaries.
Original-text alignment is used only for the offline recovery analyses below.
The method/baseline, topology and LLaDA2.1 comparisons freeze each detector's
cutoff on its own unedited matched-clean reference before attacking.
The separate \textit{chain} recovery study below retains its condition-specific
clean calibration.
\par
}

\subsubsection{Detection and quality under attacks}
\label{sec:attack-detection-quality}

\begin{figure}[ht]

\centering
\includegraphics[width=\linewidth]{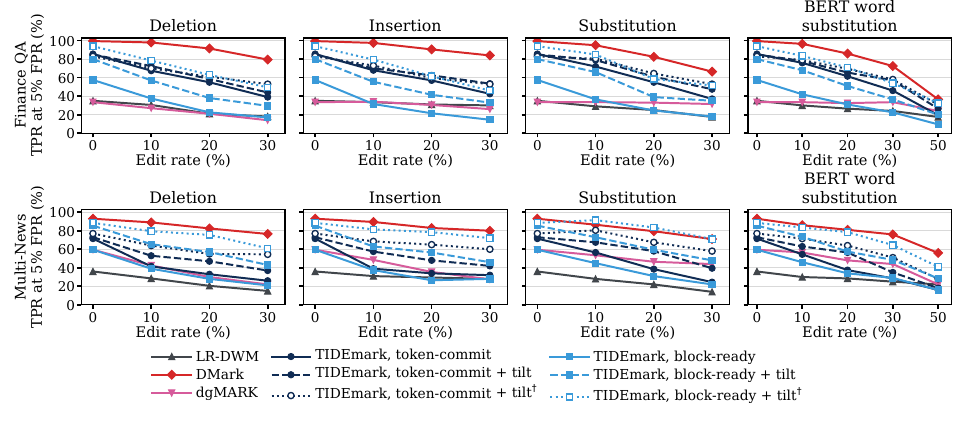}
\caption{\textbf{Detection across edit rates.} Curves show TPR at 5\% FPR
for random deletion, insertion, substitution, and BERT word substitution
on Finance QA and Multi-News. Each curve includes its source-matched
unedited result at a 0\% edit rate.
+tilt denotes post-MOT tilting.
All \methodname{} variants use previous-block MinHash.
Mild tilting uses $\delta=0.25$ for token-commit and $\delta=0.5$ for block-ready
\textit{chain}. Stronger tilting, marked $\dagger$, uses $\delta=0.5$ and
$\delta=1.0$, respectively.
Solid, dashed and dotted lines distinguish untilted, tilted and
stronger-tilt configurations, respectively; the stronger variants have
hollow markers.}
\label{fig:attack-edit-rates}
\end{figure}

{
We compare token-commit and block-ready \methodname{} with token-commit
baselines on LLaDA-Instruct.
The main \methodname{} comparison uses previous-block MinHash.
We use the \textit{chain} topology for block-ready \methodname{}.
DMark uses predict-bidirectional watermarking. dgMARK embeds its signal
through the unmasking order and proposes a sliding-window detector to
handle edits \citep{hong2026dgmark}.
Untilted \methodname{} uses $\delta=0$.
The +tilt variants apply original-score post-MOT tilting with $\delta=0.25$
for token-commit and $\delta=0.5$ for block-ready, retaining their corresponding
base configurations' remaining settings, as described in
Appendix~\ref{sec:appendix-experimental-settings}.
\par

Table~\ref{tab:attack-20-summary} reports TPR at cutoffs targeting 5\% FPR
on each detector's unedited matched-clean reference, held fixed under attack.
FPR on attacked clean text may differ.
For the MinHash configurations across the six attacks and all three tilt levels,
attacked-clean FPR ranges
from 3.5\% to 7.0\% for token-commit \methodname{} and 3.0\% to 6.0\% for
block-ready \methodname{} on Finance QA.
On Multi-News, the ranges are 4.0\% to 8.0\% and 5.0\% to 10.0\%,
respectively.
\par

We additionally evaluate stronger-tilt MinHash configurations,
marked with $\dagger$ in Tables~\ref{tab:attack-20-summary}
and~\ref{tab:attack-quality} (rows labeled (Co+tilt)$^\dagger$ and
(Ch+tilt)$^\dagger$).
These use $\delta=0.5$ for token-commit \methodname{} and $\delta=1$ for
block-ready \methodname{} with \textit{chain} on both datasets.
Like the unmarked variants, they derive their states from the preceding
block's text using MinHash.
\par

Table~\ref{tab:attack-20-summary} shows that stronger-tilt block-ready
\methodname{} retains 58.5\% and 83.5\% TPR under 20\% substitution on
Finance QA and Multi-News, respectively.
Rewriting remains damaging, reducing Finance QA TPR to 10.5\%.
Table~\ref{tab:attack-quality} shows the accompanying unedited-quality
costs: ROUGE-L decreases by 0.41 and 1.94 points, while CE increases by
0.584 and 0.619, respectively.
The stronger-tilt token-commit configuration incurs a larger Multi-News
ROUGE-L loss of 2.25 points.
\par

Table~\ref{tab:attack-quality} reports signed unedited quality changes
relative to each source's matched-clean reference.
$\Delta$ROUGE-L is clean minus watermarked, and $\Delta$CE is watermarked
minus clean. Negative values indicate improvement.
\par

Figure~\ref{fig:attack-edit-rates} compares the MinHash configurations
across the random-edit and BERT
word-substitution severity sweeps, including each source's unedited reference.
For stronger-tilt block-ready \methodname{}, 30\% deletion reduces TPR to
49.5\% on Finance QA and 61.0\% on Multi-News, whereas 30\% substitution
retains 51.5\% and 71.0\%, respectively.
\par
}
\begingroup
\begin{table}[t]
\centering
\begingroup

\setlength{\tabcolsep}{1.5pt}
\renewcommand{\arraystretch}{1.10}
\caption{\textbf{Detection under text attacks on LLaDA-Instruct.}
TPR at 5\% FPR, reported in percent.
+tilt denotes post-MOT tilting.
All \methodname{} variants use previous-block MinHash.
Mild tilting uses $\delta=0.25$ for token-commit and $\delta=0.5$ for block-ready
\textit{chain}. Stronger tilting, marked $\dagger$, uses $\delta=0.5$ and
$\delta=1.0$, respectively.}
\label{tab:attack-20-summary}
\begin{minipage}[t]{0.49\linewidth}
\centering
\vspace{0pt}
{\fontsize{6.6}{8}\selectfont
\begin{tabular}{@{}lrrrrrr@{}}
\toprule
Method & Del. & Ins. & Subst. & \shortstack{BERT\\sub.}
& \shortstack{Back-\\trans.} & Rewrite \\
\midrule
\multicolumn{7}{@{\hspace{2pt}}l}{\emph{Token-commit}} \\
LR-DWM & 21.0 & 31.0 & 25.0 & 26.5 & 23.5 & 2.5 \\
DMark & 91.5 & 90.5 & 82.5 & 86.0 & 71.0 & 15.5 \\
dgMARK & 21.0 & 30.5 & 33.0 & 32.5 & 30.0 & 10.5 \\
\methodname{} (Co) & 55.0 & 57.0 & 55.0 & 61.5 & 36.5 & 17.0 \\
\methodname{} (Co+tilt) & 58.5 & 62.5 & 61.0 & 65.5 & 47.0 & 19.5 \\
\methodname{} (Co+tilt)$^\dagger$ & 61.0 & 59.5 & 64.5 & 69.5 & 47.0 & 24.0 \\
\midrule
\multicolumn{7}{@{\hspace{2pt}}l}{\emph{Block-ready}} \\
\methodname{} (Ch) & 22.5 & 21.5 & 24.5 & 31.0 & 16.0 & 4.5 \\
\methodname{} (Ch+tilt) & 38.0 & 41.5 & 39.0 & 51.0 & 22.5 & 9.5 \\
\methodname{} (Ch+tilt)$^\dagger$ & 63.0 & 61.5 & 58.5 & 71.0 & 35.5 & 10.5 \\
\bottomrule
\end{tabular}}
\end{minipage}%
\hfill
\begin{minipage}[t]{0.49\linewidth}
\centering
\vspace{0pt}
{\fontsize{6.6}{8}\selectfont
\begin{tabular}{@{}lrrrrrr@{}}
\toprule
Method & Del. & Ins. & Subst. & \shortstack{BERT\\sub.}
& \shortstack{Back-\\trans.} & Rewrite \\
\midrule
\multicolumn{7}{@{\hspace{2pt}}l}{\emph{Token-commit}} \\
LR-DWM & 20.5 & 29.5 & 22.0 & 28.5 & 20.5 & 1.0 \\
DMark & 82.5 & 83.0 & 79.5 & 81.0 & 70.0 & 11.0 \\
dgMARK & 30.0 & 35.5 & 46.5 & 48.0 & 26.0 & 14.5 \\
\methodname{} (Co) & 33.0 & 34.0 & 38.5 & 37.5 & 20.0 & 8.0 \\
\methodname{} (Co+tilt) & 47.0 & 48.0 & 58.0 & 56.5 & 27.0 & 13.0 \\
\methodname{} (Co+tilt)$^\dagger$ & 55.0 & 65.0 & 67.5 & 64.0 & 27.5 & 10.0 \\
\midrule
\multicolumn{7}{@{\hspace{2pt}}l}{\emph{Block-ready}} \\
\methodname{} (Ch) & 28.0 & 26.5 & 31.0 & 34.0 & 15.5 & 9.5 \\
\methodname{} (Ch+tilt) & 57.0 & 56.5 & 59.5 & 57.5 & 24.5 & 8.0 \\
\methodname{} (Ch+tilt)$^\dagger$ & 75.5 & 78.5 & 83.5 & 78.5 & 34.5 & 9.5 \\
\bottomrule
\end{tabular}}
\end{minipage}
\par\vspace{2pt}
\makebox[0.49\linewidth]{\textbf{(a) Finance QA}}\hfill
\makebox[0.49\linewidth]{\textbf{(b) Multi-News}}
\endgroup
\end{table}
\endgroup
\begingroup
\begin{table}[t]
\centering
\begingroup

\setlength{\tabcolsep}{4pt}
\renewcommand{\arraystretch}{1.10}
\caption{\textbf{Unedited quality changes for the sources evaluated in
Table~\ref{tab:attack-20-summary}.} Reported as $\Delta$ROUGE-L and $\Delta$CE.
+tilt denotes post-MOT tilting.
All \methodname{} variants use previous-block MinHash.
Mild tilting uses $\delta=0.25$ for token-commit and $\delta=0.5$ for block-ready
\textit{chain}. Stronger tilting, marked $\dagger$, uses $\delta=0.5$ and
$\delta=1.0$, respectively.}
\label{tab:attack-quality}
{\fontsize{6.6}{8}\selectfont
\begin{tabular}{@{}lrrrr@{}}
\toprule
& \multicolumn{2}{c}{Finance QA} & \multicolumn{2}{c}{Multi-News} \\
\cmidrule(lr){2-3}\cmidrule(lr){4-5}
Method & $\Delta$ROUGE-L $\downarrow$ & $\Delta$CE $\downarrow$
& $\Delta$ROUGE-L $\downarrow$ & $\Delta$CE $\downarrow$ \\
\midrule
\multicolumn{5}{@{\hspace{2pt}}l}{\emph{Token-commit}} \\
LR-DWM & +0.33 & $-0.008$ & +0.66 & +0.160 \\
DMark & +1.37 & +0.280 & +3.70 & +0.677 \\
dgMARK & +0.66 & +0.081 & +0.64 & +0.287 \\
\methodname{} (Co) & $-0.10$ & +0.269 & +0.80 & +0.213 \\
\methodname{} (Co+tilt) & $-0.12$ & +0.033 & +1.21 & +0.247 \\
\methodname{} (Co+tilt)$^\dagger$ & $-0.04$ & $-0.086$ & +2.25 & +0.395 \\
\midrule
\multicolumn{5}{@{\hspace{2pt}}l}{\emph{Block-ready}} \\
\methodname{} (Ch) & +0.25 & +0.280 & +0.07 & +0.179 \\
\methodname{} (Ch+tilt) & +0.28 & +0.330 & +0.54 & +0.328 \\
\methodname{} (Ch+tilt)$^\dagger$ & +0.41 & +0.584 & +1.94 & +0.619 \\
\bottomrule
\end{tabular}
}
\endgroup
\end{table}
\endgroup

\subsubsection{State and edge recovery under edits}
\label{sec:attack-state-edge-recovery}

{
Tables~\ref{tab:attack-fq-old-chain} and
\ref{tab:attack-mn-old-chain} jointly report detection and reconstruction
for LLaDA 1.0 block-ready \textit{chain} on Finance QA and Multi-News, using the
previous-block MinHash construction.
Watermarked outputs undergo random deletion, insertion, and substitution
at rates of 10\%, 20\%, and 30\%
\citep{hong2026dgmark,raban2026lrdwm}.
This study calibrates thresholds separately on each condition's clean
reference, unlike the frozen unedited-clean cutoffs used above.
State recovery measures whether a received block recovers the original
block's watermark state.
Edge recall measures what fraction of the original graph's token-pair
connections also appear in the graph reconstructed from corrupted text.
\par

Detection persists despite incomplete reconstruction.
Under 20\% insertion on Finance QA, edge recall is 41.1\%, while
TPR@5\% reaches 82.5\%. However, robustness differs across datasets:
at 30\% edits, TPR@5\% ranges from 60.0\% to 67.0\% on Finance QA,
compared with 39.5\% to 45.0\% on Multi-News.
These results show that complete graph reconstruction is not necessary
for detection, while the retained detection rate depends on the dataset
and attack.
\par
}
\begingroup
\begin{table}[ht]
\centering
\begingroup

{\fontsize{6.6}{8}\selectfont
\setlength{\tabcolsep}{3pt}
\renewcommand{\arraystretch}{1.05}
\caption{\textbf{Robustness and reconstruction under attacks: Finance QA.}
\methodname{} (Ch) on LLaDA 1.0, using the previous-block shared-MinHash
seed from the original attack run. Detection and recovery values are in
percent, except AUC and mean $-\log_{10}p$.}
\label{tab:minhash-attacks}
\label{tab:attack-chain}
\label{tab:attack-fq-old-chain}
\begin{tabular}{@{}lcccccccc@{}}
\toprule
Attack & Edit (\%) & AUC & \shortstack{Mean\\$-\log_{10}p$} &
TPR@1\% & TPR@5\% & TPR@10\% &
\shortstack{State\\Recon. (\%)} & \shortstack{Edges\\Recon. (\%)} \\
\midrule
Unedited & -- & 0.983 & 5.103 & 64.5 & 91.9 & 94.9 & 93.7 & 94.2 \\
\cmidrule(lr){1-9}
\multirow{3}{*}{Deletion}
& 10 & 0.946 & 3.426 & 77.0 & 85.5 & 86.5 & 56.3 & 55.0 \\
& 20 & 0.902 & 2.652 & 60.5 & 71.0 & 78.5 & 39.4 & 34.3 \\
& 30 & 0.856 & 2.168 & 21.0 & 60.0 & 65.5 & 32.3 & 21.7 \\
\cmidrule(lr){1-9}
\multirow{3}{*}{Insertion}
& 10 & 0.954 & 3.589 & 77.0 & 84.5 & 88.0 & 57.5 & 59.8 \\
& 20 & 0.932 & 3.039 & 75.5 & 82.5 & 84.5 & 42.2 & 41.1 \\
& 30 & 0.884 & 2.506 & 48.5 & 67.0 & 77.5 & 32.3 & 30.4 \\
\cmidrule(lr){1-9}
\multirow{3}{*}{Substitution}
& 10 & 0.951 & 4.218 & 82.0 & 86.0 & 89.5 & 73.7 & 76.5 \\
& 20 & 0.876 & 2.996 & 45.5 & 69.0 & 73.5 & 54.7 & 59.7 \\
& 30 & 0.860 & 2.289 & 51.0 & 62.0 & 70.0 & 45.8 & 45.8 \\
\bottomrule
\end{tabular}
}
\endgroup
\end{table}

\begin{table}[ht]
\centering
\begingroup

{\fontsize{6.6}{8}\selectfont
\setlength{\tabcolsep}{3pt}
\renewcommand{\arraystretch}{1.05}
\caption{\textbf{Robustness and reconstruction under attacks: Multi-News.}
\methodname{} (Ch) on LLaDA 1.0, using the previous-block shared-MinHash
seed from the original attack run. Detection and recovery values are in
percent, except AUC and mean $-\log_{10}p$.}
\label{tab:attack-mn-old-chain}
\begin{tabular}{@{}lcccccccc@{}}
\toprule
Attack & Edit (\%) & AUC & \shortstack{Mean\\$-\log_{10}p$} &
TPR@1\% & TPR@5\% & TPR@10\% &
\shortstack{State\\Recon. (\%)} & \shortstack{Edges\\Recon. (\%)} \\
\midrule
Unedited & -- & 0.963 & 4.693 & 66.2 & 84.3 & 89.9 & 98.3 & 98.5 \\
\cmidrule(lr){1-9}
\multirow{3}{*}{Deletion}
& 10 & 0.889 & 2.322 & 34.5 & 59.0 & 71.5 & 41.9 & 34.2 \\
& 20 & 0.837 & 1.783 & 26.5 & 49.0 & 57.5 & 25.9 & 18.3 \\
& 30 & 0.802 & 1.429 & 23.5 & 39.5 & 59.5 & 20.7 & 11.3 \\
\cmidrule(lr){1-9}
\multirow{3}{*}{Insertion}
& 10 & 0.895 & 2.513 & 34.0 & 61.5 & 72.0 & 40.5 & 37.1 \\
& 20 & 0.827 & 1.971 & 27.0 & 48.5 & 56.5 & 27.5 & 21.3 \\
& 30 & 0.799 & 1.631 & 34.5 & 42.5 & 52.5 & 20.6 & 14.9 \\
\cmidrule(lr){1-9}
\multirow{3}{*}{Substitution}
& 10 & 0.924 & 3.352 & 59.5 & 73.0 & 77.0 & 76.8 & 78.6 \\
& 20 & 0.876 & 2.427 & 36.5 & 63.0 & 71.5 & 61.7 & 61.3 \\
& 30 & 0.824 & 1.636 & 36.5 & 45.0 & 57.5 & 48.6 & 46.1 \\
\bottomrule
\end{tabular}
}
\endgroup
\end{table}
\endgroup

{
Table~\ref{tab:attack-recovery} compares source-aware state agreement and
reference-edge recall for each method against its own unedited reference.
State agreement compares finite-state labels on aligned, originally scored
tokens. Edge recall counts aligned original edges present in the
received-text detector graph.
\mbox{Percentages} are pooled event ratios, and unedited identity is 100\% by
construction.
Original-text alignment is used only for this offline analysis.
\par

Table~\ref{tab:attack-recovery}(a) evaluates block-ready \methodname{} with
\textit{chain}, \textit{balanced tree} and \textit{middle-star} topologies, using the same temperatures
with no tilting.
All three derive their watermark state from the block index and key,
sharing that state across tokens within the block.
\par

Table~\ref{tab:attack-recovery}(b) reports token-commit \methodname{}
with a state derived from the previous block's MinHash.
This configuration is separate from the topology comparison.
Its first generated block is unwatermarked and supplies the MinHash context
for the next block, so state agreement is measured from the second block
onward.
For the block-index-seeded topologies, state agreement includes the first
block.
\par

State recovery also declines alongside token-pair edge recall as edit
rates increase, indicating that edits disrupt both the watermark state
and the scored connections. This motivates reconstruction-aware detection
for structures tied to specific positions, such as \textit{middle-star} and
\textit{balanced tree}, where edits can change the designated center or hierarchical
connections. Such structures may require mechanisms to recover the original connections.
By contrast, token-commit span scores public adjacent pairs without requiring
reconstruction of the private graphs used for per-update optimization or commit set.
\par
}
\begingroup\begin{table}[t]
\centering
\begingroup

{\fontsize{6.6}{8}\selectfont
\setlength{\tabcolsep}{3pt}
\renewcommand{\arraystretch}{1.05}
\caption{\textbf{State agreement and reference-edge recall under attacks:
Finance QA.} (a) Block-ready topologies using block-index states.
(b) Token-commit using previous-block MinHash.
All configurations are untilted. Values are percentages.}
\label{tab:attack-recovery}
\begin{minipage}[t]{0.64\linewidth}
\centering
\vspace{0pt}
\textbf{(a) Block-ready topologies}\\
\textbf{Block-index states}
\par\vspace{2pt}
\begin{tabular}{@{}l*{6}{r}@{}}
\toprule
& \multicolumn{2}{c}{Chain}
& \multicolumn{2}{c}{Balanced tree}
& \multicolumn{2}{c}{Middle-star} \\
\cmidrule(lr){2-3}\cmidrule(lr){4-5}\cmidrule(lr){6-7}
Attack
& State $\uparrow$ & Edge $\uparrow$
& State $\uparrow$ & Edge $\uparrow$
& State $\uparrow$ & Edge $\uparrow$ \\
\midrule
Unedited & 100.0 & 100.0 & 100.0 & 100.0 & 100.0 & 100.0 \\
Deletion 20\% & 39.1 & 52.5 & 40.7 & 10.7 & 42.7 & 1.2 \\
Insertion 20\% & 36.9 & 69.2 & 38.7 & 14.5 & 38.9 & 1.9 \\
Substitution 20\% & 89.6 & 52.9 & 90.1 & 32.4 & 91.4 & 32.6 \\
BERT 20\% & 90.5 & 63.7 & 90.6 & 34.9 & 92.4 & 33.3 \\
Back-translation & 63.8 & 56.3 & 63.1 & 14.4 & 66.6 & 7.3 \\
Rewriting & 56.0 & 29.6 & 57.2 & 7.6 & 57.3 & 2.5 \\
\bottomrule
\end{tabular}
\end{minipage}%
\hfill
\begin{minipage}[t]{0.34\linewidth}
\centering
\vspace{0pt}
\textbf{(b) Token-commit}\\
\textbf{Previous-block MinHash}
\par\vspace{2pt}
\begin{tabular}{@{}lrr@{}}
\toprule
& \multicolumn{2}{c}{Token-commit} \\
\cmidrule(lr){2-3}
Attack & State $\uparrow$ & Edge $\uparrow$ \\
\midrule
Unedited & 100.0 & 100.0 \\
Deletion 20\% & 51.9 & 51.3 \\
Insertion 20\% & 49.2 & 68.2 \\
Substitution 20\% & 61.1 & 54.2 \\
BERT 20\% & 70.5 & 63.0 \\
Back-translation & 54.2 & 55.9 \\
Rewriting & 40.0 & 27.4 \\
\bottomrule
\end{tabular}
\end{minipage}
}
\endgroup
\end{table}
\endgroup

\subsubsection{Topology sensitivity}
\label{sec:attack-topology-sensitivity}

{
Table~\ref{tab:attack-topology} reports detection under the same attacks for
the three block-index-seeded topologies on LLaDA-Instruct.
All three use the same temperatures with no tilting.
Each topology uses its own cutoff targeting 5\% FPR on unedited clean text,
held fixed under attack.
\textit{chain} has the highest observed TPR under each attack, but unedited detection
also differs across topologies.
Under rewriting, TPR falls to 12.0\% for \textit{chain} and 7.0\% for both \textit{balanced tree} and \textit{middle-star}.
\par
}
\begingroup\begin{table}[ht]
\centering
\begingroup

\setlength{\tabcolsep}{3pt}
\renewcommand{\arraystretch}{1.05}
\caption{\textbf{Detection under text attacks across topologies: Finance QA.}
TPR at 5\% FPR, reported in percent.
All topologies use block-index states, without tilting.}
\label{tab:attack-topology}
{\fontsize{6.6}{8}\selectfont
\begin{tabular}{@{}lrrr@{}}
\toprule
Attack & \methodname{} (Ch) $\uparrow$ & \methodname{} (B) $\uparrow$ & \methodname{} (S) $\uparrow$ \\
\midrule
Unedited & 74.0 & 49.5 & 41.5 \\
Deletion 20\% & 33.0 & 13.0 & 9.5 \\
Insertion 20\% & 33.5 & 9.0 & 7.0 \\
Substitution 20\% & 55.0 & 29.5 & 30.0 \\
BERT word substitution 20\% & 54.0 & 32.5 & 22.0 \\
Back-translation & 30.5 & 8.5 & 9.5 \\
Rewriting & 12.0 & 7.0 & 7.0 \\
\bottomrule
\end{tabular}
}
\endgroup
\end{table}
\endgroup

\subsubsection{Transfer to LLaDA2.1}
\label{sec:attack-llada21-transfer}

{
We also evaluate robustness on LLaDA2.1-mini using its native Quality sampler.
Table~\ref{tab:attack-llada21} compares token-commit \methodname{} and
block-ready \methodname{} with \textit{chain}, using
previous-block MinHash at $\delta=0$ and with post-MOT tilt strengths
$\delta=0.25$ and $\delta=0.5$, respectively, with DMark's
predict-bidirectional watermarking under the same attack types.
Each method retains its own generation recipe and cutoff targeting 5\% FPR
on unedited matched-clean text, held fixed under attack.
\par

Unedited quality changes are measured against each method's matched-clean
reference, using the definitions in Table~\ref{tab:attack-quality}.
DMark has higher TPR under every tested attack and a larger unedited CE
increase. Under rewriting, TPR is 19.5\% for token-commit \methodname{},
11.5\% for block-ready \methodname{} with \textit{chain} and 51.0\% for DMark.
Post-MOT tilting raises rewriting TPR from 19.5\% to 29.5\% for token-commit
\methodname{} and from 11.5\% to 24.5\% for \textit{chain}.
The improvement is not uniform: under back-translation, token-commit TPR
decreases from 25.5\% to 24.0\%.
\par
}
\begingroup\begin{table}[ht]
\centering
\begingroup

\setlength{\tabcolsep}{3pt}
\renewcommand{\arraystretch}{1.05}
\caption{\textbf{Detection under text attacks on LLaDA2.1-mini: Finance QA.}
(a) TPR at 5\% FPR, in percent.
(b) Unedited quality changes, reported as $\Delta$ROUGE-L and $\Delta$CE.
Co+tilt and Ch+tilt use post-MOT strengths $\delta=0.25$ and $\delta=0.5$,
respectively.
All \methodname{} variants use previous-block MinHash.}
\label{tab:attack-llada21}
{\fontsize{6.6}{8}\selectfont
\begin{tabular}{@{}lrrrrrrr@{}}
\toprule
Method & Unedited & Deletion & Insertion & Substitution
& \shortstack{BERT word\\substitution}
& \shortstack{Back-\\translation} & Rewriting \\
\midrule
\methodname{} (Co) & 53.5 & 36.5 & 36.5 & 24.5 & 34.5 & 25.5 & 19.5 \\
\methodname{} (Co+tilt) & 66.5 & 50.0 & 51.0 & 32.0 & 47.0 & 24.0 & 29.5 \\
\methodname{} (Ch) & 22.5 & 20.0 & 14.5 & 12.5 & 10.5 & 9.5 & 11.5 \\
\methodname{} (Ch+tilt) & 78.5 & 65.5 & 53.5 & 41.0 & 50.0 & 26.5 & 24.5 \\
DMark & 97.5 & 90.0 & 90.0 & 87.5 & 91.5 & 77.0 & 51.0 \\
\bottomrule
\end{tabular}
\par\vspace{2pt}
\textbf{(a) Detection}
\par\vspace{5pt}
\begin{tabular}{@{}lrr@{}}
\toprule
Method & $\Delta$ROUGE-L $\downarrow$ & $\Delta$CE $\downarrow$ \\
\midrule
\methodname{} (Co) & +0.30 & +0.095 \\
\methodname{} (Co+tilt) & +0.08 & +0.169 \\
\methodname{} (Ch) & +0.10 & +0.070 \\
\methodname{} (Ch+tilt) & +0.30 & +0.349 \\
DMark & +0.13 & +0.386 \\
\bottomrule
\end{tabular}
\par\vspace{2pt}
\textbf{(b) Unedited quality}
}
\endgroup
\end{table}
\endgroup

\subsubsection{Seeding comparison}
\label{sec:attack-seeding-comparison}

{
Table~\ref{tab:attack-seeding-comparison} compares MinHash and block-index
configurations for block-ready \methodname{} with \textit{chain}.
The block-index configurations achieve higher detection rates in most tested
conditions, but also incur larger unedited quality losses.
Deletion results are mixed: block-index seeding performs better on Finance QA,
while MinHash performs better on Multi-News.
\par

Although block-index seeding gives stronger detection in these experiments,
reconstructing its states depends on the original block indices.
That information may be unavailable when a user removes a whole page of earlier
context or submits only an excerpt.
Restarting block numbering at zero can then assign different states to the
retained text.
MinHash instead derives states from the preceding block's content, avoiding
dependence on the original global block index.
It still requires sufficient local context and compatible block boundaries.
\par
}
\begingroup\begin{table}[ht]
\centering
\begingroup

\setlength{\tabcolsep}{2.5pt}
\renewcommand{\arraystretch}{1.05}
\caption{\textbf{Seeding comparison for block-ready \textit{chain} on
LLaDA-Instruct.} All configurations use $\delta=1$.
TPR values are percentages at cutoffs targeting 5\% FPR on each configuration's
unedited clean reference, held fixed under attack, with 200 requested texts
per configuration.
Unedited quality changes use the definitions in Table~\ref{tab:attack-quality}.}
\label{tab:attack-seeding-comparison}
{\fontsize{6.6}{8}\selectfont
\begin{tabular}{@{}llrrrrrr@{}}
\toprule
Dataset & Seeding & \shortstack{Unedited\\TPR}
& \shortstack{Deletion\\20\%}
& \shortstack{Substitution\\20\%}
& Rewriting & $\Delta$ROUGE-L $\downarrow$ & $\Delta$CE $\downarrow$ \\
\midrule
Finance QA & MinHash & 94.0 & 63.0 & 58.5 & 10.5 & +0.41 & +0.584 \\
Finance QA & Block index & 95.5 & 74.5 & 95.5 & 20.5 & +0.87 & +0.698 \\
\midrule
Multi-News & MinHash & 88.5 & 75.5 & 83.5 & 9.5 & +1.94 & +0.619 \\
Multi-News & Block index & 98.0 & 70.0 & 97.5 & 18.0 & +2.36 & +0.835 \\
\bottomrule
\end{tabular}
}
\endgroup
\end{table}
\endgroup

\subsection{Regularization tradeoff}
\label{sec:appendix-regularization-tradeoff}

\begin{figure}[htb]

\centering
\definecolor{epsilonBlue}{HTML}{3A9AD9}
\definecolor{epsilonMuted}{HTML}{687783}
\begin{tikzpicture}
\begin{axis}[
    width=0.78\linewidth,
    height=0.36\linewidth,
    font=\small,
    tick label style={font=\scriptsize},
    xlabel={Mean conditional token KL (nats) $\downarrow$},
    ylabel={Mean detector score $z$ $\uparrow$},
    xmin=-0.012, xmax=0.56,
    ymin=-0.12, ymax=4.3,
    xtick={0,0.1,0.2,0.3,0.4,0.5},
    ytick={0,1,2,3,4},
    scaled ticks=false,
    grid=major,
    major grid style={gray!22},
    axis background/.style={fill=white},
    tick align=outside,
    legend columns=2,
    legend style={
        at={(0.5,1.04)}, anchor=south,
        font=\scriptsize, draw=none, fill=none,
        inner sep=0pt, cells={anchor=west}
    }
]
\addplot+[
    restrict expr to domain={\coordindex}{1:5},
    color=epsilonBlue, dashed, line width=0.8pt,
    mark=*, mark size=1.8pt,
    mark options={solid,fill=epsilonBlue},
    point meta=explicit symbolic,
    nodes near coords={$\epsilon=\pgfplotspointmeta$},
    every node near coord/.append style={
        font=\scriptsize,text=black,anchor=south west,xshift=3pt,yshift=2pt
    },
    error bars/.cd,
    y dir=both, y explicit,
    error bar style={solid,line width=0.5pt},
    error mark=|,
    error mark options={rotate=0,mark size=1.8pt,line width=0.5pt}
] table[
    col sep=comma,
    x=mean_token_kl_nats, y=mean_z, meta=epsilon,
    y error minus expr={\thisrow{mean_z}-\thisrow{z_mc_ci95_low}},
    y error plus expr={\thisrow{z_mc_ci95_high}-\thisrow{mean_z}}
] {data/epsilon_z_kl.csv};
\addlegendentry{Converged solver runs}
\addplot+[
    restrict expr to domain={\coordindex}{0:0},
    only marks, color=epsilonMuted,
    mark=x, mark size=2.8pt, line width=0.8pt,
    point meta=explicit symbolic,
    nodes near coords={$\epsilon=\pgfplotspointmeta$},
    every node near coord/.append style={
        font=\scriptsize,text=epsilonMuted,anchor=north west,xshift=3pt,yshift=-5pt
    },
    error bars/.cd,
    y dir=both, y explicit,
    error bar style={line width=0.5pt},
    error mark=|,
    error mark options={rotate=0,mark size=1.8pt,line width=0.5pt}
] table[
    col sep=comma,
    x=mean_token_kl_nats, y=mean_z, meta=epsilon,
    y error minus expr={\thisrow{mean_z}-\thisrow{z_mc_ci95_low}},
    y error plus expr={\thisrow{z_mc_ci95_high}-\thisrow{mean_z}}
] {data/epsilon_z_kl.csv};
\addlegendentry{Unconverged}
\addplot[only marks,mark=square*,mark size=2pt,color=black,forget plot]
  coordinates {(0,0)};
\node[font=\scriptsize,anchor=south west,xshift=30pt,yshift=1pt]
  at (axis cs:0,0) {Input reference};
\end{axis}
\end{tikzpicture}
\caption{\textbf{Regularization, distribution shift, and detector score.}
\methodname \textit{chain} on one outcome-selected 32-token LCC block from
LLaDA 1.0-Instruct.
Conditional token KL is averaged over states and positions.
The cross marks an unconverged solver run.}
\label{fig:epsilon-z-kl-tradeoff}
\end{figure}
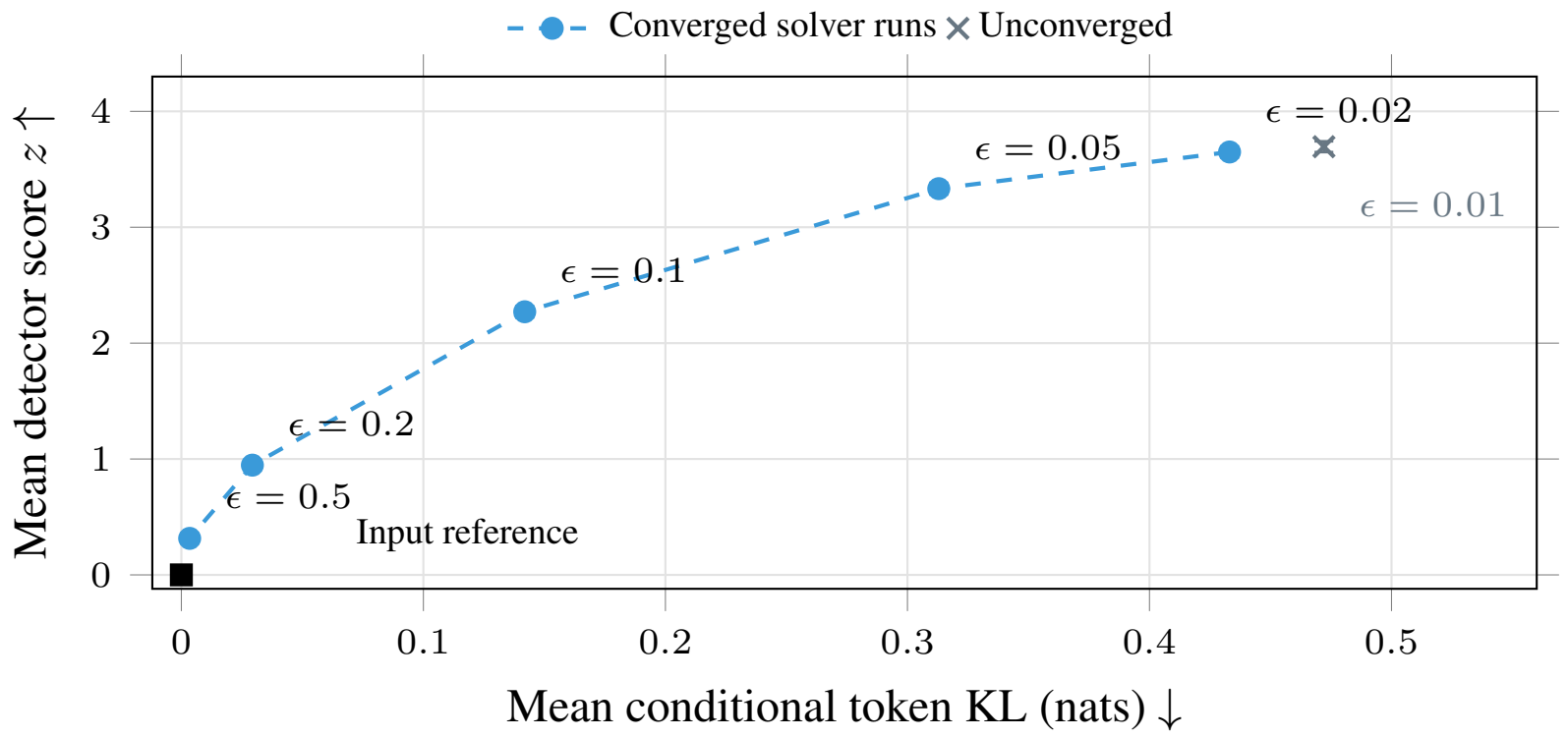

Figure~\ref{fig:epsilon-z-kl-tradeoff} illustrates how entropy regularization
affects conditional token distributions and detector scores on one fixed block.
Among converged solver runs, increasing $\epsilon$ from $0.02$ to $0.5$ reduces mean
conditional token KL from $0.433$ to $0.0034$ nats and mean detector score $z$
from $3.649$ to $0.315$.
Stronger regularization therefore keeps the state-conditioned token
distributions closer to the input marginals, but weakens the detection signal
on this block.
The $\epsilon=0.01$ solver run did not converge within the iteration budget
and is shown separately.
These measurements characterize a single selected block, rather than text
quality or detection rates across a dataset.

\subsection{Runtime cost}
\label{sec:results-cost}

{
Runtime measurements quantify the computational cost of adding watermarking
to LLaDA-Instruct generation on LCC.
Timings cover synchronized generation, MOT optimization and token sampling on NVIDIA
RTX 6000 Ada GPUs, excluding model loading, warmup, logging, I/O and
external scoring.
All methods use the same request-local KV-cache procedure.
Relative runtime is calculated against a common reference obtained by
averaging the native-clean timings across configurations.
\par
}

\begin{figure}[ht]

\centering
\includegraphics[width=0.82\linewidth]{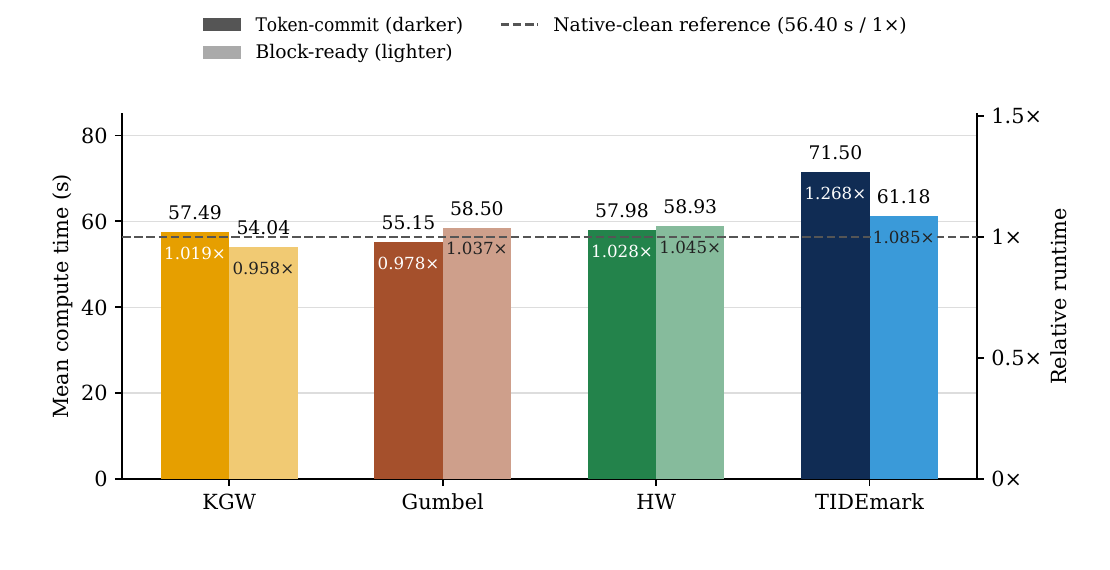}
\caption{\textbf{Generation compute time on LCC.}
Darker and lighter bars show token-commit and block-ready timings.
Labels give seconds above each bar and relative runtime inside.
The dashed line marks the native-clean reference.}
\label{fig:runtime-compute}
\end{figure}

{
Figure~\ref{fig:runtime-compute} reports relative runtimes of
$1.268\times$ for token-commit \methodname{} and
$1.085\times$ for block-ready \textit{chain}.
Block-ready insertion reduces the frequency of watermarking calls by
operating once per completed block rather than at each denoising update.
KGW, Gumbel and HW range from $0.958\times$ to $1.045\times$ the common reference.
The reference combines controls measured across different devices
and generation configurations, so these ratios describe compute time
relative to that reference.
Ratios below one do not establish that watermarking accelerates generation.
\par
}

\FloatBarrier
\subsection{Limitations and Scope}
\label{sec:limitations}

{
The distortion-free guarantee applies to each recorded token
distribution under the ideal coupling, after averaging over the secret side
information. It does not establish marginal preservation during adaptive
token-commit generation, preservation of the full text distribution, or
unchanged task performance. The practical implementation uses restricted
candidate sets and a finite solver iteration budget, so its numerical accuracy must
be assessed separately.
\par

Block-ready watermarking waits for the entire block to finish decoding before
applying the watermark and releasing its tokens. In revision-based diffusion
models, tokens can still change during decoding, so waiting for the block to
stabilize is part of the generation process rather than a requirement
introduced by watermarking. Watermarking nevertheless adds processing time
after block completion and before release.
\par

Solving MOT and sampling watermarked tokens also add computational cost.
Token-commit \methodname{} solves MOT repeatedly during denoising and takes
$1.268\times$ the native-clean reference.
Block-ready \textit{chain} solves MOT once per completed block and takes $1.085\times$,
compared with $1.045\times$ for block-ready HW and $1.037\times$ for
block-ready Gumbel.
These measurements use LLaDA-Instruct on LCC and the common reference
described in Appendix~\ref{sec:results-cost} and
Figure~\ref{fig:runtime-compute}.
\par
}

\end{document}